\documentclass[3p,anonymous]{elsarticle}
\usepackage{amsmath}
\usepackage{amssymb}
\usepackage{url}
\usepackage{amsthm}

\usepackage{booktabs,tabularx}

\usepackage[mode=text, math-rm=\text, detect-all, binary-units=true, per-mode=symbol, range-phrase=\,--\,, range-units=single, detect-weight=true, detect-family=true]{siunitx}
\DeclareSIUnit{\bit}{bit}
\DeclareSIUnit{\bps}{\bit\per\s}

\usepackage{tikz}
\usepackage{pgfplots}
\pgfplotsset{compat=1.17}
\usetikzlibrary{calc,arrows.meta}
\usepackage{subcaption}
\usepackage{dblfloatfix} 

\usepackage{csquotes}

\graphicspath{{figures}}

\usepackage[nice]{nicefrac} 
\usepackage{xcolor}         
\usepackage{xifthen}
\usepackage{mathtools}
\usepackage{multicol}
\usepackage{placeins}

\usepackage{paralist}

\usepackage[capitalize]{cleveref}

\crefname{section}{\S}{Sections}
\Crefname{section}{Section}{Sections}
\crefformat{section}{\S#2#1#3}
\Crefformat{section}{Section~#2#1#3}
\crefformat{footnote}{#2\footnotemark[#1]#3}
\def\appendixname{Appendix}

\usepackage{enumitem}       
\newcommand{\enumfont}{\bfseries\sffamily}
\SetEnumitemKey{myenum}{
}
\newlist{benenum}{enumerate}{1}
\setlist[benenum]{myenum, label={\enumfont B\arabic*}, ref=B\arabic*}
\crefname{benenumi}{benefit}{benefits}

\usepackage[normalem]{ulem}
\newcommand{\gennote}[5][blue]{{\color{#1}%
		$\rule{8pt}{8pt}_\textsf{\bfseries #2}^\textsf{\bfseries #3}$
		\textcolor{gray}{\emph{\sout{#4}}}#5}%
}

\newcommand{\authorcomment}[3]{%
	\expandafter\newcommand\csname#1\endcsname[1]{\gennote[#3]{#2}{}{}{##1}}%
	\expandafter\newcommand\csname#1S\endcsname[2][]{\gennote[#3]{#2}{}{##1}{##2}}%
	\expandafter\newcommand\csname#1Q\endcsname[1]{\gennote[#3]{#2}{Question}{}{##1}}%
	\expandafter\newcommand\csname#1N\endcsname[1]{\gennote[#3]{#2}{Note}{}{##1}}%
	\expandafter\newcommand\csname#1C\endcsname[1]{\gennote[#3]{#2}{Comment}{}{##1}}%
	\expandafter\newcommand\csname#1T\endcsname[1]{\gennote[#3]{#2}{TODO}{}{##1}}%
}
\authorcomment{stefan}{StS}{blue}
\authorcomment{simon}{SiS}{orange}
\authorcomment{adrian}{AP}{red}
\authorcomment{ml}{ML}{purple}

\newtheorem{theorem}{Theorem}
\newtheorem{definition}{Definition}

\newcommand*\circled[1]{\tikz[baseline=(char.base)]{
            \node[shape=circle,draw,inner sep=0pt] (char) {#1};}}

\makeatletter
\def\ps@pprintTitle{%
  \let\@oddhead\@empty
  \let\@evenhead\@empty
  \let\@oddfoot\@empty
  \let\@evenfoot\@oddfoot
}
\makeatother

\usepackage{scalerel}[2016/12/29]
\newcommand\Oslash{\scaleobj{0.7}{\boldsymbol{\oslash}}}

\begin{document}

\title{Quality Competition Among Internet Service Providers}

\author[add1]{Simon Scherrer}
\author[add1]{Seyedali Tabaeiaghdaei}
\author[add1]{Adrian Perrig}
\address[add1]{ETH Zurich}



\begin{abstract}

Internet service providers (ISPs) have a variety of quality attributes 
that determine their attractiveness for data transmission, ranging from
quality-of-service metrics such as jitter to security properties
such as the presence of DDoS defense systems.
ISPs should optimize these attributes in line with their profit objective,
i.e., maximize revenue from attracted traffic
while minimizing attribute-related cost, all in the context of alternative offers
by competing ISPs. 
However, this attribute optimization is difficult not least
because many aspects of ISP competition are barely understood
on a systematic level, e.g., the multi-dimensional and cost-driving 
nature of path quality, and the distributed decision making
of ISPs on the same path.

In this paper, we improve this understanding by analyzing how
ISP competition affects path quality and ISP profits. 
To that end, we develop a game-theoretic model
in which ISPs (i) affect path quality via multiple attributes that entail costs,
(ii) are on paths together with other selfish ISPs,
and (iii) are in competition with alternative paths when attracting traffic.
The model enables an extensive theoretical analysis, surprisingly showing 
that competition can have both positive and negative effects
on path quality and ISP profits, depending on the network topology and
the cost structure of ISPs. However, a large-scale simulation, 
which draws on real-world data to instantiate the model,
shows that the positive effects will likely prevail in practice:
If the number of selectable paths towards any destination increases 
from 1 to 5, the prevalence of quality attributes increases
by at least 50\%, while 75\% of ISPs improve their profit.
\end{abstract}


\maketitle

\section{Introduction}
\label{sec:intro}

In today's Internet, the Border Gateway Protocol (BGP)
supplies ISPs with potentially multiple paths
towards an IP prefix. When selecting among these paths,
ISPs decide on the basis of price and quality of the available paths.
This path quality is determined by multiple \emph{quality attributes}
of potentially multiple on-path ISPs.
Such quality attributes may include conventional performance metrics 
(e.g., bandwidth, latency, loss rate, jitter) or security features 
(e.g., presence of security middleboxes),
but also properties that traditionally receive less attention, e.g., 
environmental, social, and governmental (ESG) properties such as
carbon emission from data transmission~\cite{zilberman2022toward} or 
geopolitical concerns regarding on-path ISPs~\cite{davidson2022tango}.
Transit ISPs invest in their attributes,
communicate them in path announcements,
and thereby attract traffic from selecting ISPs. However, 
improving these attributes comes at a cost, which may exceed 
the additional revenue from attracted traffic,
especially if ISPs on competing paths also
raise their quality level.

Given this competitive setting, ISPs today face
complex strategic questions when optimizing profit:
What quality attributes
should be invested in, and to what extent?
How should prices be determined?
And how are these decisions affected by ISPs on competing paths \emph{and} 
ISPs elsewhere on the provided paths?

Well-informed strategic decisions thus require 
a fundamental understanding of
ISP competition under path selection,
not only on an intuitive, but also on a rigorous
analytical level. 
While such an understanding has been 
furthered by previous academic 
research~\cite{marentes2015exploring,nagurney2014cournot,shakkottai2006economics,teixeira2017economic,zhang2010interactions}, 
many open questions of practical relevance 
remain, e.g.,
regarding the multi-attribute nature of path quality,
the dependence of fixed and variable ISP cost 
on provided quality, the feasibility of
cooperation among ISPs on the same path,
and the impact of 
differing degrees of competition intensity
(cf.~\cref{sec:related-work}).

To address these questions, we present a new game-theoretic model, 
enabling a rigorous investigation of quality competition among ISPs. 
We perform this investigation through theoretical analysis and simulation:

\paragraph{Theoretical analysis}
We conduct an extensive theoretical analysis to systematically understand
the effect of ISP competition on path quality, and ISP profits
(Path price constitutes a quality attribute in our model).
In particular, we identify closed-form solutions for the Nash equilibria
of the competition dynamics, prove the stability of these equilibria,
and contrast them for varying degrees of path diversity and ISP heterogeneity.
On the one hand, this theoretical analysis confirms intuitive insights,
namely that competition tends to raise
the prevalence of valuable attributes. 
On the other hand, our model reveals counter-intuitive insights, 
namely that the cooperation between ISPs on the same path suffers
from a prisoner's dilemma, that ISP profits can increase under intensified competition, 
and that additional paths may decrease the 
prevalence of quality attributes if unchangeable path attributes are starkly different.

\paragraph{Simulation-based case study}
To determine which competition effects are significant in practice,
we leverage our model for a simulation-based case study.
In this case study, we investigate the competition dynamics in the Internet core
with respect to two attributes (internal bandwidth and the share of clean
energy used by an ISP).
This simulation requires a numerical instantiation of the model,
based on real-world data.
For this model instance, our simulations yield robust evidence
that competition raises the prevalence of valuable attributes,
the quality of available paths, and the profits of most ISPs. 

\medskip
In summary, our work includes the following contributions:

\begin{compactitem}
    \item \textbf{Game-theoretic competition model:}
    Our new ISP-competition model (\cref{sec:model}) departs from 
    previous competition models by representing
    both inter-path competition and intra-path cooperation,
    accommodating a multi-faceted notion of path quality,
    revealing the effect of path diversity, and 
    reflecting realistic ISP cost structures (\cref{sec:related-work}).
    
    \item \textbf{Theoretical analysis:}
    We conduct a rigorous theoretical
    analysis by reasoning from
    basic competition scenarios that showcase the fundamental
    effects in ISP competition (\cref{sec:theo}).
    In particular, we contrast monopolistic
    and competitive scenarios in ISP path selection,
    investigate networks with varying similarity in
    ISP profit functions, and identify asymptotically stable
    equilibria and social optima of the competition. 
    Our analysis suggests that ISP competition
    has nuanced effects on ISP profits and path quality,
    going beyond the predictions of basic economic theory. 
    
    \item \textbf{Large-scale simulation:} 
    We demonstrate how to instantiate our model
    based on real-world data, with the goal of predicting
    competition effects in the Internet core (\cref{sec:example}).
    These predictions are generated with simulations, which
    rely on randomization to achieve robust results,
    represent the competition behavior with better-response
    dynamics, and are executed for varying path diversity.
    The simulation results suggest that competition, 
    induced by path diversity, has positive effects for a majority
    of ISPs on multiple tiers of the Internet, i.e., raises ISP profits and
    path quality (\cref{sec:simulation}).
\end{compactitem}

\section{Model and First Insights}
\label{sec:model}

In the following, we present a game-theoretic model,
which we employ to investigate the competition dynamics
under attribute-oriented path selection.
While our model reflects common characterizations
of inter-domain network economics, it is more general
than previous models (cf.~\cref{sec:related-work}).

\paragraph{Network and paths} We abstract the network as a set~$N$ of
ISPs, which represent the players in the competition game.
Each ISP~$n \in N$ is assumed to be fully rational. 
The ISPs form paths, where each path~$r \subseteq N$ is a set of ISPs.
All usable paths in a network are collected in the path set~$R$,
and all usable paths between selecting ISP~$n_1$ and destination ISP~$n_2$
constitute the set~$R(n_1,n_2)$. 
Throughout this work, we study how ISPs affect the quality of 
paths as given by path set~$R$, not how
ISPs strategically adapt the set~$R$ of usable paths 
via interconnection agreements and announcements, 
which is a related but distinct problem~\cite{meirom2014network,scherrer2021enabling}.

\paragraph{Attributes} We consider a network with a set~$K$ of
ISP attributes, $|K| \geq 1$, that are relevant in path selection.
Hence, each ISP~$n$ is associated with an attribute vector~$\mathbf{a}_n \in \mathbb{R}_{\geq0}^K$,
where~$a_{nk} \in \mathbb{R}_{\geq0}$ denotes the prevalence of attribute~$k \in K$
in ISP~$n$. As a player in the competition game, each ISP~$n$ strives to choose
its attributes~$\mathbf{a}_n$ in order to optimize its profit (see below).
Since the lowest possible degree of attribute prevalence is attained
if an ISP does not possess an attribute at all, we restrict the attribute 
values to non-negative real numbers: $\forall n\in N, \forall k \in K.\ a_{nk} \geq 0$.
For convenience of notation, we also define an attribute matrix~$\mathbf{A} \in \mathbb{R}_{\geq0}^{|N|\times |K|}$,
with the~$n$-th row being~$\mathbf{a}_n$.

\paragraph{Path valuations} The attributes of an ISP~$n$ determine the attractiveness 
of using paths including that ISP. Hence, we define the attractiveness
of available options on the level of paths, specifically by \emph{valuation 
functions}~$\{v_r\}_{r\in R}$. The valuation~$v_r$ for path~$r$ then depends on all 
attributes~$\mathbf{a}_n$ of all on-path ISPs~$n \in r$.
Since we consider desirable attributes in our model, every function 
that is monotonically increasing in all attribute values
is a suitable valuation function. Throughout this paper, we use affine
functions:
\begin{equation}
    v_r(\mathbf{A}) = \sum_{n\in r} \sum_{k \in K} \alpha_{rnk} a_{nk} + \alpha_{r0},
    \label{eq:model:valuation}
\end{equation} where each~$\alpha_{rnk} > 0$ determines how strongly attribute~$k$ of
ISP~$n$ affects the valuation of path~$r$, and~$\alpha_{r0} \geq 0$ is the \emph{base valuation}
of path~$r$. This formulation captures several real-world aspects of path valuations,
as not all attributes are equally important and not all ISPs on a path
equally affect the valuation, e.g., ISPs providing a large segment of the path 
might be more relevant for the valuation.
The linear formulation might be counter-intuitive given
that the marginal utility of attribute prevalence 
is likely decreasing; 
we rely on the formulation for path-selection probability 
below to capture that
the volume of attracted demand on a path is sub-linear in path attributes.
Moreover, we show by simulation that the model predictions
do not strongly rely on the affine formulation (cf.~\cref{sec:simulation:results:nonlinearity}).

\paragraph{Path-selection probability} Path valuations inform the path selection at the selecting ISP, and thus determine the probability of each path being selected.
More precisely, when a selecting ISP~$n_1$ selects a
path towards a prefix hosted by ISP~$n_2$,
each path~$r$ among the available paths~$R(n_1, n_2)$
is selected for transit with probability~$p_r(\mathbf{A})$.
Inspired by the popular
logit-demand model~\cite{besanko1998logit}, 
we consider the selection probability~$p_r$ to
be proportional to the \emph{relative attractiveness} of path~$r$ 
compared to alternative paths:
\begin{equation}
    \forall (n_1, n_2) \in N \times N,\ r \in R(n_1, n_2).\quad p_r(\mathbf{A}) = \frac{v_r(\mathbf{A})}{1 + \sum_{r'\in R\left(n_1, n_2\right)} v_{r'}(\mathbf{A})}.
    \label{eq:model:demand:probability}
\end{equation}
Crucially, the addition term~1 in the fraction denominator captures
\emph{demand elasticity}, i.e., selecting ISP~$n_1$ 
might not select any path in~$R(n_1, n_2)$ at all
if the available paths are generally unattractive.
Instead, selecting ISP~$n_1$ might not offer its customers
any path to~$n_2$, create a new path to~$n_2$ by concluding a peering agreement, 
or obtain the desired content from another destination 
ISP than~$n_2$.

\paragraph{Demand} The path-selection probabilities above
determine the demand volume~$D_n$ that is obtained
by any ISP~$n$, which we formalize by:
\begin{equation}
    D_n(\mathbf{A}) = \sum_{r \in R.\ n \in r} \delta_r(\mathbf{A})
    = \sum_{\substack{r \in R.\ n \in r\\r \in R(n_1, n_2)}} p_r(\mathbf{A}) \cdot d_{(n_1, n_2)}.
    \label{eq:model:demand}
\end{equation}
Due to the elasticity of demand, actual total demand 
(i.e., $\delta_{r}(\mathbf{A})$ summed over all~$r\in R\left(n_1,n_2\right)$)
is strictly below the \emph{demand limit}~$d_{(n_1,n_2)}$.

The practical interpretation of path demand~$\delta_r$
in~\cref{eq:model:demand} depends on the transit behavior 
of the selecting ISP~$n_1$. 
If~$n_1$ is a stub AS, then traffic originates within~$n_1$
and can be split across multiple paths towards a given prefix.
If~$n_1$ thus selects multiple paths, 
$\delta_r$ denotes the actual demand allocated to path~$r$. 
In contrast, if ISP~$n_1$ is a transit AS,
the traffic transited by~$n_1$ must follow 
the single path announced by ISP~$n_1$ to neighboring ISPs,
as BGP transit loops might arise otherwise.
If~$n_1$ thus selects only a single path for transit,
$\delta_r$ corresponds to the \emph{expected} demand
allocated on path~$r$.

\paragraph{Profit} Given the demand model, an ISP~$n$ can affect attracted demand~$D_n$ with an 
appropriate choice of~$\mathbf{a}_n$. However, the profit~$\pi_n$ depends not
only on the volume of attracted demand, but also on the cost for provision of the
attributes. We thus consider the profit function~$\pi_n$
of ISP~$n$ to have three components. First, the ISP profit is increased by a revenue function~$\mathcal{R}_n$,
which is a function of~$D_n$. Second, the profit is reduced by a demand-dependent cost
function~$\Phi_n$, which as well depends on~$D_n$, but also directly on~$\mathbf{a}_n$, as
the cost of transmitting a unit of demand depends on the chosen attributes.
Third, the profit is reduced by a demand-independent cost function~$\Gamma_n$, which depends
only directly on the chosen attributes~$\mathbf{a}_n$ and thus corresponds to the `fixed cost'
of possessing certain attributes.

While these component functions could in principle be any monotonically increasing
function, we use the following formulations for the component functions throughout
this paper:
\begin{align}
    \mathcal{R}_n(\mathbf{A}) = \rho_n \cdot D_n(\mathbf{A}) \hspace{17pt}
    \Phi_n(\mathbf{A}) = \left(\sum_{k\in K} \phi_{nk}a_{nk} + \phi_{n0}\right) \cdot D_n(\mathbf{A}) \hspace{17pt}
    \Gamma_n(\mathbf{a}_n) = \sum_{k \in K} \gamma_{nk}a_{nk} + \gamma_{n0}
    \label{eq:model:revenue-cost}
\end{align} where the coefficient~$\rho_n > 0$ corresponds to the per-unit base revenue
of ISP~$n$ (transit prices are additionally subsumed within quality
attributes, as described below). 
Furthermore, the coefficients~$\phi_{nk} \geq 0$ and~$\gamma_{nk} \geq 0$
determine the attribute-specific increase in demand-dependent and demand-independent cost, respectively,
and the intercepts~$\phi_{n0} \geq 0$ and~$\gamma_{n0} \geq 0$ express the attribute-independent
basic values for the respective cost terms. Throughout the paper, we assume~$\rho_n \geq \phi_{n0}$ for all
ISPs~$n \in N$, as a rational ISP that loses money by attracting demand even with the most cost-saving
strategy~(i.e., $\mathbf{a}_n = \boldsymbol{0}$) would in fact go out of business.
The affine formulations of~$\Phi_n$ and~$\Gamma_n$ predict qualitatively similar competition effects
as quadratic functions,
as we demonstrate by simulation in~\cref{sec:simulation:results:nonlinearity}.

In summary, the profit function~$\pi_n$ is of the following form in our investigations:
\begin{equation}
    \pi_n(\mathbf{A}) = \mathcal{R}_n(\mathbf{A}) - \Phi_n(\mathbf{A}) - \Gamma_n(\mathbf{a}_n) = D_n(\mathbf{A}) \cdot \left(\rho_n - \sum_{k\in K} \phi_{nk}a_{nk} - \phi_{n0}\right) - \sum_{k \in K} \gamma_{nk}a_{nk} - \gamma_{n0}.
    \label{eq:model:profit}
\end{equation}

\paragraph{Undesirable attributes}
So far, our model formulation
assumes \emph{desirable} attributes, which
increase path attractiveness in high quantities
and are costly to increase, 
e.g., bandwidth. However, many relevant ISP attributes
are \emph{undesirable} in 
high quantities and are challenging to decrease, 
e.g., transit price or latency.
To accommodate undesirable attributes
in the model, a naive approach would consist of
allowing negative coefficients~$\alpha_{rnk}$, $\phi_{nk}$,
and~$\gamma_{nk}$ for any undesirable attribute~$k$.
However, such negativity would entail nonsensical
model predictions, such as potentially negative
path-selection probabilities from~\cref{eq:model:demand:probability},
or infinite profit given an undesirable
attribute~$(n,k)$ with~$a_{nk} = \infty$ and~$\gamma_{nk} < 0$
(cf.~\cref{eq:model:profit}).

To avoid such nonsensical predictions and preserve model tractability, 
we suggest to convert undesirable attributes into their desirable counterparts.
For example, the actual transit-price attribute~$a_{nk}' \in [0, P^{\max}_n]$
could be translated to a non-negative 
\emph{cheapness} attribute~$a_{nk} = P^{\max}_{n} - a_{nk}' \in [0, P^{\max}_{n}]$,
where~$P^{\max}_{n}$ is the maximum price that ISP~$n$ can
reasonably ask. The cheapness attribute 
then formally contributes to the costs in~$\Phi$ and~$\Gamma$, 
while actually quantifying foregone revenue in the profit function~$\pi$.

\paragraph{Nash Equilibrium} The competition dynamics in attribute-oriented
path selection can be characterized by their Nash equilibria.
In our setting, such a Nash equilibrium is a choice of attributes
in which each ISP has optimal attributes (w.r.t. its profit)
given the attributes of all other ISPs:
\begin{definition}
    A choice of attributes~$\mathbf{A}^+$
form a \textbf{Nash equilibrium} if and only if
\begin{equation}
    \forall n\in N.\ \mathbf{a}_n^+ = {\arg\max}_{\mathbf{a}_n \in \mathbb{R}_{\geq 0}^{|K|}}\ \pi_n\big(\mathbf{a}_n \oplus_n \mathbf{A}^+_{-n}\big)
    \label{eq:model:equilibrium}
\end{equation} where~$\oplus_n$ combines the attribute
choice~$\mathbf{a}_n$ of ISP~$n$ with the equilibrium attribute choices ~$\mathbf{A}^+_{-n}$
of the remaining ISPs.
\label{def:eq}
\end{definition}

In this abstract form, the Nash equilibria offer little opportunity for analytical characterization.
However, if the attributes~$a_{nk}$ are restricted to~$[0, a_{\max}]$ rather than to~$\mathbb{R}_{\geq0}$
(e.g., if there is an upper bound~$a_{\max}$ on all attribute values),
the existence of Nash equilibria is guaranteed by Brouwer's fixed-point theorem~\cite{brouwer1910abbildungen}. 
To gain a deeper understanding of Nash equilibria beyond that special case, we concretize equilibria in this work, and investigate these equilibria with respect
to existence, uniqueness, stability, and efficiency.

\paragraph{Social Optimum} To assess the efficiency of Nash equilibria,
we compare these equilibria to \emph{social optima}.
In our setting, such a social optimum optimizes a metric that aggregates the perspectives of
all agents involved in the competitive dynamics.
Our model contains two types of agents, namely \emph{selecting ISPs} and \emph{transit ISPs},
with non-aligned interests, which warrants two different formalizations
of the social optimum.

First, selecting ISPs are interested in path quality.
Hence, the social efficiency for selecting ISPs
is simply measured by the aggregate valuation~$V$ of all paths
in the network, given a choice of attributes~$\mathbf{A}$:
\begin{equation}
    V(\mathbf{A}) = \sum_{r \in R} v_r(\mathbf{A}).
\end{equation}
Since the valuation functions~$v_r$ are assumed to be 
linear and therefore unbounded in this paper, a finite social optimum
for selecting ISPs only exists if all attributes
are restricted to a finite domain.

Second, transit ISPs are interested in profit.
To characterize the social optimum from the perspective
of transit ISPs, we rely on the conditions
of the \emph{Nash bargaining solution (NBS)}, i.e.,
the conditions that a global attribute choice~$\mathbf{A}$
would have to fulfill if ISPs had to agree on it in
cooperative bargaining~\cite{nash1950bargaining}.
The two most important NBS conditions are
\emph{Pareto-optimality}, i.e., no ISP
can increase its profit without any other ISP
experiencing a decrease in its profit,
and \emph{symmetry}, i.e., among
Pareto-optimal profit distributions, 
the fairest distribution is preferred.
These conditions are achieved if the
attribute choice~$\mathbf{A}$ optimizes
the \emph{Nash bargaining product}:

\begin{definition}
    A choice of attributes~$\mathbf{A}^{\circ}$
forms a \textbf{social optimum} from the perspective of transit ISPs 
if it corresponds to the \textbf{Nash bargaining solution (NBS)}, i.e.,
\begin{equation}
    \mathbf{A}^{\circ} = {\arg\max}_{\mathbf{A} \in \mathbb{R}_{\geq 0}^{|N|\times |K|}}\ \Pi_{n\in N}\ \pi_n(\mathbf{A}).
    \label{eq:model:equilibrium}
\end{equation}
\label{def:eq}
\end{definition}
\section{Theoretical Analysis}
\label{sec:theo}

In this section, we theoretically analyze the competition dynamics
in path selection.
For that purpose, we focus on an individual \emph{market} in isolation, 
i.e., the competition between transit ISPs for traffic
between a single source-destination pair~$(n_1, n_2)$.
As a result, we write~$R = R(n_1, n_2)$ and~$d = d_{(n_1, n_2)}$ throughout this section.


\subsection{Optimal Attribute}
\label{sec:theo:sapp}

To analyze the competition dynamics, we first investigate
how any single ISP~$n$ should choose its attribute~$k$ 
in response to the attribute choices of all other ISPs.
This optimal attribute is given by the following
closed-form solution:

\begin{theorem}
    \textbf{Best-Response Attribute.}
    In an individual market with arbitrarily overlapping paths, 
    the optimal admissible attribute~$a_{nk}^{\ast}$ of ISP~$n$
given the remaining attributes~$\mathbf{A}_{-nk}$ is
    \begin{equation}
        a^{\ast}_{nk}(\mathbf{A}_{-nk}) = \begin{cases}
            \hat{a}_n^{\ast}(\mathbf{A}_{-nk}) & \text{if } \hat{a}_n^{\ast}(\mathbf{A}_{-nk}) \in \mathbb{R} \text{ and }
            \hat{a}_n^{\ast}(\mathbf{A}_{-nk}) \geq 0,\\
            0 & \text{otherwise,}
        \end{cases}
        \label{eq:theo:a_opt:restricted}
    \end{equation}
where~$\hat{a}^{\ast}_{nk}(\mathbf{A}_{-nk})$ is the optimal unrestricted (i.e., potentially negative) attribute:
\begin{equation}
    \begin{split}
    \hat{a}_{nk}^{\ast}(\mathbf{A}_{-nk}) =\ \frac{1}{\alpha_{nk}} \Bigg( &\sqrt{\frac{d\big(1+v_{-r(n)}(\mathbf{A}_{-nk})\big)}{d\phi_{nk}+\gamma_{nk}} \big(\phi_{nk} (1+v_{-nk}(\mathbf{A}_{-nk})) + \alpha_{nk} (\rho_n - \Phi_{-nk}(\mathbf{A}_{-nk})) \big) }\\
    &-\big(1+v_{-nk}(\mathbf{A}_{-n})\big)\Bigg)
    \label{eq:theo:a_opt}
    \end{split}
\end{equation}
\cref{eq:theo:a_opt} uses the following abbreviations:
\begin{flalign}
    &\hspace{52pt} \alpha_{nk} = \sum_{\substack{r \in R.\\n \in r}} \alpha_{rnk} \hspace{30pt} & v_{-r(n)}(\mathbf{A}) &= \sum_{\substack{r'\in R.\\n \notin r'}} v_{r'}(\mathbf{A})&\\
    &\hspace{30pt}v_{-nk}(\mathbf{A}) = \sum_{r \in R} \alpha_{r0} + \sum_{\substack{(n',k') \in N\times K.\\(n',k')\neq(n,k)}} \alpha_{n'k'}a_{n'k'} \hspace{10pt}
    &\Phi_{-nk}(\mathbf{A}) &= \sum_{k' \in K\setminus k} \phi_{nk'} a_{nk'} + \phi_{n0}&
    \label{eq:theo:sapp:shorthands}
\end{flalign}
    \label{thm:sapp:a_opt}

\end{theorem}

The proof of~\cref{thm:sapp:a_opt} is provided in~\ref{sec:app:proofs:parallel--optimal-attr}. To provide intuition about the formula in~\cref{eq:theo:a_opt}, 
we transform it to a simplified version:
\begin{equation}
    \hat{v}^{*}_{r(n)} = \sum_{\substack{r \in R.\\n \in r}}
    \hat{v}^{*}_{r}
    = 
    \sqrt{
    \Big(
        \underbrace{
            \frac{d\phi_{nk} (1+v_{-nk})}
            {d\phi_{nk}+\gamma_{nk}}
        }_{
            \circled{1}
        } + 
        \underbrace{
            \frac{d\alpha_{nk} (\rho_n - \Phi_{-nk})}
            {d\phi_{nk}+\gamma_{nk}}
        }_{
            \circled{2}
        }
        \Big)
        \underbrace{
        \big(1+v_{-r(n)}\big)
    }_{
        \circled{3}
    }
        }
    -
    \underbrace{
        \big(1+v_{-r(n)}\big)
    }_{
        \circled{3}
    }
\end{equation}
where~$\hat{v}^{*}_{r(n)}$ contains the sum of unrestricted valuations 
of all paths containing ISP~$n$
that would be optimal for~$n$ given~$\mathbf{A}_{-nk}$.
This~$\hat{v}^{*}_{r(n)}$ (and thus also~$\hat{a}_{nk}$) 
correlates positively with
term~\circled{1}, which relates to the share of demand-dependent
cost ($\propto d\phi_{nk}$) among total cost ($\propto d\phi_{nk} + \gamma_{nk}$) with respect to attribute~$(n,k)$.
This correlation suggests that ISPs should champion attributes with 
low demand-independent cost compared to demand-dependent cost. 
Moreover,~$\hat{v}^{*}_{r(n)}$ correlates with term~\circled{2}, 
which relates to the revenue from attribute~$(n,k)$
per unit of cost from the attribute, i.e., the `return' on attribute~$(n,k)$.
Term~\circled{3}, which describes the attractiveness of paths 
avoiding~$n$, can have a positive effect on~$\hat{v}^{*}_{r(n)}$
up to a point, as competition incentivizes ISP~$n$
to raise its attribute values. However, from a certain point onwards, term~\circled{3} has a negative effect on~$\hat{v}^{*}_{r(n)}$,
as detracting traffic from highly attractive alternatives becomes
too costly compared to the achievable revenue.

For an individual market, the equilibrium condition from~\cref{def:eq}
can thus be concretized based on~\cref{thm:sapp:a_opt}:
A choice of attributes~$\mathbf{A}^+$ is a Nash equilibrium if and only if
\begin{equation}
    \forall n \in N, \forall k \in K. \quad a_{nk}^+ = a_{nk}^{\ast}(\mathbf{A}_{-nk}^+).
    \label{eq:theo:sapp:equilibrium}
\end{equation}

For the general case, we find that deriving equilibria based on this condition is intractable. 
For example, when considering a market with two
disjoint paths, a single attribute, and a single ISP with arbitrary
parameters on each path, 
the equilibrium must be found by solving a quartic equation, 
which  impedes an analysis even for that simple network.
Fortunately, we identify two types of markets 
that allow deriving closed-form equilibria and therefore analytic insights, 
while still capturing the fundamental characteristics of ISP competition,
i.e., inter-path competition, intra-path cooperation, and ISP heterogeneity.
Concretely, we separately analyze homogeneous markets~(cf.~\cref{sec:theo:sapp-homogeneous})
and heterogeneous markets with attribute-independent traffic-unit cost (cf.~\cref{sec:theo:sapp-constant}),
both with disjoint paths.

\subsection{Homogeneous Markets}
\label{sec:theo:sapp-homogeneous}

By homogeneous markets, we refer to topologies 
of~$Q := |R| > 0$ disjoint paths in competition,
each of which accommodates the same number~$I = |N|/Q$ of ISPs.
All ISPs are identical and all attributes are identically
valuable and costly, i.e., for all ISPs~$n \in N$,
it holds that~$\rho_n = \rho$ and~$\phi_{n0} = \phi_0$, 
and $\forall k \in K$,
it holds that~$\alpha_{nk} = \alpha_{1}$, $\phi_{nk} = \phi_{1}$,
and~$\gamma_{nk} = \gamma_1$. Moreover, the
path-valuation functions for all paths are identical as well,
i.e., $\forall r\in R$.\ $\alpha_{r0} = \alpha_0$.
While artificial, such competition among
completely equal goods (here: paths) and firms (here: ISPs) 
is common in competition models, as homogeneity
allows isolating pure competition effects
that are not due to variety in offers~\cite{cournot1838recherches,edgeworth1925pure}.

In our case, the homogeneity also permits to identify the Nash equilibria
of the competition dynamics~(\cref{sec:theo:sapp-homogeneous:eq}),
to investigate the convergence to these equilibria~(\cref{sec:theo:sapp-homogeneous:stability}),
to compare these equilibria to social optima~(\cref{sec:theo:sapp-homogeneous:intra-path}),
and to evaluate the effect of competition intensity~(\cref{sec:theo:sapp-homogeneous:competition}).

\subsubsection{Equilibria}
\label{sec:theo:sapp-homogeneous:eq}

The symmetry of the homogeneous markets 
allows finding a competition
equilibrium:

\begin{theorem}
    \textbf{Nash Equilibrium in Homogeneous Markets.}
    The Nash equilibrium of a homogeneous market is given
    by an attribute sum~$a^+$ such that~$\sum_{k} a_{nk}^+ = a^+$ $\forall n \in N$, 
    where~$a^+ = \max(0,\ \hat{a}^+)$ with
    \begin{equation}
        \hat{a}^+ = \frac{\sqrt{T_2^2 - 4 T_1 T_3} - T_2}{2 T_1}.
        \label{eq:theo:sapp-homogeneous:equilibrium:raw}
    \end{equation} \cref{eq:theo:sapp-homogeneous:equilibrium:raw} uses the following abbreviations:
    \begin{align}
        T_1 = &\ Q^2 I^2 \alpha_1^2 - \frac{d}{d\phi_1 + \gamma_1} (QI-1) (Q-1) I \alpha_1^2 \phi_1, \\
        T_2 = &\ 2QI\alpha_1\left(1+Q\alpha_{0}\right) -\frac{d}{d\phi_1+\gamma_1} \cdot\\\
        &\ \Big( \alpha_1\phi_1\left(QI-1\right)\big(1+(Q-1)\alpha_0\big)+I\alpha_1\left(Q-1\right)\big(\phi_1\left(1+Q\alpha_{0}\right)+\alpha_1\left(\rho-\phi_{0}\right)\big) \Big), \text{ and}\nonumber\\
         T_3 = &\ (1 + Q\alpha_0)^2 - \frac{d}{d\phi_1 + \gamma_1} (1 + (Q-1)\alpha_0) \big(\phi_1(1+Q\alpha_0)+\alpha_1(\rho-\phi_0)\big).
    \end{align}
    \label{thm:theo:sapp-homogeneous:equilibrium}
\end{theorem}
The proof of~\cref{thm:theo:sapp-homogeneous:equilibrium} is provided
in~\ref{sec:app:proofs:homogeneous--equilibrium}.
Note that the equilibrium in~\cref{thm:theo:sapp-homogeneous:equilibrium}
is only unique with respect to the attribute sum~$a^+$ of any ISP
and hence also with respect to path valuations,
but not necessarily with respect to individual attribute values~$a_{nk}$.

\subsubsection{Stability}
\label{sec:theo:sapp-homogeneous:stability}

The Nash equilibrium from~\cref{thm:theo:sapp-homogeneous:equilibrium} is an interesting
fixed point of the competitive dynamics in homogeneous markets.
However, the equilibrium is only relevant if the distributed profit optimization by the ISPs 
converges to it.
Hence, the equilibrium must be additionally investigated with respect to its \emph{stability}, i.e.,
its attractive effect on the competition dynamics. To investigate this stability, we
formally describe the competition by the following system of ordinary differential equations (ODEs):
\begin{equation}
    \forall n \in N. \quad \dot{a}_n(t) = a_n^{\ast}\big(\mathbf{A}_{-n}(t)\big) - a_n(t)
    \label{eq:theo:sapp-homogeneous:dynamics}
\end{equation} 
Intuitively, one ODE in this system describes 
the behavior of an ISP~$n$ which continuously adjusts its 
attribute-sum value~$a_n$ towards the optimal choice~$a_n^{\ast}$
given the contemporary attribute values of all other ISPs. 
Given this dynamic process, we can show the following property:
\begin{theorem}
    \textbf{Stability of Homogeneous Equilibrium.}
    The Nash equilibrium from~\cref{thm:theo:sapp-homogeneous:equilibrium} is
    an asymptotically stable equilibrium of the competition dynamics in
    ~\cref{eq:theo:sapp-homogeneous:dynamics}.
    \label{thm:theo:sapp-homogenous:stability}
\end{theorem}
The proof of~\cref{thm:theo:sapp-homogenous:stability} is provided
in~\ref{sec:app:proofs:homogeneous--stability}.

\subsubsection{Intra-Path Dynamics}
\label{sec:theo:sapp-homogeneous:intra-path}

The equilibrium formalization from~\cref{thm:theo:sapp-homogeneous:equilibrium}
also applies to the case where an ISP pair is only connected by a single
usable path. Such a single-path scenario represents
a monopoly in economic terms.
Crucially, the ISPs on the same single path are supposed to
\emph{cooperate} rather than compete, 
as the decisions by each ISP
contribute to path attractiveness, which in turn benefits all ISPs.
For that single-path case, we can make the following interesting observation
about the cooperation among on-path ISPs:

\begin{theorem}
    \textbf{Suboptimality of Homogeneous Equilibrium.}
    On a single path with~$I$ identical ISPs, 
    the equilibrium attribute sum~$a^+$ 
    is generally lower than the NBS attribute sum~$a^{\circ}$, i.e.,
    $\forall I \in \mathbb{N}, I \geq 1.\ a^+ \leq a^{\circ}$.
    \label{thm:theo:sapp-homogeneous:intra-path:equilibrium-optimum}
\end{theorem}
The proof of~\cref{thm:theo:sapp-homogeneous:intra-path:equilibrium-optimum} is provided
in~\ref{sec:app:proofs:homogeneous--intra-path}.

Intuitively,~\cref{thm:theo:sapp-homogeneous:intra-path:equilibrium-optimum} states
that the cooperation by on-path ISPs suffers from
inefficiency caused by individual selfishness,
similar to a prisoner's dilemma~\cite{rapoport1989prisoner}. 
More precisely, the NBS attribute sum~$a^{\circ}$, which would 
optimize every ISP's profit if chosen universally, is not a rational 
choice for an individual ISP.
In particular, if an ISP~$n$ chooses~$a_n = a^{\circ}$,
ISP~$n$ enables another ISP~$m$ to optimize
its profit~$\pi_m$ by choosing a lower attribute sum~$a_m < a^{\circ}$,
and thus to free-ride on the path attractiveness created
by ISP~$n$. Because of this selfish deviation 
from the global optimum, the on-path ISPs converge
to the equilibrium attribute prevalence~$a^+$,
which is generally lower than the NBS attribute sum~$a^{\circ}$,
prevents the transit ISPs from reaping optimal profit, and
also reduces the path attractiveness for the path-selecting ISP.

\subsubsection{Competition Effects}
\label{sec:theo:sapp-homogeneous:competition}

After investigating intra-path cooperation in the preceding section,
we now investigate the effect of \emph{inter-path competition} on attribute prevalence.
In particular, we are interested in the dependence of the equilibrium attribute~$a^+$
on the number of available paths between an origin-destination pair.

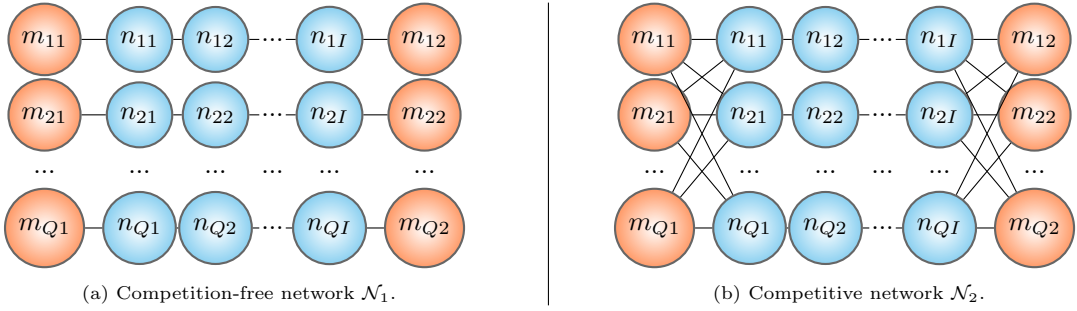
\begin{figure*}
    \begin{subfigure}{0.48\linewidth}
        \centering
        \begin{tikzpicture}[
	ASnode/.style={circle, draw=black!60, shading=radial,outer color={rgb,255:red,137;green,207;blue,240},inner color=white, thick, minimum size=8mm},
	endnode/.style={circle, draw=black!60, shading=radial,outer color={rgb,255:red,255;green,153;blue,102},inner color=white, thick, minimum size=5mm},
]
    \foreach \y/\ylabel in {4/1, 3/2, 1.5/Q} {
        \node[endnode] (m\ylabel1) at (-0.25, \y) {$m_{\ylabel1}$};
        \node[endnode] (m\ylabel2) at (4.75, \y) {$m_{\ylabel2}$};
        \foreach \x/\xlabel in {1/1, 2/2, 3.5/I} {
            \node[ASnode] (n\ylabel\xlabel) at (\x, \y) {$n_{\ylabel\xlabel}$};
        }
        \node at (2.75, \y) {...};
        \node (pseudo\ylabel1) at (2.7, \y) {};
        \node (pseudo\ylabel2) at (2.8, \y) {};
        \foreach \fromnode/\tonode in {m\ylabel1/n\ylabel1, n\ylabel1/n\ylabel2, n\ylabel I/m\ylabel2, n\ylabel2/pseudo\ylabel1, pseudo\ylabel2/n\ylabel I} {
            \draw[-] (\fromnode) -- (\tonode);
        }
    }
    \node at (-0.25, 2.25) {...};
    \node at (4.75, 2.25) {...};
    \foreach \x in {1, 2, 2.75, 3.5} {
        \node at (\x, 2.25) {...};
    }

\end{tikzpicture}
        \caption{Competition-free network~$\mathcal{N}_1$.}
        \label{fig:theo:sapp-homogeneous:competition:1}
    \end{subfigure}
    \vrule
    \begin{subfigure}{0.48\linewidth}
    \centering
        \begin{tikzpicture}[
	ASnode/.style={circle, draw=black!60, shading=radial,outer color={rgb,255:red,137;green,207;blue,240},inner color=white, thick, minimum size=8mm},
	endnode/.style={circle, draw=black!60, shading=radial,outer color={rgb,255:red,255;green,153;blue,102},inner color=white, thick, minimum size=5mm},
]
    \foreach \y/\ylabel in {4/1, 3/2, 1.5/Q} {
        \node[endnode] (m\ylabel1) at (-0.25, \y) {$m_{\ylabel1}$};
        \node[endnode] (m\ylabel2) at (4.75, \y) {$m_{\ylabel2}$};
        \foreach \x/\xlabel in {1/1, 2/2, 3.5/I} {
            \node[ASnode] (n\ylabel\xlabel) at (\x, \y) {$n_{\ylabel\xlabel}$};
        }
        \node at (2.75, \y) {...};
        \node (pseudo\ylabel1) at (2.7, \y) {};
        \node (pseudo\ylabel2) at (2.8, \y) {};
        \foreach \fromnode/\tonode in {n\ylabel1/n\ylabel2, n\ylabel2/pseudo\ylabel1, pseudo\ylabel2/n\ylabel I} {
            \draw[-] (\fromnode) -- (\tonode);
        }
    }
    \node at (-0.25, 2.25) {...};
    \node at (4.75, 2.25) {...};
    \foreach \x in {1, 2, 2.75, 3.5} {
        \node at (\x, 2.25) {...};
    }

    \foreach \y/\ylabel in {4/1, 3/2, 1.5/Q} {
        \foreach \yy/\yylabel in {4/1, 3/2, 1.5/Q} {
            \draw[-] (m\ylabel1) -- (n\yylabel 1);
            \draw[-] (n\ylabel I) -- (m\yylabel 2);
        }
    }

\end{tikzpicture}
        \caption{Competitive network~$\mathcal{N}_2$.}
        \label{fig:theo:sapp-homogeneous:competition:2}
    \end{subfigure}
    \vspace{-5pt}
    \caption{Homogeneous markets with and without inter-path competition.}
    \label{fig:theo:sapp-homogeneous:competition}
\end{figure*}

To characterize this dependence, we compare the Nash equilibria in two homogeneous markets.
First, we consider the competition-free network~$\mathcal{N}_1$ in~\cref{fig:theo:sapp-homogeneous:competition:1},
which is partitioned between~$Q$ origin-destination pairs~$\{(m_{q1},m_{q2})\}_{q=1,...,Q}$, 
each connected by a single path with~$I$ ISPs and obtaining the same demand limit~$d'$. 
Effectively, this network is a
set of homogeneous markets, each with only one available path and no
competition.
Second, we consider the competitive network~$\mathcal{N}_2$ in~\cref{fig:theo:sapp-homogeneous:competition:2},
where each of the~$Q$ origin-destination pairs can use all~$Q$ available paths.

By identifying the equilibrium attribute value~$a^+(\mathcal{N})$
for each network~$\mathcal{N}$, we find that
competition has a consistently positive effect on attribute prevalence:
\begin{theorem}
    \textbf{Attribute Improvement under Homogeneous Competition.}
    The equilibrium attribute prevalence is never lower in the competitive network~$\mathcal{N}_2$
    than in the competition-free network~$\mathcal{N}_1$, i.e., $a^+(\mathcal{N}_2) \geq a^+(\mathcal{N}_1)$
    for all~$Q \in \mathbb{N} \geq 1$.
    \label{thm:beneficial-competition}
\end{theorem}
The proof of~\cref{thm:beneficial-competition} is provided
in~\ref{sec:app:proofs:homogeneous--competition-attributes}.

Surprisingly, the higher attribute values in the competitive equilibrium 
do not necessarily come at the cost of lower ISP profits.
Instead, the profit of the ISPs may even increase under competition, 
which is important because the ISPs partially control whether they engage 
in competition at all, namely through path announcements.
For example, an increase in  equilibrium profits through competition happens 
if competition causes only a modest increase in attribute values:
\begin{theorem}
    \textbf{Profit Improvement under Homogeneous Competition.}
    The equilibrium profit~$\pi^+(\mathcal{N}_2)$ in the competitive network 
    preserves or exceeds the equilibrium profit~$\pi^+(\mathcal{N}_1)$ of the uncompetitive network
    if~$a^+(\mathcal{N}_2) \in [a^+(\mathcal{N}_1), a^{\circ}(\mathcal{N}_1)]$,
    i.e., the equilibrium attribute sum~$a^+(\mathcal{N}_2)$ from the competitive network
    is between the equilibrium attribute sum~$a^+(\mathcal{N}_1)$ of the uncompetitive
    network and the corresponding NBS attribute sum~$a^{\circ}(\mathcal{N}_1)$.
    \label{thm:theo:sapp-homogeneous:competition:profits}
\end{theorem}
The proof of~\cref{thm:theo:sapp-homogeneous:competition:profits} is provided
in~\ref{sec:app:proofs:homogeneous--competition-profits}.
In summary, we conclude that inter-path competition in homogeneous markets
is always desirable from the perspective of path-selecting ISPs, and
potentially desirable from the perspective of transit ISPs.

\subsection{Heterogeneous Markets}
\label{sec:theo:sapp-constant}

The homogeneous markets discussed in the previous section
can reflect competition dynamics among arbitrarily many paths.
However, these market models cannot represent differences between paths 
that go beyond attribute values. 
In reality, on-path ISPs may differ in their importance for the path valuation,
in their revenue per traffic unit, and in their attribute-specific costs.
In this section, we study competition among such heterogeneous ISPs,
i.e., every ISP~$n$ has arbitrary parameters~$\alpha_{nk}\ \forall k \in K$, $\rho_n$, $\phi_{n0}$,
$\gamma_{nk}\ \forall k \in K$, and~$\gamma_{n0}$.
To achieve tractability despite this additional complexity,
we restrict our analysis to markets with 
at most two paths. Moreover, since traffic-unit cost
is commonly considered negligible for ISPs~\cite{varian1994some},
we consider especially the attribute-dependent
part of this traffic-unit cost to be negligible, i.e., $\phi_{nk} = 0\ \forall n \in N, k \in K$.

\subsubsection{Intra-Path Dynamics}
\label{sec:theo:sapp-constant:single-path}

To characterize the attribute-choice dynamics among ISPs
in a heterogeneous market, we first consider a single path
in isolation, i.e., a monopoly scenario.
As before, ISPs on a path collectively
determine the attractiveness of the path, but optimize only
their individual profit. This selfishness may lead to 
a sub-optimal global outcome, both regarding
ISP profits and path valuations. 
To quantify this shortfall, we first
identify the Nash bargaining solution (NBS) for
the attribute choices, i.e., the attribute values
that all on-path ISPs would agree on if they
collectively negotiated and if they were bound
by the result of the negotiation.
This Nash bargaining solution represents the global
optimum with respect to the ISP profits.

\begin{theorem}
    \textbf{Profit Optimum on Heterogeneous Paths.}
    On a path~$r$ with heterogeneous ISPs, the attributes~$\mathbf{A}^{\circ}$
    form a Nash bargaining solution if and only if these attributes
    optimize the product of all ISP profits while
    \begin{enumerate}
        \item leading to the NBS path valuation~$v_r^{\circ}$,
        \item containing non-zero attribute values only for attributes~$(n,k)$ with maximal ratio~$\frac{\alpha_{nk}}{\gamma_{nk}}$,
        \item and containing zero attribute values for all other attributes.
    \end{enumerate}
    
    More formally, the conditions on~$\mathbf{A}^{\circ}$ can be stated as follows:
    \begin{equation}
        \mathbf{A}^{\circ} =\ {\arg\max}_{\mathbf{A} \in \mathbb{R}_{\geq 0}} \Pi_{n \in r} \pi_n(\mathbf{A})\\
    \end{equation}
    \begin{equation}
         \text{subject to } \hspace{25pt} v_r(\mathbf{A}) = v_r^{\circ} \hspace{25pt} \forall (n, k) \in K^{\circ}_r.\ a_{nk} \geq 0
            \hspace{25pt} \forall (n, k) \notin K^{\circ}_r.\ a_{nk} = 0 \nonumber
    \end{equation}
    \begin{flalign}
        &\hspace{30pt}\text{where}\ &v_r^{\circ} =\ &\max\left(\alpha_{r0},\ \max_{(n',k') \in r\times K} \sqrt{\frac{\alpha_{n'k'}}{\gamma_{n'k'}}} \sqrt{d\sum_{n\in r} (\rho_n - \phi_{n0})} - 1\right), \text{ and}& \label{eq:heterogeneous:single-path:optimum:valuation}\\
        &&K^{\circ}_r =\ &\left\{(n,k)\ |\ (n,k) = {\arg\max}_{(n',k') \in r\times K} \frac{\alpha_{n'k'}}{\gamma_{n'k'}}\right\}.& \label{eq:heterogeneous:single-path:optimum:attributes}
    \end{flalign}
    
    \label{thm:heterogeneous:single-path:optimum}
\end{theorem}
The proof of~\cref{thm:heterogeneous:single-path:optimum} is provided
in~\ref{sec:app:proofs:heterogeneous--intra-path-optimum}.

To optimize aggregate profit, the on-path ISPs should thus only 
upgrade the attribute(s) with maximal `return'~$\alpha_{nk}/\gamma_{nk}$
while minimizing the prevalence of all other attributes.
This return ratio~$\alpha_{nk}/\gamma_{nk}$ yields the
valuation for attribute~$k$ of ISP~$n$, compared to the 
cost that ISP~$n$ incurs for adopting that attribute.
The return ratio also correlates with the optimal path
valuation (cf.~\cref{eq:heterogeneous:single-path:optimum:valuation}).

However, aggregate profit is not the objective of selfish ISPs
when determining attribute values. Instead,
selfish ISPs optimize their individual profit,
and eventually arrive at the following equilibrium
by their non-aligned optimization behavior:
\begin{theorem}
    \textbf{Nash Equilibrium on Heterogeneous Paths.}
    On a path~$r$ with heterogeneous ISPs, the attributes~$\mathbf{A}^{+}$
    form a Nash equilibrium if and only if these attributes
    \begin{enumerate}
        \item lead to the equilibrium path valuation~$v_r^{+}$,
        \item contain non-zero attribute values only for attributes~$(n,k)$ with maximal ratio~$\frac{\alpha_{nk}(\rho_n - \phi_{n0})}{\gamma_{nk}}$,
        \item and contain zero attribute values for all other attributes.
    \end{enumerate}
    
    More formally, the conditions on~$\mathbf{A}^{+}$ can be stated as follows:
    \begin{equation}
         v_r(\mathbf{A}^+) = v_r^{+}
         \hspace{25pt} 
         \forall (n, k) \in K^{+}_r.\ a_{nk}^{+} \geq 0
         \hspace{25pt}
         \forall (n, k) \notin K^{+}_r.\ a_{nk}^{+} = 0
    \end{equation}
    \begin{flalign}
        &\hspace{30pt}\text{where} &v_r^{+} =&\ \max\left(\alpha_{r0},\ \max_{(n',k') \in r\times K} \sqrt{\frac{\alpha_{n'k'}}{\gamma_{n'k'}}} \sqrt{d (\rho_{n'} - \phi_{n'0})}- 1 \right) , \text{ and} & \label{eq:heterogeneous:single-path:equilibrium:valuation} \\
        & &K^{+}_r =&\ \left\{(n,k)\ |\ (n,k) = {\arg\max}_{(n',k') \in r\times K} \frac{\alpha_{n'k'}(\rho_{n'} - \phi_{n'0})}{\gamma_{n'k'}}\right\}. & 
        \label{eq:heterogeneous:single-path:equilibrium:attributes}
    \end{flalign}
    \label{thm:heterogeneous:single-path:equilibrium}
\end{theorem}
The proof of~\cref{thm:heterogeneous:single-path:equilibrium} is provided
in~\ref{sec:app:proofs:heterogeneous--intra-path-equilibrium}.

Interestingly, the equilibrium in~\cref{thm:heterogeneous:single-path:equilibrium}
is similar to the Nash-bargaining solution in~\cref{thm:heterogeneous:single-path:optimum},
but contains one crucial difference: The return ratio
associated with cultivated attributes includes the
net revenue per unit of traffic~$\rho_n - \phi_{n0}$ of ISP~$n$
(\cref{eq:heterogeneous:single-path:optimum:attributes} vs.\ \cref{eq:heterogeneous:single-path:equilibrium:attributes}).
This inclusion reflects that each ISP~$n$ optimizes its
individual profit rather than the aggregate profit:
When optimizing an attribute~$(n,k)$ for individual profit, 
an ISP~$n$ only considers its individual net revenue per traffic unit,
not the \emph{aggregate} net revenue per traffic unit,
which would be relevant for aggregate profit.

This difference, albeit subtle, generally leads to different attribute choices
in equilibrium than postulated by the Nash-bargaining solution,
meaning that the transit ISPs generate sub-optimal profits.
Unfortunately, also the path-selecting ISP 
suffers from this selfishness,
as the individual-profit optimization leads to less valuable paths:
\begin{theorem}
    \textbf{Suboptimality of Heterogeneous Equilibrium.}
    On a path~$r$ with heterogeneous ISPs, the equilibrium path valuation~$v_r(\mathbf{A}^+)$
    never exceeds the NBS path valuation~$v_r(\mathbf{A^{\circ}})$, i.e.,
    $v_r(\mathbf{A}^+) \leq v_r(\mathbf{A^{\circ}})$.
    \label{thm:heterogeneous:single-path:valuation}
\end{theorem}
The proof of~\cref{thm:heterogeneous:single-path:valuation} is provided
in~\ref{sec:app:proofs:heterogeneous--intra-path-valuation}.

\subsubsection{Two-Path Equilibria}
\label{sec:theo:sapp-constant:two-path}

In the preceding section, the social optimum and the Nash equilibrium
are characterized for a single-path scenario, which is exclusively 
informed by (failing) intra-path cooperation among selfish ISPs.
Since we are also interested in the effect of inter-path competition,
we now consider heterogeneous markets in which the path-selecting ISP
can select between two disjoint paths.
For these networks, the single-path equilibrium in~\cref{thm:heterogeneous:single-path:equilibrium}
can be adjusted as follows:
\begin{theorem}
    \textbf{Nash Equilibrium in Heterogeneous Markets.}
    In a two-path heterogeneous market,
    the attribute values~$\mathbf{A}^+$ form a Nash equilibrium if and only if
    the attribute values~$\mathbf{A}^+$ satisfy the conditions from~\cref{thm:heterogeneous:single-path:equilibrium},
    but with modified equilibrium path valuation~$v_r^+$:
    \begin{equation}
        v_r^+ = \max\left(\alpha_{r0},\ \hat{v}_r^{*}\left(\max\left(\alpha_{\overline{r}0}, \hat{v}_{\overline{r}}^+\right)\right)\right)
    \end{equation}
    \begin{flalign}
        &&\text{where}\quad& \text{\ $\overline{r}$ is the alternative path to $r$,} \quad
        \hat{v}_r^{*}(v_{\overline{r}}) = \psi_r \sqrt{d}\sqrt{1+v_{\overline{r}}} - (1+v_{\overline{r}}),&&\nonumber\\
        &&&\hat{v}_r^+ = \frac{\psi_r^3\psi_{\overline{r}}}{(\psi_r^2 + \psi_{\overline{r}}^2)^2} \left(\sqrt{d\left(\psi_r^{2}+\psi_{\overline{r}}^{2}\right)+\frac{1}{4}\psi_r^{2}\psi_{\overline{r}}^{2}d^{2}}+\frac{d}{2}\psi_r\psi_{\overline{r}}\right)-\frac{\psi_{\overline{r}}^{2}}{\psi_r^{2}+\psi_{\overline{r}}^{2}} \text{, and} \label{eq:theo:sapp-constant:competition:eq-path-valuation}&&\\
         &&&\psi_r = \max_{\substack{n \in r\\k \in  K}} \sqrt{ \frac{\alpha_{nk}(\rho_n - \phi_{n0})}{\gamma_{nk}} }. \label{eq:theo:sapp-constant:competition:psi}&&
    \end{flalign}
    \label{thm:heterogeneous:equilibrium:two}
\end{theorem}
The proof of~\cref{thm:heterogeneous:equilibrium:two} is provided
in~\ref{sec:app:proofs:heterogeneous--two-path-equilibrium}.
We note that~$\psi_r$ from~\cref{eq:theo:sapp-constant:competition:psi}
is the square root of the maximum individual return ratio 
discussed in the previous section, albeit only among the attributes
of path~$r$.
In the following, we refer to~$\psi_r$
as the \emph{characteristic ratio} of path~$r$.

Similar to~\cref{sec:theo:sapp-homogeneous},
we are again interested in the stability of the equilibrium
w.r.t. the process:
\begin{equation}
    \forall n \in N, k \in  K. \quad \dot{a}_{nk}(t) =
    a^{*}_{nk}(\mathbf{A}_{-nk}(t)) - a_{nk}(t).
    \label{eq:theo:sapp-constant:process}
\end{equation}
However, stability analysis in the case of heterogeneous
two-path networks is complicated by the fact
that the equilibrium from~\cref{thm:heterogeneous:equilibrium:two}
is only unique in the path valuations~$\{v_r\}_{r\in R}$,
but not necessarily unique in the attribute 
choices~$\mathbf{A}$ by the ISPs.
Therefore, if the equilibrium is not unique in~$\mathbf{A}$,
no single equilibrium~$\mathbf{A}^+$ is asymptotically stable
in a narrow sense, as the process in~\cref{eq:theo:sapp-constant:process} 
does not converge to~$\mathbf{A}^+$ from~$\mathbf{A}(t)$ 
in case~$\mathbf{A}(t)$
already represents a different equilibrium.

Therefore, we focus on the stability of unique equilibria:
\begin{theorem}
    \textbf{Stability of Heterogeneous Equilibrium.}
    The Nash equilibrium~$\mathbf{A}^+$ from~\cref{thm:heterogeneous:equilibrium:two} is
    an asymptotically stable equilibrium of the competition dynamics 
    in~\cref{eq:theo:sapp-constant:process}
    if the equilibrium~$\mathbf{A}^+$ is unique, i.e.,
    if there is only one attribute on every path
    which has potentially non-zero prevalence
    ($|K_r^+| = |K_{\overline{r}}^+| = 1$).
    \label{thm:heterogeneous:equilibrium:stability}
\end{theorem}
The proof of~\cref{thm:heterogeneous:equilibrium:stability} is provided
in~\ref{sec:app:proofs:heterogeneous--two-path-stability}.

\subsubsection{Competition Effects}
\label{sec:theo:sapp-constant:competition}

Based on the equilibria for single-path and two-path markets,
we now investigate the effect of inter-path competition in heterogeneous markets.
For this investigation, we use a similar approach as in~\cref{sec:theo:sapp-homogeneous:competition}:
We contrast a competition-free network~$\mathcal{N}_3$, which consists 
of two paths~$r$ and~$\overline{r}$, each connecting one origin-destination pair,
with a competitive network~$\mathcal{N}_4$, where both origin-destination pairs
are connected by both paths. The origin-destination pair connected by path~$r$
in the competition-free network~$\mathcal{N}_3$
has demand limit~$d_r$; hence, the total demand limit~$d = d_r + d_{\overline{r}}$
is distributed over both paths in the competitive network~$\mathcal{N}_4$.
The networks $\mathcal{N}_3$
and~$\mathcal{N}_4$ thus differ in the same manner as the networks~$\mathcal{N}_1$
and~$\mathcal{N}_2$ from~\cref{fig:theo:sapp-homogeneous:competition},
except that the different paths may have different length in ISPs,
each ISP may have different parameters, and each origin-destination pair
may have a different demand limit.
When contrasting these two networks, we gain the following insight:
\begin{theorem}
    \textbf{Attribute Improvement under Heterogeneous Competition.}
    For any competition-free network~$\mathcal{N}_3$ and the corresponding competitive network~$\mathcal{N}_4$,
    a demand limit~$d$ exists such that the competitive network~$\mathcal{N}_4$ has a higher equilibrium
    valuation than the competition-free network~$\mathcal{N}_3$ 
    independent of the demand distributions~$(d_r, d_{\overline{r}})$, i.e.,
    \begin{equation}
        \exists d \text{ s.t. } \forall d_r, d_{\overline{r}} \text{ with } 
        d_r + d_{\overline{r}} = d.\ V^{+}(\mathcal{N}_4) \geq V^{+}(\mathcal{N}_3)
    \end{equation}
    \label{thm:theo:sapp-constant:competition:good}
\end{theorem}
The proof of~\cref{thm:theo:sapp-constant:competition:good} is provided
in~\ref{sec:app:proofs:heterogeneous--competition-beneficial}.

In simplified terms, inter-path competition thus affects the attribute values
and the path valuations positively for high-enough demand, given
the remaining network parameters.
This condition on demand, however, raises the question whether
competition reduces
the network valuation in some circumstances.
Indeed, we find that such a counter-intuitive effect can
arise at every demand level if the remaining network parameters
are unfavorable:
\begin{theorem}
    \textbf{Attribute Decline under Heterogeneous Competition.}
    For every demand distribution~$(d_r, d_{\overline{r}})$, there exist 
    characteristic ratios~$(\psi_r, \psi_{\overline{r}})$ and
    path base valuations~$(\alpha_{r0}, \alpha_{\overline{r}0})$ such that 
    the competitive network~$\mathcal{N}_4$ has a lower equilibrium valuation
    than the competition-free network~$\mathcal{N}_3$, i.e.,
    \begin{equation}
            \forall d_r, d_{\overline{r}}.\ \exists \psi_r, \psi_{\overline{r}}, \alpha_{r0}, \alpha_{\overline{r}0} \text{ s.t. }  V^{+}(\mathcal{N}_4) < V^{+}(\mathcal{N}_3).
    \end{equation}
    \label{thm:theo:sapp-constant:competition:bad}
\end{theorem}
The proof of~\cref{thm:theo:sapp-constant:competition:bad} is provided
in~\ref{sec:app:proofs:heterogeneous--competition-bad}.

To understand this effect intuitively, we note that 
an ISP~$n$ optimizes its profit by balancing
the \emph{marginal revenue} and the \emph{marginal cost}
with respect to attribute prevalence, i.e., adjusts
attribute prevalence as long as the adjustment generates more
revenue than cost. In the competition-free scenario
of~$\mathcal{N}_3$, the marginal revenue and cost
of an ISP~$n$ with respect to attribute~$(n,k)$ are:
\begin{equation}
    \frac{\partial \mathcal{R}_n}{\partial a_{nk}} =
    \frac{d_r \alpha_{nk}}{\left(1 + v_r\right)^2} \cdot \rho_n
    \hspace{25pt}
    \frac{\partial}{\partial a_{nk}}\left(\Phi_n+ \Gamma_n \right)
    = \frac{d_r \alpha_{nk}}{\left(1 + v_r\right)^2} \cdot \phi_{n0} + \gamma_{nk}
\end{equation}
In contrast, the corresponding terms for the competitive scenario
in network~$\mathcal{N}_4$ are as follows:
\begin{equation}
    \frac{\partial \mathcal{R}_n}{\partial a_{nk}} =
    \frac{d \alpha_{nk}\cdot(1+v_{\overline{r}})}{\left(1 + v_r + v_{\overline{r}}\right)^2} \cdot \rho_n
    \hspace{25pt}
    \frac{\partial}{\partial a_{nk}}\left(\Phi_n + \Gamma_n \right)
    = \frac{d \alpha_{nk}\cdot(1+v_{\overline{r}})}{\left(1 + v_r + v_{\overline{r}}\right)^2} \cdot \phi_{n0} + \gamma_{nk}
\end{equation}
On the one hand, competition has a positive effect on
marginal revenue~$\partial\mathcal{R}_n/\partial a_{nk}$ by increasing
the total amount of attractable demand from~$d_r$ to~$d > d_r$.
On the other hand, the new competition embodied by the
alternative-path valuation~$v_{\overline{r}}$ has a negative
effect on marginal revenue.
The negative effect predominates if the alternative-path
valuation~$v_{\overline{r}}$ is relatively large and unresponsive
to competition, as the proof of~\cref{thm:theo:sapp-constant:competition:bad} demonstrates.
If marginal revenue in fact decreases, marginal cost 
decreases less strongly as~$\rho_n \geq \phi_{n0}$.
Given negative marginal profit, the profit
of ISP~$n$ is thus optimized by a lower attribute 
prevalence~$a_{nk}$, which translates into decreasing path value.
\section{A Model Instance Based on Real-World Data}
\label{sec:example}

In this section, we demonstrate how to instantiate 
our competition model from~\cref{sec:model} 
to investigate a large-scale network
containing multiple intertwined markets.
To that end, we construct a topology approximating the Internet
core and a corresponding traffic
matrix in~\cref{sec:simulation:network}.
Furthermore, we consider two ISP attributes in the competitive
dynamics, namely \emph{internal bandwidth} and 
\emph{clean-energy share}, and estimate appropriate model
parameters in~\cref{sec:simulation:attribute-1} and~\cref{sec:simulation:attribute-2},
respectively. 
Attribute-independent parameters are estimated in~\cref{sec:simulation:independent}. 

Importantly, we note that estimating highly realistic parameters for 
the model goes beyond the scope of this paper, as the scarcity of publicly 
available data and the complexity of real-world business practices considerably
complicates this estimation. Therefore, the goal of the
following parameter estimation is to place the parameters in the
right order of magnitude, especially in relation to each other,
rather than to determine each parameter highly realistically.
Interestingly, our sensitivity analysis in~\cref{sec:simulation} suggests
that such an approximate estimation might be sufficient to
yield useful predictions.

\subsection{Network Topology and Demand}
\label{sec:simulation:network}

To investigate the effects of competition in practically interesting, 
large-scale settings while keeping the complexity of the simulation manageable, 
we extract a network topology that roughly approximates
the Internet core from a public dataset.
In particular, we rely on a CAIDA
dataset containing \num{12300} autonomous systems (ASes), 
their economic relationships, 
and the geolocation of their interconnections 
(i.e., inter-domain interfaces)~\cite{CAIDA-Data-Geo}.
From this dataset,  we extract the topology of the \num{2000} most
interconnected ASes by iteratively removing the lowest-degree ASes.

In this reduced topology, we aim at finding the 5 
shortest paths between every origin-destination pair of ASes.
For scalability, we can only consider AS paths
with at most 4 AS hops, which is not
a strong limitation: The paths in our topology
only represent the core-traversing segments
of whole Internet paths, which have an average
length of around 5 hops (and decreasing)~\cite{huston2021bgp}.
Moreover, for both scalability and practical relevance, 
we only consider paths that are Gao-Rexford-compliant~\cite{gao2001stable}, 
i.e., are compatible with the economic self-interest of ASes
regarding monetization of traffic.
With these constraints, we can identify 5 paths for $\sim52.4$\%
of AS pairs in the topology.

While only a subset of all AS pairs, these pairs of closely
located ASes are disproportionately relevant for the competition dynamics,
as they account for a substantial share of traffic 
given the gravity-like nature of Internet traffic~\cite{roughan2005simplifying}.
Generally, gravity models predict that the
traffic demand~$d_{(n_1,n_2)}$ between two ISPs~$n_1$ and~$n_2$
is proportional to the product of the `masses'~$m_1 \cdot m_2$
of the two ISPs divided by the squared distance~$r_{12}^2$
between the ISPs:
\begin{equation}
    d_{(n_1,n_2)} \propto G_{12} = \frac{m_1 \cdot m_2}{r_{12}^2}.
    \label{eq:simulation:gravity}
\end{equation}

In order to synthesize a traffic matrix
for our purpose, we
concretize this gravity model as follows.
First, we calculate the mass~$m_n$ of an AS~$n$
as the number of distinct IPs in all prefixes
owned by AS~$n$ and by the ASes in the customer
cone of AS~$n$. This information is available
via the datasets `Routeviews Prefix-to-AS Mapping'~\cite{pfx2as}
and `AS Relationships'~\cite{CAIDA-Data-AS-rel},
both from CAIDA.
Second, we determine the distance~$r_{12}$ for each 
AS pair~$(n_1,n_2)$ as the average number of hops
in the 5~paths connecting the AS pair.
Third, we calculate the gravity~$G_{12}$
according to~\cref{eq:simulation:gravity} for
every AS pair~$(n_1,n_2)$. 
Finally, we allocate the total Internet traffic volume
of 170 Tbps~\cite{miller2021global} to the AS pairs~$(n_1,n_2)$
according to the relative size of~$G_{12}$.

\subsection{Attribute 1: Internal Bandwidth}
\label{sec:simulation:attribute-1}

To instantiate the model, we define
the ISP attributes~$K$ that are affected by the
competitive dynamics, and the corresponding model parameters.
As an intuitive example of desirable ISP attributes, we consider the \emph{internal bandwidth}
of an ISP (in Gbps) the first such attribute~($k = 1$). 
If the ISPs along a path have
a large bandwidth capacity, these ISPs 
are likely able to absorb sudden traffic surges,
tolerate equipment failures,
handle large traffic flows, and
in general deliver a high quality of service;
hence, the internal bandwidth of on-path ISPs
correlates with the attractiveness of the given path.

\subsubsection{Valuation} 

This valuation by path-selecting ISPs is quantified
by the valuation function~$v_{r1}$, quantifying
the valuation of a bit traversing path~$r$ given
the internal bandwidth of on-path ISPs.
This valuation function~$v_{r1}$
is characterized by the parameters~$\alpha_{rn1}$,
giving the valuation of a bit traversing ISP~$n$ on path~$r$
for a unit of the internal bandwidth of ISP~$n$.
For this quantification, we rely on two empirical findings.
First, the average US consumer transmits 536.3 GB of data per month~\cite{openvault2021ovbi},
and is willing to pay 94 USD per month for a 1Gbps connection~\cite{liu2018distinguishing}.
Hence, we arrive at a monthly 
willingness-to-pay of around~$w = 0.17$ USD per GB 
at the quality of a 1Gbps connection.
With this willingness-to-pay~$w$, we determine the bandwidth valuation 
parameters~$\alpha_{rn1}$, namely by defining~$\alpha_{rn1} = w / (|r| \cdot m_n)$, 
where~$|r|$ is the number of ASes on path~$r$
(averaging the internal bandwidth across on-path ISPs)
and~$m_n$ is the number of IPs in the customer cone 
of AS~$n$ (correcting for the number of end-points sharing the bandwidth).
Multiplied with the internal bandwidth~$a_{n1}$, 
these valuation parameters thus approximate
the valuation per bit traversing ISP~$n$ given the
internal bandwidth of ISP~$n$.
Furthermore, the bandwidth valuation function~$v_{r1}$
is also characterized by the base valuation~$\alpha_{r0}'$ 
of path~$r$. However, since a path only has value
in terms of bandwidth if the on-path ASes have non-zero
internal bandwidth, we choose~$\alpha_{r0}' = 0$.

\subsubsection{Cost}

Apart from increasing valuation by path-selecting ISPs, 
providing bandwidth also has a cost. 
However, it is difficult to quantify the cost
of providing a Gbps of internal bandwidth,
as this cost heavily depends 
on the way of provision (leasing
or physically installing new capacity),
on the necessary installation procedures
(e.g., length of cables to be newly laid),
on the location where capacity should be added,
and on numerous other aspects. 
Hence, we rely on the simple insight
that the cost of providing a Gbps of connectivity
is likely lower than the corresponding willingness-to-pay
by consumers (94 USD per Gbps per month~\cite{liu2018distinguishing}), 
as ISPs would go out of business otherwise.
Hence, we randomly vary the cost parameter~$\gamma_{n1}$
between~0 and~$94$ USD per Gbps per month in our simulations, 
for all~$n \in N$.
Importantly, the provision of bandwidth 
only affects the demand-independent cost~$\Gamma_n$ of
an ISP~$n$, as providing a certain bandwidth capacity
causes the same cost independent of the actually experienced demand.
Hence, we can also define the demand-dependent
cost parameter for the bandwidth attribute~$k = 1$: 
$\phi_{n1} = 0$ for all~$n \in N$.

\subsubsection{Attribute Bounds.}

Using internal bandwidth as one of multiple attributes leads
to an implausible model prediction
in the case where all ASes on a path~$r$ have zero
internal bandwidth~($a_{n1} = 0\ \forall n \in r$),
but some non-zero values for other attributes.
In that case, the valuation function~$v_r$
might still assign some non-zero valuation and
some demand to path~$r$, although
the zero-bandwidth path~$r$ is clearly worthless.
To avoid this implausible case of the model,
we place a lower bound on the bandwidth attribute~$a_{n1}\ \forall n \in N$.
This lower bound is given by~$10\%$ of the demand experienced
by AS~$n$ if the demand of every origin-destination pair was equally
distributed among the available 5 paths:
\begin{equation}
    \forall n \in N.\ a_{n1} \geq \frac{0.1}{5} \cdot 
    \sum_{\substack{r \in R.\ n \in r\\
    r \in R(n_1, n_2)}} d_{(n_1,n_2)}
\end{equation}

\subsection{Attribute 2: Clean-Energy Share}
\label{sec:simulation:attribute-2}

Path-selection preferences are not exclusively related 
to transmission performance (such as internal bandwidth
of on-path ISPs), but may also reflect ESG considerations~\cite{chen2015greening,liu2011greening}.
For example, in carbon-intelligent routing~\cite{tabaeiaghdaei2023green_routing,zilberman2022toward},
path selection takes into account the carbon emission
that results from data transmission.
More precisely, the path-specific \emph{transmission carbon intensity}, i.e., 
the volume of carbon emission
per bit of transmitted data on a given path, affects path selection.
To investigate the effect of competition on this carbon intensity,
we choose the \emph{share of clean
energy} used by an ISP (in percent) as the
second attribute~($k = 2$) for our simulations,
i.e.,~$a_{n2} \in [0, 1]\ \forall n \in N$.

\subsubsection{Transmission carbon intensity}

The clean-energy share attributes of on-path ASes
determine the carbon intensity of a path as follows.
First, any AS-level path~$r$ must be transformed into a
router-level path~$s_r$, which is possible by means 
of the CAIDA ITDK dataset~\cite{CAIDA-ITDK}.
For simplicity, we assume that the intra-AS router-level path~$s_{rn}$
in AS~$n$ is the shortest router-level path between
the two AS interconnections derived from the AS-level path~$r$.
For any intra-AS
path~$s_{rn}$, we determine the energy intensity~$e_{rn}$,
i.e., the amount of consumed electricity per bit transmitted
on path~$s_{rn}$. This energy intensity~$e_{rn}$ can be calculated
from the number of routers and the covered distance of
path~$s_{rn}$, given by the CAIDA ITDK dataset, and
the energy-intensity values for various devices,
as reported by Heddeghem at al.~\cite{multilayer}.
Then, we calculate the maximum transmission
carbon intensity~$c_{rn,\max}$
of any intra-AS path~$s_{rn}$ by
multiplying the corresponding energy intensity~$e_{rn}$
with the the energy carbon intensity~$c_{\max}$ of the most carbon-intensive 
electricity,  namely 875~$\text{gCO}_2/\text{kWh}$ for coal-generated electricity~\cite{iea2022coal}.
This maximum transmission carbon intensity~$c_{rn,\max}$ thus quantifies
the carbon emission associated with the transmission of a bit
across path~$s_{rn}$ if AS~$n$ used maximally carbon-intensive 
electricity.
Finally, we derive
the \emph{actual} transmission carbon intensity~$c_{rn}$ of any intra-AS path~$s_{rn}$
as the product of the maximum transmission carbon intensity~$c_{rn,\max}$
and the dirty-energy share of ISP~$n$,
i.e., $1 - a_{n2}$.
The carbon intensity~$c_r$ of a path~$r$ is the sum
of carbon-intensity values~$c_{rn}$ of the constituting intra-AS paths~$s_{rn}\ \forall n \in r$:
\begin{equation}
    c_r(\mathbf{A}) = \sum_{n \in r} c_{rn}(\mathbf{A}) = \sum_{n \in r} c_{rn,\max} \cdot (1 - a_{n2}) = \sum_{n \in r} e_{rn} \cdot c_{\max} \cdot (1 - a_{n2}).
\end{equation}

\subsubsection{Valuation}

This carbon-intensity calculation also informs the valuation~$v_{r2}$,
which quantifies the valuation of path~$r$ exclusively with respect to carbon emissions.
In fact, we understand~$v_{r2}$ as an affine function of
the \emph{negative} carbon intensity of path~$r$:
\begin{equation}
    v_{r2}(\mathbf{A}) = \sum_{n \in r} \alpha_{rn2} a_{n2} + \alpha_{r0}'' = - \sum_{n \in r} p_{\text{CO}_2} c_{rn}(\mathbf{A}) + q_r = \sum_{n \in r} 
    \left( p_{\text{CO}_2} \cdot c_{rn,\max} \cdot a_{n2} - p_{\text{CO}_2} \cdot c_{rn,\max} \right) + q_r.
    \label{eq:simulation:carbon-valuation}
\end{equation}
where~$p_{\text{CO}_2}$ is the cost of emitted $\text{CO}_2$,
chosen as 90 USD per ton according to the EU emission-trading scheme~\cite{ember2022eucarbonprice},
and~$q_r$ is a constant that ensures the non-negativity and comparability of the valuation (see below).
From~\cref{eq:simulation:carbon-valuation}, we can determine the valuation
parameters~$\alpha_{rn2}$, describing the valuation of ISP~$n$'s clean-energy share
on path~$r$, as~$p_{\text{CO}_2} \cdot c_{rn,\max}$.
The base valuation~$\alpha_{r0}''$ is determined based on two considerations.
First, the valuation function~$v_{r2}$ must be consistently non-negative.
Second, the valuation function must allow a meaningful comparison
between paths~$R(n_1, n_2)$ connecting the same AS pair~$(n_1,n_2)$:
For example, if all ISPs use zero clean energy, a path with higher
energy intensity should still be valued less than a path with lower
energy intensity. Conversely, if all ISPs use perfectly clean energy,
all paths should be valued identically.
To achieve these properties, we determine~$\alpha_{r0}''$ as follows:
\begin{equation}
    \alpha_{r0}'' = - \sum_{n \in r} p_{\text{CO}_2} \cdot c_{rn,\max} + q_r = - \sum_{n \in r} p_{\text{CO}_2} \cdot c_{rn,\max} + \max_{\substack{r' \in R(n_1,n_2)\\ r \in R(n_1,n_2)}}\sum_{n' \in r'} p_{\text{CO}_2} \cdot c_{r'n',\max}.
\end{equation}

With such determined~$v_{r2}$, we can formalize the complete path-valuation~$v_r$
as the sum of the attribute-specific valuation functions~$v_{r1}$ and~$v_{r2}$.
Since~$\alpha_{rnk} a_{nk}$ yields a valuation per bit for both attributes~$k \in \{1,2\}$,
the attribute-specific valuation functions are compatible.

\subsubsection{Cost}

To estimate the costs associated with the clean-energy share of an ISP~$n$,
we rely on the analysis of the levelized cost of energy (LCOE) of different
electricity-generation technologies, performed
by Lazard~\cite{lazard2020levelized}. 
According to the Lazard analysis, electricity from low-carbon sources
(solar, wind, nuclear) is on average $g = 3.375$ USD per MWh more expensive
than electricity from high-carbon sources (coal, gas).
This cost penalty, together with the average energy intensity of all intra-AS paths
in AS~$n$,
yields the parameter~$\phi_{n2}$ relevant for demand-dependent cost:
\begin{equation}
    \phi_{n2} = g \cdot \frac{1}{|R(n)|} \cdot \sum_{r \in R(n)} e_{rn},
\end{equation}
where~$R(n) = \{r \in R\ |\ n \in R\}$. 
Multiplied with the clean-energy share attribute~$a_{n2}$, 
the parameter~$\phi_{n2}$ yields the extra cost per transported bit 
that AS~$n$ incurs by using clean energy.

Regarding demand-independent cost, we note that the idle-power requirement
of network devices plays an important role, as this requirement generates
electricity bills even in absence of demand. 
The idle-power consumption~$u_n$
of a complete AS~$n$ can be estimated from the number
of devices in AS~$n$~\cite{CAIDA-ITDK},
the power consumption of network devices~\cite{multilayer},
and an average idle-power requirement of 85\%~\cite{jacob2022internet}.
This idle-power consumption~$u_n$, together with the
extra cost~$g$ for clean energy, determines the 
parameter~$\gamma_{n2}$ relevant for demand-independent cost:
$\gamma_{n2} = g \cdot u_n$.

\subsection{Attribute-Independent Parameters}
\label{sec:simulation:independent}

In addition to the attribute-specific parameters 
in~\cref{sec:simulation:attribute-1,sec:simulation:attribute-2},
the attribute-independent parameters~$\rho_n$, $\phi_{n0}$ and~$\gamma_{n0}$
also appear in our model.

The parameter~$\rho_{n}$ quantifies the revenue per transported bit of AS~$n$.
To estimate this parameter, we use a top-down approach:
We divide the global annual revenue of wholesale Internet backbone 
providers (45.2 billion USD in 2019~\cite{whelan2020wholesale})
by the amount of global annual Internet traffic (433 exabyte in 2019~\cite{miller2021global}),
and arrive at an average revenue of~$\rho = 0.104$ USD per GB.
For simplicity, we use this~$\rho$ as revenue parameter~$\rho_n$ 
for every ISP~$n \in N$.

The parameter~$\phi_{n0}$ describes the marginal cost of AS~$n$ per
transported bit, excluding extra marginal cost due to clean-energy usage (cf.~\cref{sec:simulation:attribute-2}).
As this marginal cost is commonly understood to be `essentially zero'~\cite{varian1994some},
we determine~$\phi_{n0} = 0\ \forall n \in N$.

Conversely, the attribute-independent fixed cost~$\gamma_{n0}$ of AS~$n$ can be quite
substantial. However, since we are mainly interested in the attribute-optimization behavior of ASes
under competition, and~$\gamma_{n0}$ does not affect this optimization behavior,
we abstain from estimating~$\gamma_{n0}$, i.e., use~$\gamma_{n0} = 0$ in our simulations.
As a result, the absolute value of the profit function~$\pi_n$ is not meaningful,
which we take into account for the result discussion in~\cref{sec:simulation:results}.
\section{Simulation}
\label{sec:simulation}

Section~\ref{sec:theo} theoretically illustrates the
diverse results of quality competition among ISPs.
These results include both positive and negative effects of competition on attribute
prevalence and profits, depending on the concrete 
topologies of competing paths in the considered
markets. 
In this section, we investigate which types of effect are observable 
if competition is introduced in a large-scale
topology where transit ASes (ASes, corresponding
to ISPs) simultaneously compete in multiple markets.
To that end,
we run simulation experiments described in~\cref{sec:simulation:experiments}
for the instance of the competition model constructed in~\cref{sec:example}, 
and discuss the results in~\cref{sec:simulation:results}.

\subsection{Experiments}
\label{sec:simulation:experiments}

Since the parameters estimated in~\cref{sec:example} are afflicted with
considerable uncertainty, we conduct our experiments without being overly
reliant on the exact estimated parameter values.
More precisely, we generate 10 different sets of model parameters 
by randomly modifying each model parameter~$y$
such that it lies between~$0$ and~$2y$ in virtually all cases.
We achieve this modification by randomly sampling each model parameter~$y'$ 
from the following restricted normal distribution,
based on the corresponding estimated parameter~$y$:
\begin{equation}
    y' \sim \max\left(\mathcal{N}\left(y, \frac{y^2}{9}\right), 0\right).
\end{equation}

For each random sample of parameters, we investigate the effect
of increasing \emph{intensity of competition} on the attribute-value 
choice of transit ASes.
In our experiments, the intensity of competition corresponds
to the number of usable paths between any AS pair.
This number of usable paths is varied between 1 and 5 across experiments,
where the case of 1 path corresponds to a
monopoly scenario.
In each simulation experiment, we thus simulate the competitive dynamics
given a set of randomly sampled model parameters and
given a certain intensity of competition. 

Each simulation experiment amounts to computing round-robin better-response 
dynamics~\cite{cabrales2011implementation}, where all ASes consecutively
adjust their attribute values in the direction which increases their profit.
This discrete process is an approximation of the continuous ODE process
in~\cref{eq:theo:sapp-homogeneous:dynamics}.
Moreover, the process can be understood as reflecting bounded rationality~\cite{simon1990bounded},
as we assume that ASes can only identify profit-improving
rather than profit-optimal attribute values.
The simulation is terminated once the competitive dynamics have converged
upon an equilibrium, i.e., each round only causes negligible change in the
attribute values~$\mathbf{A}$.
The attribute values~$\mathbf{A}^+$ in the equilibrium then
represent the results of the experiment.

\subsection{Results}
\label{sec:simulation:results}

The results of the experiments described in~\cref{sec:simulation:experiments}
are visualized in~\cref{fig:as-improvements,fig:mean,fig:as-pair-improvements}.
The error bar of any data point in these figures
illustrates the variance of the respective aggregate result 
across the 10 random parameter samples.
Interestingly, the variance of the aggregate results is
quite limited, although the variance in individual parameters
is considerable. This observation indicates that
our results are not highly sensitive to the
model parameters from~\cref{sec:example},
and suggests that an approximate estimation of model parameters
might be sufficient to yield useful predictions.

The central question in our analysis is:
How does the intensity
of competition affect the attribute values and the ISP profits?
Our theoretical investigation in~\cref{sec:theo}
indicates that the competitive dynamics can
both increase and decrease these indicators compared to
a monopoly scenario, depending
on network properties.
Hence, we investigate which type of effect is predominant
for the large-scale network from~\cref{sec:example}.

In this analysis, we distinguish three groups of ASes
that differ in their topology rank, namely
tier-1 ASes (ASes that have no provider),
tier-2 ASes (ASes that have only tier-1 providers),
and tier-3 ASes (ASes that have only tier-1 and tier-2 providers).
Since these AS groups differ in their market power,
the effect on competition on attributes and profits
for these groups may be different.

\subsubsection{Attribute Prevalence}
\label{sec:simulation:attributes}

Regarding competition effects on attribute prevalence,
we gain the following insights:

\begin{figure}
    \centering
   \includegraphics[width=\linewidth,trim=0 7 0 0]{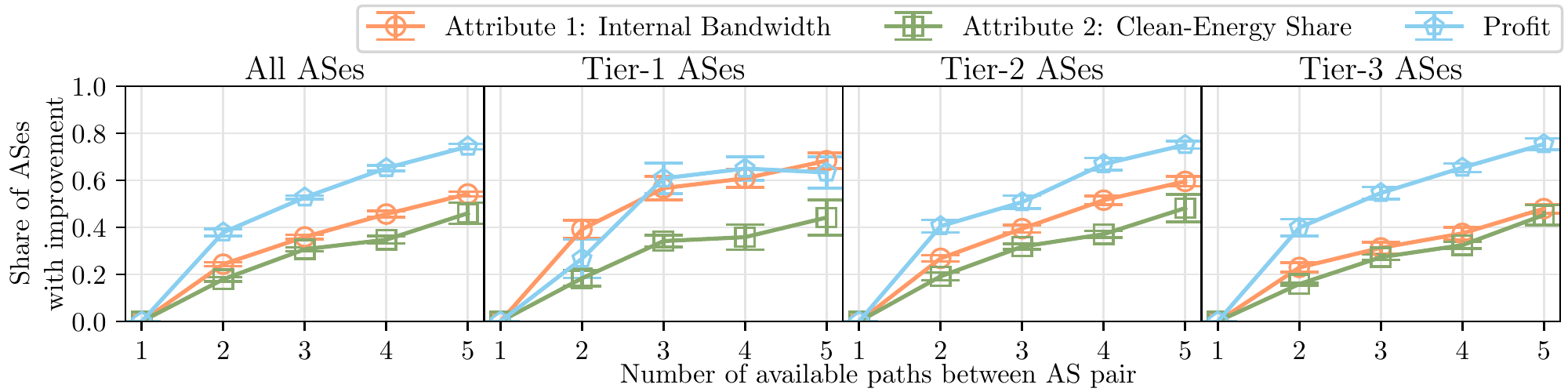}
    \caption{Under increasing competition, more ASes improve their attribute values or
    their profit compared to a single-path scenario, independent of their
    topology level.}
    \label{fig:as-improvements}
\end{figure}

\paragraph{Effects on transit ASes}
\cref{fig:as-improvements} confirms that 
an increasing number of options for the path-selecting ASes
intensifies competition between transit ASes, which then
improve their attribute values in response.
In particular, about half of ASes improve
both their attribute values given 5 available paths,
whereas only 20\% improve their attributes in a duopoly scenario
(compared to a monopoly scenario).
Note that some ASes may decrease their attribute
values under competition for the counter-intuitive
reasons explained in~\cref{sec:theo:sapp-constant:competition}.

\begin{figure}
    \centering
    \includegraphics[width=\linewidth,trim=0 7 0 0]{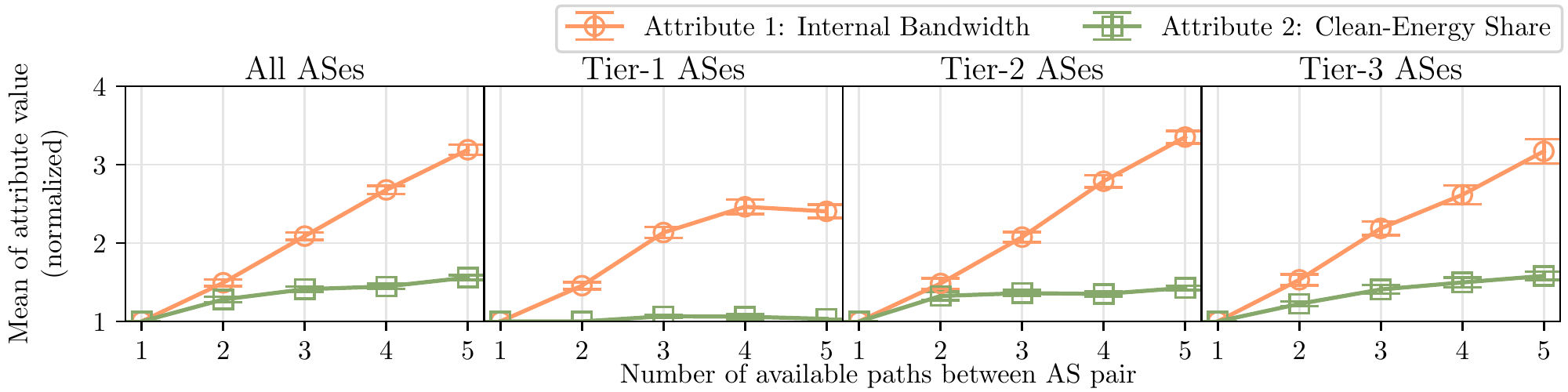}
    \caption{Competition tends to raise the mean attribute values across ASes (for all AS tier groups),
    compared to a monopoly situation. However, attributes may
    be affected differently by competition due to differences
    in valuation and cost.}
    \label{fig:mean}
\end{figure}

Moreover, the general level of attribute prevalence is raised by competition,
which is implied by the increasing global average of attribute values
in~\cref{fig:mean}.
The internal-bandwidth attribute is more strongly affected by this
average gain than the clean-energy--share attribute,
because (i) the bandwidth attribute is not upper-bounded, 
and (ii) the bandwidth attribute has zero demand-dependent cost.

\begin{figure}
    \centering
    \includegraphics[width=\linewidth,trim=0 7 0 0]{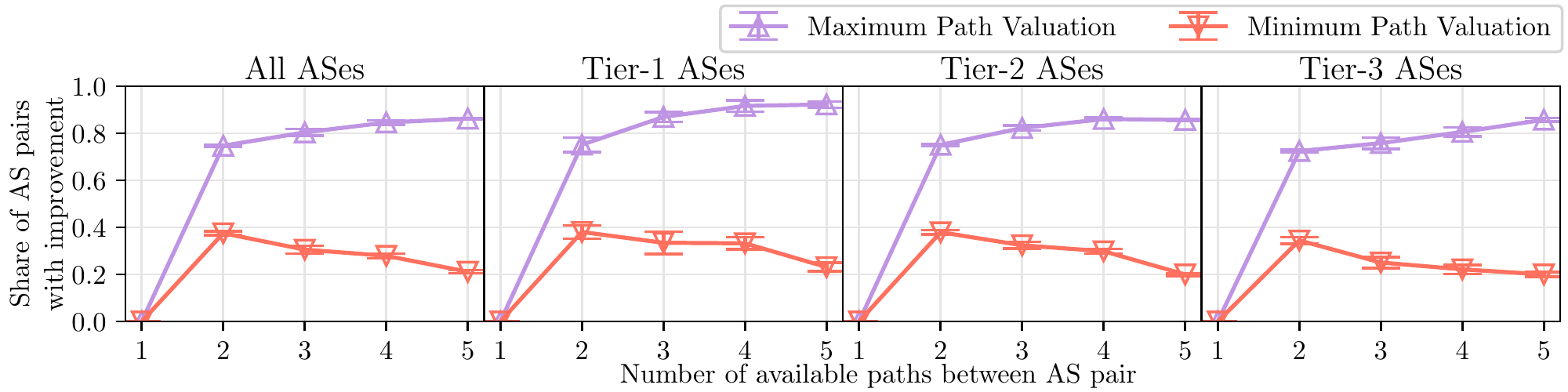}
    \caption{On average, competition raises the the attractiveness of the most
    \emph{and} the least attractive path that connect two ASes, independent of the source-AS tier.}
    \label{fig:as-pair-improvements}
\end{figure}

\paragraph{Effects on selecting ASes}
The average improvement in attributes
translates into a more attractive offer for path-selecting ASes.
More precisely, the most attractive path between
each AS pair tends to become more attractive
as competition increases:
\cref{fig:as-pair-improvements} shows 
that~75\% of AS pairs obtain access to a more
valuable path if two paths instead of a single path
are available (increasing maximum valuation),
irrespective of the tier of the path-selecting AS. 
Notably, we would expect that around 50\%
of AS pairs obtain a second path of higher quality
in the absence of dynamic competition effects.
Hence, the increasing maximum valuation
is a combined effect of multi-path availability
and competition.
Moreover, in absence of competition
effects, a second path can only decrease,
but not raise the minimum
valuation across available paths.
However, we observe that for 40\% of AS pairs,
both paths in a two-path scenario are more
highly valued than the single path in
a monopoly scenario, which suggests
that competition raises the value of
the previously monopolistic path. 
However, as the number of available paths increases, 
the tendency of additional paths to decrease 
the minimum quality becomes more visible.
Finally, these offer improvements
materialize for all tiers of path-selecting ASes.

\paragraph{Differences in market power}
Intriguingly, the higher market power of tier-1 ASes is
not visible in~\cref{fig:as-improvements}, as tier-1 ASes
are equally likely as lower-tier ASes to improve
their attributes. However, the market power of
tier-1 ASes becomes apparent in~\cref{fig:mean}, which indicates
that tier-1 ASes in competition improve their attributes
to a lower extent than ASes on lower tiers.

\subsubsection{Profits}
\label{sec:simulation:profits}

Regarding competition effects on profits,
we make the following observations:

\paragraph{Effects on transit ASes}
Increasing path diversity and competition
lead to increasing profits for a substantial share of 
ASes~(cf.~\cref{fig:as-improvements}).
At 5 available paths per AS pair, 75\% of ASes increase 
their profits.
This insight is surprising, given that
competition is traditionally expected to increase consumer
welfare and to reduce producer profits.
In an ISP market, however, profits may increase because
competition is modulated by path diversity.
Such path diversity not only allows selecting ASes to select more
different paths, but also allows transit ASes to attract and 
monetize traffic from more selecting ASes, increasing profit.
Importantly, such an increase in attracted demand for an ISP
does not necessarily reduce the attracted demand of another ISP,
as the volume of total demand is elastic in our model.

\paragraph{Differences in market power}
Interestingly, the profit increase under competition among 5 paths 
is more pronounced for tier-2 and tier-3 ASes than for tier-1 ASes.
The reason is that the tier-1 ASes become more likely to be circumvented 
as path-selecting ASes obtain additional path options, and
lower-tier ASes can thus attract more demand.

\subsubsection{Sensitivity to Model Functions}
\label{sec:simulation:results:nonlinearity}

\begin{figure*}
    \centering
        \begin{subfigure}{0.32\linewidth}
         \includegraphics[width=\linewidth,trim=0 0 0 0]{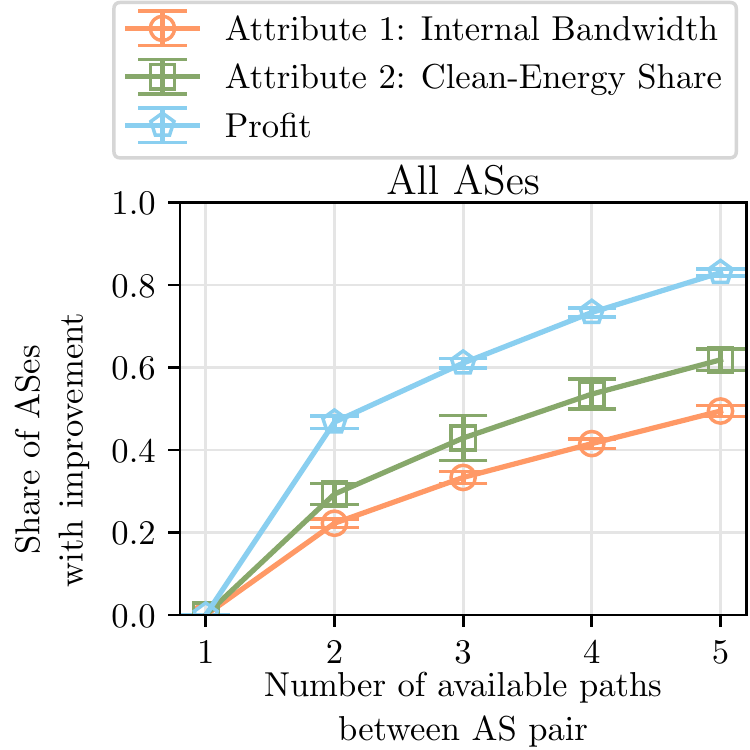}
        \caption{Improvements in transit ASes.}
        \label{fig:nonlinearity:as-improvements}
    \end{subfigure}\ \vrule\ 
    \begin{subfigure}{0.32\linewidth}
         \includegraphics[width=\linewidth,trim=0 0 0 0]{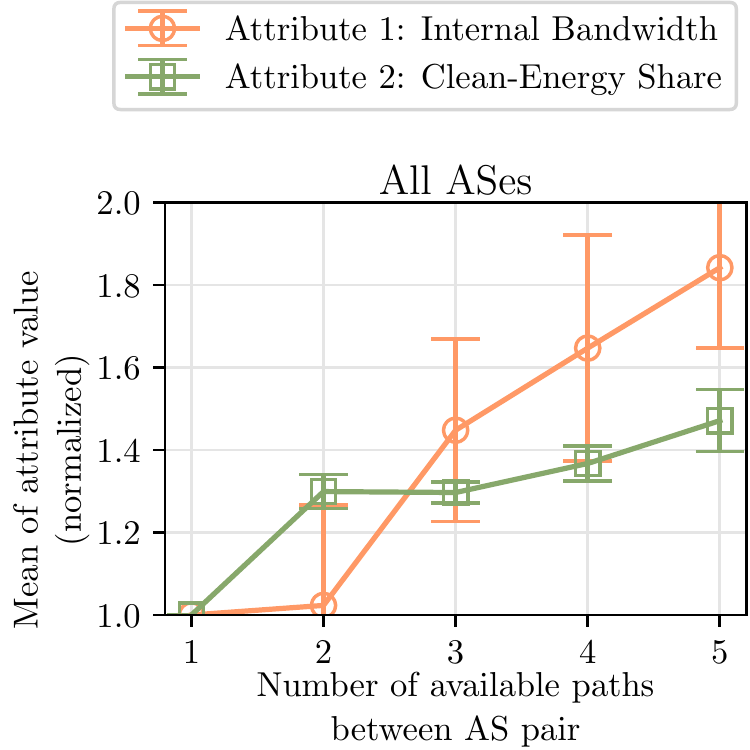}
        \caption{Mean of attribute values.}
        \label{fig:nonlinearity:mean}
    \end{subfigure}\ \vrule\ 
    \begin{subfigure}{0.32\linewidth}
         \includegraphics[width=\linewidth]{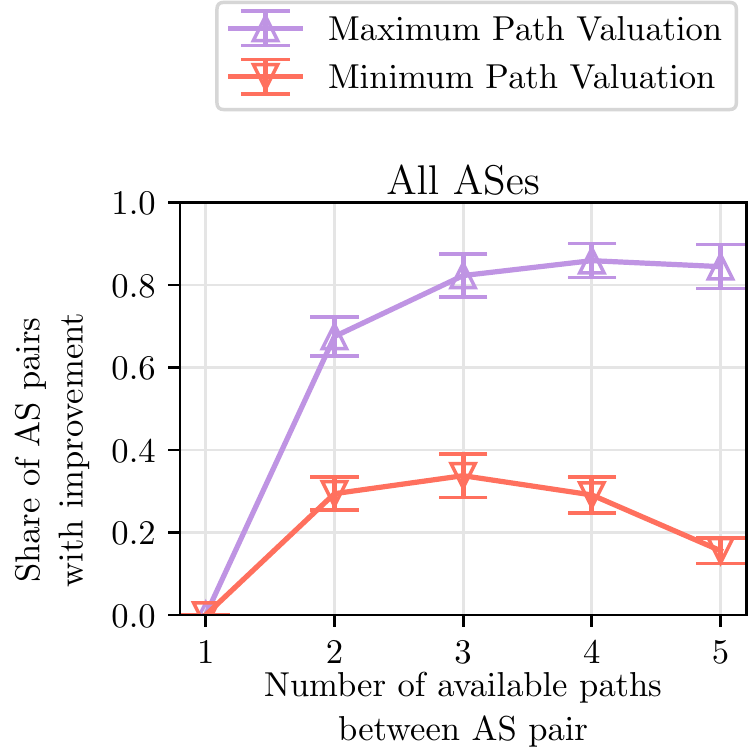}
        \caption{Improvements for selecting ASes.}
        \label{fig:nonlinearity:as-pair-improvements}
    \end{subfigure}
    \caption{Equilibrium results for competitive dynamics with non-affine functions
    for valuation and cost.}
    \label{fig:simulation:nonlinearity}
\end{figure*}

In our model, we assume affine functions for both the 
path-valuation functions~$v_r$ and the cost functions~$\Phi_n$
and~$\Gamma_n$. To assess the impact of this assumption,
we rerun the simulations by replacing linear dependencies
within these functions. In particular, we
replace~$a_{nk}$ in~$v_r$ by sub-linear~$\sqrt{a_{nk}}$ (cf.~\cref{eq:model:valuation}),
and replace~$a_{nk}$ in~$\Phi_n$ and~$\Gamma_n$ by super-linear~$a_{nk}^2$
(cf.~\cref{eq:model:revenue-cost}). The results are presented
in~\cref{fig:simulation:nonlinearity}.

Intriguingly, the results for the non-affine functions
closely match the results for affine functions in a qualitative
sence, i.e., competition improves the attribute values, profits,
and path options for a large share of ASes.
Quantitatively, the largest difference concerns the mean increase
in attribute values (cf.~\cref{fig:nonlinearity:mean} vs.~\cref{fig:mean}),
which is considerably lower for the bandwidth attribute
for the non-affine functions. However, this effect is to be
expected because the non-affine functions make attributes
both less valuable and more costly (if $a_{nk} > 1$, as for the
bandwidth attribute), and thus less attractive
to invest in.

\section{Related Work}
\label{sec:related-work}

\paragraph{General competition models}
Internet competition is frequently studied
by means of the three fundamental competition models
from the economic literature.
First, Cournot competition~\cite{cournot1838recherches} describes
multiple firms that produce the same homogeneous good,
individually determine the quantity to be produced,
and thus indirectly determine the market price.
Cournot competition suggests that in comparison
with a monopoly, a duopoly
increases the quantity of the good and reduces its
price, indicating that competition benefits consumers.
In the second competition model, Bertrand competition~\cite{edgeworth1925pure},
firms also produce a homogeneous good, but directly set a price instead of a quantity.
Moreover, the firm with the lowest price
acquires the whole market. Hence, Bertrand competition is
considered more suitable to analyze highly competitive markets.
Lastly, Stackelberg competition~\cite{von2010market} 
is similar to Cournot competition, but is suitable for hierarchical
markets in which follower firms determine their production quantity
after observing the quantity produced by a leader firm.
All competition models have been adapted to \emph{networked markets},
i.e., markets in which each segment of consumers can only
be served by a corresponding subset of 
firms~\cite{anshelevich2015price,bimpikis2019cournot,motalleb2019networked}.

Our competition model is more strongly inspired
by the logit-demand model~\cite{besanko1998logit}, 
which originates from econometrics, can more directly
represent competition between goods with different
characteristics, and has been used in research
on Internet transit pricing~\cite{valancius2011many}.
Still, the market in our model is networked, as
determined by the network topology.

\paragraph{Internet competition models}
To study Internet competition in particular, Shakkottai and Srikant~\cite{shakkottai2006economics} leverage Bertrand and Stackelberg competition 
to theoretically analyze the effect
of competition in different levels of the Internet,
i.e., for tier-1, tier-2, and local ISPs.
Their model shows that competition
may exert downwards pressure on prices, and
an assimilating pressure on the quality-of-service (QoS) levels
offered by different ISPs.
These insights are confirmed by a subsequent line
of simulation-based research~\cite{marentes2015exploring,teixeira2017economic}.
These studies analyze the ISP competition 
induced by virtualized access networks and
by the ChoiceNet proposal~\cite{wolf2014choicenet}, 
which describes a marketplace for transit services.
Using both theoretical analysis and simulation,
Nagurney and Wolf~\cite{nagurney2014cournot}
expand on a study by Zhang et al.~\cite{zhang2010interactions} to
investigate the intertwined competition dynamics
among service providers (in Bertrand competition)
and among network providers (in Cournot competition).
In this analysis, the offers of service providers and
network providers differ in both price and quality,
and converge to an equilibrium describable by
variational-inequality conditions.

In this paper, we extend the previous work in
a number of aspects.
First, our model acknowledges 
that path quality may be determined collectively by
multiple selfish on-path ISPs, and
reveals the inefficient cooperation within a
path due to selfishness.
In contrast, previous work assumed that
network service is bought from a single
access/transit provider, and is thus
effectively limited to one-hop paths.
Second, path quality in our model depends
on multiple underlying attributes, whereas
previous work abstracts path quality in a single
attribute. This fine-grained view of quality attributes
is valuable, as it reveals how different attributes
are affected differently by competition (cf.~\cref{sec:theo:sapp-constant}).
Third, our model represents the internal cost structure
of ISPs in a detailed manner, as it (i)
distinguishes demand-dependent and demand-independent cost,
and (ii) formalize the cost dependence on
quality attributes, unlike previous work.
Fourth, we make an effort to find realistic parameters
for our large-scale simulations, whereas the parameters
in previous simulation-based works are arbitrary.
Finally, we explicitly investigate the differences between
differing degrees of competition, and find
network examples in which more intense competition lead
to previously undocumented effects, namely
increasing profits and decreasing path quality.

\paragraph{ISP cooperation}
The economic dynamics between between network entities 
that collectively provide connectivity
has been studied with the lens of cooperative game theory~\cite{branzei2008models},
i.e., assuming that agents within a group choose rules
which are enforced thereafter.
Such considerations can inform financial settlement
among ISPs in a coalition, where settlement mechanisms
based on the Shapley value~\cite{ma2007internet} or
ISP characteristics~\cite{zhou2017cooperative} have been
proposed.
In our work, we discuss intra-path dynamics using
non-cooperative game theory, as setting up a binding contract 
among the ISPs on every path is difficult in practice. 
Moreover, our model also reflects that multiple coalitions (paths) 
are in competition, which is missing from previous work.

\section{Conclusion}
\label{sec:conclusion}

ISPs determine the quality attributes of their connectivity offer
(e.g., performance metrics, security features, sustainability properties)
in line with their profit objective and alternative
offers by other ISPs.
The presence of such alternative offers (i.e., competition)
tends to improve path quality, 
as we demonstrate in this paper.
We provide evidence for this positive
effect of ISP competition with an extensive
theoretical analysis, based
on a new game-theoretic model, 
and a large-scale simulation,
based on empirical data. 
Our theoretical analysis suggests that
an augmented path choice incentivizes
transit ISPs to improve path quality, especially if ISPs
have similar cost structures
(\cref{thm:beneficial-competition,thm:theo:sapp-constant:competition:good}). 
Interestingly,
this higher investment in quality attributes
does not necessarily reduce transit ISP profits, 
as entering competition (by connecting to new customers) 
also allows transit ISPs to attract revenue-generating traffic from new customer segments 
(\cref{thm:theo:sapp-homogeneous:competition:profits}).
While these positive effects do not materialize 
in unfavorable circumstances (\cref{thm:theo:sapp-constant:competition:bad}), 
our simulation-based
case study indicates largely positive effects
of competition in practice.

Our analysis does not only reveal the macroscopic effects
of competition, but also formalizes the rational attribute choice
for ISPs, which can inform ISP business strategies.
In particular, we obtain three main recommendations for
quality investment from our analysis:
\begin{itemize}
    \item \textit{Invest in attributes with low fixed cost:}
    \Cref{thm:sapp:a_opt} suggests that the optimal extent of a quality
    attribute correlates with the ratio of demand-dependent attribute-specific 
    cost to total attribute-specific cost.
    Hence, the lower the demand-independent (fixed) cost of
    an attribute, the higher the optimal investment in the attribute.
    For example, renting internal bandwidth on-demand tends to
    improve profit more than a fixed bandwidth installation.
    
    \item \textit{Invest exclusively in attributes with high return:}
    \Cref{thm:sapp:a_opt} also shows a correlation between the
    optimal attribute extent and the attribute return, i.e.,
    the attribute-specific net revenue per traffic unit,
    divided by the attribute-specific cost. 
    \Cref{thm:heterogeneous:equilibrium:two}
    even suggests \emph{specialization} in heterogeneous markets
    with negligible demand-dependent cost, 
    i.e., only the path attribute with the highest return should be 
    invested in, while all less attractive attributes should be abandoned.

    \item \textit{Engage in competition and on-path cooperation:}
    Both our theoretical analysis (\cref{thm:theo:sapp-homogeneous:competition:profits}) and
    our simulations (\cref{sec:simulation}) show that engaging
    in competition by connecting to new customers tends
    to increase transit ISP profit, as the revenue from newly attracted traffic
    generally outweighs the costs of competing in the new markets.
    Furthermore, transit ISPs should also engage in cooperation with
    other on-path ISPs by coordinating attribute
    investment. Such coordination leads to achievement of the Nash
    bargaining solution, and therefore to higher profits and
    higher path quality, which also benefits path users 
    (\cref{thm:theo:sapp-homogeneous:intra-path:equilibrium-optimum,thm:heterogeneous:single-path:valuation}).
    However, to achieve stable cooperation among selfish ISPs,
    additional work based on mechanism design is needed.
    
\end{itemize}


Finally, we emphasize that our new competition model 
is not only applicable to ISP competition,
but in general to settings in which coalitions of
selfish entities stand in competition.
While ISPs and paths represent the entities and coalitions
in the ISP market, competition in other markets
arises between firms that form value chains.
Our model allows investigating complex economic phenomena
such as the interaction between firms along a value chain,
or the effect of overlapping value chains.
Hence, we are intrigued to investigate whether 
our findings for the ISP market translate to other
economic sectors.

\bibliographystyle{ACM-Reference-Format}
\bibliography{ref}

\newpage
\appendix



\section{Proof of~\cref{thm:sapp:a_opt}}
\label{sec:app:proofs:parallel--optimal-attr}

\subsection{Unrestricted best response in~\cref{eq:theo:a_opt}}
    For an individual source-destination pair~$(n_1, n_2)$, we can simplify~\cref{eq:model:profit} as follows:
    \begin{equation}
        \pi_n = \frac{dv_{r(n)}}{1 + \sum_{r'\in R} v_{r'}} \cdot \left(\rho_n - \sum_{k \in K} \phi_{nk}a_{nk} - \phi_{n0}\right) - \sum_{k\in K}\gamma_{nk}a_{nk} - \gamma_{n0}
        \label{eq:theo:sapp:profit}
    \end{equation} where the argument~$\mathbf{A}$ has been omitted and
    \begin{equation}
        v_{r(n)} = \sum_{r \in R.\ n \in r} v_r.
    \end{equation}
    Differentiating~\cref{eq:theo:sapp:profit} with respect to attribute~$a_{nk}$ of ISP~$n$ yields:
    \begin{equation}
        \frac{\partial\pi_n}{\partial a_{nk}} = \frac{d\alpha_{nk}\left(1+v_{-r(n)}\right)}{\left(1 + \sum_{r'\in R} v_{r'}\right)^2} \cdot \left(\rho_n - \sum_{k \in K} \phi_{nk}a_{nk} - \phi_{n0}\right) - \frac{d \phi_{nk}  v_{r(n)}}{1 + \sum_{r'\in R} v_{r'}}  - \gamma_{nk}
        \label{eq:theo:sapp:profit:d1}
    \end{equation} where the abbreviations from~\cref{eq:theo:sapp:shorthands} have been used.
    Setting~\cref{eq:theo:sapp:profit:d1} to 0 allows the following rewriting:
    \begin{equation}
        d\alpha_{nk} \left(1 + v_{-r(n)}\right) \cdot \left(\rho_n -\sum_{k \in K} \phi_{nk}a_{nk} - \phi_{n0}\right) - d \phi_{n1} v_{r(n)} \cdot \left(1 + \sum_{r'\in R} v_{r'}\right) - \gamma_{n1} \cdot \left(1 + \sum_{r'\in R} v_{r'}\right)^2 = 0
    \end{equation} where we can substitute
    \begin{equation}
        v_{r(n)} = \alpha_{nk}a_{nk} + \sum_{k' \in K\setminus k} \alpha_{nk'} a_{nk'} + \alpha_{n0} \hspace{20pt}
        \sum_{r' \in R} v_{r'} = \alpha_{nk}a_{nk} + v_{-nk}
    \end{equation} and thus obtain a quadratic equation in~$a_{nk}$. This quadratic equation has the solutions:
    \begin{equation}
        a_{nk} = \frac{1}{\alpha_{nk}}\left(\pm\sqrt{\frac{d\big(1+v_{-r(n)}\big)}{d\phi_{nk}+\gamma_{nk}} \big(\phi_{nk} (1+v_{-nk}) + \alpha_{nk} (\rho_n - \Phi_{-nk}) \big) \Big) } -\big(1+v_{-nk}\big)\right)
        \label{eq:theo:sapp:quadratic-eq-solution} 
    \end{equation} where only the upper solution (i.e., with the positive coefficient of the square-root term) is potentially valid
    given the non-negativity of attributes.
    Hence, we arrive at~$\hat{a}_{nk}^\ast(\mathbf{A}_{-nk})$ as in~\cref{eq:theo:a_opt}.

\subsection{Confirmation of maximum}

    This solution is a maximum of~$\pi_n$ if~$\pi_n$ 
    is concave in~$a_{nk}$, which can be
    demonstrated by means of the second derivative:
    \begin{equation}
         \frac{\partial^2\pi_n}{\partial a_{nk}^2} = \frac{-2d\alpha_{kn}^2\left(1+v_{-r(n)}\right)}{\left(1 + \sum_{r'\in R} v_{r'}\right)^3} \cdot \left(\rho_n - \sum_{k \in K} \phi_{nk}a_{nk} - \phi_{n0}\right) -   \frac{2d\alpha_{n}\phi_{nk}\left(1+v_{-r(n)}\right)}{\left(1 + \sum_{r'\in R} v_{r'}\right)^2}
         \label{eq:theo:sapp:profit:d2}
    \end{equation}
    Clearly, $\pi_n$ may only be non-concave under the following condition:
    \begin{equation}
        \frac{\partial^2\pi_n}{\partial a_{nk}^2} > 0 \implies \rho_n - \sum_{k \in K} \phi_{nk}a_{nk} - \phi_{n0} < 0
        \label{eq:theo:sapp:profit:non-concavity}
    \end{equation}

    However, if the condition in~\cref{eq:theo:sapp:profit:non-concavity} is true for some~$\mathbf{a}_n \in \mathbb{R}^{K}_{\geq 0}$,
    then the profit function has a negative slope at that point (cf.~\cref{eq:theo:sapp:profit:d1}).
    Hence, the profit function has no extrema in the non-concave regions, 
    and thus any extremum, in particular~$\hat{a}_{nk}^\ast(\mathbf{A}_{-nk})$,
    is guaranteed to be located in the concave regions and to be a maximum.

\subsection{Restricted best response in \cref{eq:theo:a_opt:restricted}}

    Since attribute values must be non-negative, we now investigate
    how the unrestricted best-response~$\hat{a}_{nk}^\ast$
    informs the best-response~$a_{nk}^{\ast}$ on the restricted domain~$\mathbb{R}_{\geq 0}$.
    Clearly, if~$\hat{a}_{nk}^\ast \geq 0$, then $a_{nk}^{\ast} = \hat{a}_{nk}^\ast$.
    Otherwise, if~$\hat{a}_{nk}^\ast < 0$, the boundary point~$a_{nk} = 0$ constitutes
    a local maximum on the restricted domain~$\mathbb{R}_{\geq 0}$.
    To confirm this statement, we have to distinguish the cases~$\phi_{nk} > 0$
    and~$\phi_{nk} = 0$. 
    If~$\phi_{nk} > 0$, we note that the profit function can only be 
    non-concave for high enough~$a_{nk}$:
    \begin{equation}
        \rho_n - \sum_{k \in K} \phi_{nk}a_{nk} - \phi_{n0} < 0 \iff a_{nk} > \frac{\rho_n - \Phi_{-nk}}{\phi_{nk}} =: \overline{a}_{nk}
        \label{eq:theo:sapp:profit:concavity-threshold}
    \end{equation}
    Hence,~$\pi_{n}$ is guaranteed to be concave for all~$a_{nk} \leq \overline{a}_{nk}$. Furthermore,
    as argued in the previous paragraph, $\pi_n$ is strictly decreasing for all~$a_{nk} > \overline{a}_{nk}$.
    Hence,~$a_{nk} = 0$ is the maximum on the restricted domain independent of~$\overline{a}_{nk}$,
    given that~$\hat{a}^{\ast}_{nk} < 0$.
    If~$\phi_{nk} = 0$, $\pi_n$ can only be non-concave 
    if~$\rho_n - \Phi_{-nk} < 0$, independent of~$a_{nk}$.
    Hence,~$\pi_n$ is either guaranteed to be consistently concave in~$a_{nk}$
    (if~$\rho_n - \Phi_{-nk} \geq 0$), or guaranteed to be decreasing
    for all~$a_{nk} \geq 0$. In both cases, the boundary point~$a_{nk} = 0$
    constitutes a local maximum.

    Finally, we investigate the case where the unrestricted
    best response~$\hat{a}^{\ast}_{nk}$ is \emph{undefined} on~$\mathbb{R}$, 
    i.e., if the term under the square root is negative.
    A necessary condition for this negativity is that~$\rho_n - \Phi_{-nk}$
    is negative, which according to~\cref{eq:theo:sapp:profit:concavity-threshold}
    implies that~$\overline{a}_{nk} < 0$. Hence, in this case~$\pi_n$ is decreasing
    for~$a_{nk} \geq 0$, which again makes~$a_{nk} = 0$ a local maximum
    on the restricted domain~$\mathbb{R}_{\geq 0}$.

\section{Proof of~\cref{thm:theo:sapp-homogeneous:equilibrium}}
\label{sec:app:proofs:homogeneous--equilibrium}
\subsection{Homogeneous profit function}

    In a homogeneous network, all attributes are equally valuable and costly,
    i.e., $\alpha_{nk}$, $\phi_{nk}$ and~$\gamma_{nk}$ are equal across
    all attributes~$k \in K$. This homogeneity allows to understand
    the profit function~$\pi_n$ of a ISP~$n$ as a function of
    the attribute sums~$a_n = \sum_{k \in K} a_{nk}$:
    \begin{equation}
        \pi_n(\mathbf{A}) = \frac{\alpha_1 a_{n} + \alpha_1 \sum_{n' \in r(n)\setminus\{n\}} a_{n'} + \alpha_0 }{1 + \alpha_1 a_{n} + \alpha_1 \sum_{n'\in N\setminus\{n\}} a_{n'} + Q\alpha_0} 
        d (\rho - \phi_1 a_n - \phi_0) - \gamma_1 a_n - \gamma_0
    \end{equation}
    In the following, we thus treat the attribute sum~$a_n$ like a single attribute of ISP~$n$. 

\subsection{Unrestricted equilibrium in~\cref{eq:theo:sapp-homogeneous:equilibrium:raw}}
    
    The equilibrium conditions in~\cref{eq:theo:sapp:equilibrium} suggest that
    the equilibrium for a homogeneous parallel-path network satisfies
    the following equation for every~$n\in N$:
    \begin{equation}
        a_n^+ = \max\left(0,\ \frac{\sqrt{\frac{d\big(1+v_{-r(n)}(\mathbf{A}_{-n}^+)\big)}{d\phi_{1}+\gamma_{1}} \big(\phi_{1} (1+v_{-n}(\mathbf{A}_{-n}^+)) + \alpha_1 (\rho - \phi_{0}) \big) \Big) } -\big(1+v_{-n}(\mathbf{A}_{-n}^+)\big)}{\alpha} \right)
        \label{eq:the:sapp-homogeneous:equilibrium:best-response}
    \end{equation} where
    \begin{align}
        v_{-r(n)}(\mathbf{A}_{-n}^+) &= \alpha_1 \sum_{n' \in N\setminus r(n)} a_{n'}^+ + (Q-1) \alpha_0\\
        v_{-n}(\mathbf{A}_{-n}^+) &= v_{-r(n)}(\mathbf{A}_{-n}^+) + \alpha_1 \sum_{n' \in r(n)\setminus\{n\}} a_{n'}^+ + \alpha_0
    \end{align}
    Note that we effectively consider a single attribute in the
    style of the attribute sum, which simplifies~$\Phi_{-nk} = \phi_0 \leq \rho$.
    Hence, the undefined case from~\cref{thm:sapp:a_opt} does not arise because
    the term under the square root in~\cref{eq:the:sapp-homogeneous:equilibrium:best-response}
    is always non-negative.
    
    To solve the equation system \cref{eq:the:sapp-homogeneous:equilibrium:best-response} $\forall n \in N$, 
    we first consider the unrestricted system (i.e., without required non-negativity
    of solutions) in~$\mathbf{\hat{A}}^+$:
    \begin{equation}
        \forall n \in N.\quad \hat{a}_n^+ = \frac{\sqrt{\frac{d\big(1+v_{-r(n)}(\mathbf{\hat{A}}_{-n}^+)\big)}{d\phi_{1}+\gamma_{1}} \big(\phi_{1} (1+v_{-n}(\mathbf{\hat{A}}_{-n}^+)) + \alpha_1 (\rho - \phi_{0}) \big) \Big) } -\big(1+v_{-n}(\mathbf{\hat{A}}_{-n}^+)\big)}{\alpha}
    \end{equation} This equation system can be transformed such that the LHS is constant
    across all equations:
    \begin{equation}
        \forall n\in N.\quad 1 + \alpha_1 \sum_{r\in R}\sum_{n' \in r} \hat{a}^+_n + Q\alpha_0 =  \sqrt{\frac{d\big(1+v_{-r(n)}(\mathbf{\hat{A}}_{-n}^+)\big)}{d\phi_{1}+\gamma_{1}} \big(\phi_{1} (1+v_{-n}(\mathbf{\hat{A}}_{-n}^+)) + \alpha_1 (\rho - \phi_{0}) \big) \Big) }
    \end{equation} which implies that all~$a_n^+\ \forall n\in N$ are equal to a value~$\hat{a}^+$. 
    This value~$\hat{a}^+$ can be found by solving the following single equation:
    \begin{equation}
        \hspace{-6pt}1 + QI \alpha_1 \hat{a}^+ + Q\alpha_0 =  \sqrt{\frac{d\big(1+(Q-1)(I\alpha_1\hat{a}^+ + \alpha_0)\big)}{d\phi_{1}+\gamma_{1}} \big(\phi_{1} (1+(QI-1)\alpha_1\hat{a}^+ + Q\alpha_0) + \alpha_1 (\rho - \phi_{0}) \big) \Big) }
    \end{equation}
    This equation is solved by~$\hat{a}^+$ as defined in~\cref{thm:theo:sapp-homogeneous:equilibrium}.

\subsection{Restricted equilibrium}

    It remains to show how the solution~$\hat{a}^+$ of the unrestricted system can be used to derive
    the actual solution~$a^+$ of the restricted system. If~$\hat{a}^+ \geq 0$, the unrestricted-system
    solution~$\hat{a}^+$ is clearly also a solution to the restricted system, i.e., $a^+ = \hat{a}^+$.
    However, if~$\hat{a}^+ < 0$, the attribute values as suggested by the unrestricted system are negative,
    which is invalid for the restricted system. In this case, we can show that a solution of the
    restricted system is given by~$a^+ = 0$, i.e., $0 = a_n^{\ast}(\mathbf{0})$
    (where~$a_n^{\ast}$ is the optimal choice in parallel-path networks according to \cref{thm:sapp:a_opt}).
    To show this property, we first observe that the condition for~$\hat{a}^+ < 0$
    can be simplified to a condition on sub-term~$T_3$:
    \begin{align}
        \hat{a}^+ < 0 &\implies \frac{\sqrt{T_2^2 - 4 T_1 T_3} - T_2}{2 T_1} < 0\\
        \text{For $T_1 > 0$:} & \quad T_2^2 - 4 T_1 T_3 < T_2^2 \implies - 4 T_1 T_3 < 0 \implies T_3 > 0\\
        \text{For $T_1 < 0$:} & \quad T_2^2 - 4 T_1 T_3 > T_2^2 \implies - 4 T_1 T_3 > 0 \implies T_3 > 0\\
        &\implies T_3 = (1 + Q\alpha_0)^2 - \frac{d(1 + (Q-1)\alpha_0)}{d\phi_1 + \gamma_1} \big(\phi_1(1+Q\alpha_0)+\alpha_1(\rho-\phi_0)\big) > 0
    \end{align}
    Furthermore, the inequality on~$T_3$ allows the following conclusion:
    \begin{align}
        &(1 + Q\alpha_0)^2 - \frac{d(1 + (Q-1)\alpha_0)}{d\phi_1 + \gamma_1}  \big(\phi_1(1+Q\alpha_0)+\alpha_1(\rho-\phi_0)\big) > 0\\
        \iff\ &(1 + Q\alpha_0) - \sqrt{\frac{d(1 + (Q-1)\alpha_0)}{d\phi_1 + \gamma_1} \big(\phi_1(1+Q\alpha_0)+\alpha_1(\rho-\phi_0)\big)} > 0 \label{eq:theo:sapp-homogeneous:0-equilibrium}
    \end{align}
    due to the equivalence~$x^2 - y > 0 \iff x^2 > y \iff x > \sqrt{y} \iff x - \sqrt{y} > 0$ (if $x, y \geq 0$).
    
    \cref{eq:theo:sapp-homogeneous:0-equilibrium} has a striking similarity to~$\hat{a}_n^{\ast}(\mathbf{0})$
    for a homogeneous parallel-path network:
    \begin{equation}
        \hat{a}_n^{\ast}(\mathbf{0}) = \frac{\sqrt{\frac{d\big(1+(Q-1)\alpha_0\big)}{d\phi_{1}+\gamma_{1}} \big(\phi_{1} (1+Q\alpha_0) + \alpha_1 (\rho - \phi_{0}) \big) \Big) } -\big(1+Q\alpha_0\big)}{\alpha}
    \end{equation} More precisely, \cref{eq:theo:sapp-homogeneous:0-equilibrium} implies 
    that~$\hat{a}_n^{\ast}(\mathbf{0})$ is always below 0 if~$\hat{a}^+ < 0$,
    and that~$a_n^{\ast}(\mathbf{0})$ thus always equals 0 for~$\hat{a}^+ < 0$. This insight concludes the proof.

\section{Proof of~\cref{thm:theo:sapp-homogenous:stability}}
\label{sec:app:proofs:homogeneous--stability}
\subsection{Linearization of dynamic system}

    In order to prove asymptotic stability of the given Nash equilibrium,
    we leverage the indirect Lyapunov method~\cite{pukdeboon2011review}. 
    This method requires that the equilibrium of an ODE system is asymptotically
    stable if the Jacobian matrix of the ODE system, evaluated at the equilibrium
    point, has exclusively negative eigenvalues. 
    More formally, given the Jacobian matrix~$\mathbf{J}(\mathbf{A}^+) \in \mathbb{R}^{|N|\times|N|}$,
    it must hold that~$\forall \lambda \in \mathbb{R}$ that $\exists \mathbf{x} \in \mathbb{R}^{|N|}, \mathbf{x} \neq \mathbf{0}.\ \mathbf{J}(\mathbf{A}^+)\mathbf{x} = \lambda\mathbf{x} \implies \lambda < 0$.
    This matrix~$\mathbf{J}(\mathbf{A}^+)$ is defined as follows for the dynamic system
    from~\cref{eq:theo:sapp-homogeneous:dynamics}:
    \begin{alignat}{4}
        &J_{nn} &= \frac{\partial \dot{a}_n}{\partial a_n}(\mathbf{A}^+)  &= -1 \\
        n\neq m,\ r(n) = r(m):\quad &J_{nm} &= \frac{\partial \dot{a}_n}{\partial a_m}(\mathbf{A}^+) &= 
        \begin{cases}
            \frac{T_4}{T_5} - 1 & \text{if } \hat{a}^+ \geq 0\\
            0 & \text{otherwise}
        \end{cases}\\
        n\neq m,\ r(n) \neq r(m):\quad & J_{nm} &= \frac{\partial \dot{a}_n}{\partial a_m}(\mathbf{A}^+)  &=
        \begin{cases}
            \frac{T_4+T_6}{T_5} - 1 & \text{if } \hat{a}^+ \geq 0\\
            0 & \text{otherwise}
        \end{cases}
    \end{alignat}
    where~$\hat{a}^+$ is the unrestricted equilibrium attribute value according to \cref{thm:theo:sapp-homogeneous:equilibrium}, and
    \begin{align}
        T_4 &= d\phi_1\big(1+v_{-r(n)}(\mathbf{\hat{A}}^+)\big),\\
        T_5 &= 2(d\phi_1+\gamma_1) \sqrt{\frac{d\big(1+v_{-r(n)}(\mathbf{\hat{A}}_{-n}^+)\big)}{d\phi_{1}+\gamma_{1}} \big(\phi_{1} (1+v_{-n}(\mathbf{\hat{A}}_{-n}^+)) + \alpha_1 (\rho - \phi_{0}) \big) }, \text{ and}\\
        T_6 &= d\big(\phi_{1} (1+v_{-n}(\mathbf{\hat{A}}_{-n}^+)) + \alpha_1 (\rho - \phi_{0}) \big).
    \end{align}

\subsection{Case 1: Non-negative unrestricted equilibrium ($\hat{a}^+ \geq 0$)}
    
    We first consider the case of a non-negative unrestricted equilibrium value~$\hat{a}^+$ 
    such that~$\mathbf{A}^+ = \mathbf{\hat{A}}^+$. In that case, the eigenvalue condition
    induces the following system of equations:
    \begin{equation}
        \forall n \in N. \quad (-\lambda - 1) x_n + \left(\frac{T_4}{T_5}-1\right) \sum_{n'\in r(n)\setminus\{n\}} x_{n'} + \left(\frac{T_4+T_6}{T_5} - 1\right) \sum_{n'\in N\setminus r(n)} x_{n'}= 0
    \end{equation}
    This system has a number of solutions~$(\lambda, \mathbf{x})$. First, for~$\lambda_1 = -T_4/T_5$, 
    the equation system is reduced from~$|N|$ ISP-specific to~$|R|$ path-specific equations:
    \begin{equation}
        \forall r \in R. \quad \left(\frac{T_4}{T_5}-1\right) X_r + \left(\frac{T_4+T_6}{T_5} - 1\right) \sum_{r'\in R\setminus \{r\}} X_{r'} = 0 \quad \text{where} \quad X_r = \sum_{n'\in r} x_{n'}
    \end{equation}
    Equation systems of this form may have three types of solutions in $\mathbf{x}$. 
    For~$T_6 = 0$ and $T_4 = T_5$, any~$\mathbf{x}$ is a solution, as the coefficients
    of the variables~$X_r$ are 0.
    For~$T_6 = 0$ and $T_4 \neq T_5$, any~$\mathbf{x}$ with entries summing up to 0 is a solution,
    as the sum of all~$X_r$ has a single non-zero coefficient.
    For~$T_6 \neq 0$, any~$\mathbf{x}$ with~$X_r = 0\ \forall r \in R$ is a solution. 
    More importantly for the proof,~$\lambda_1 = -T_4/T_5$ is consistently negative
    given the parameter ranges, except for the case where~$\phi_1 = 0$ and hence~$\lambda_1 = 0$.
    The case of~$\phi_1 =0$ is indeed an interesting special case for which the equilibrium
    is not unique, as we will show in~\cref{sec:theo:sapp-constant}; 
    for this special case, $\lambda_1 = 0$ describes the fact that the dynamics
    do not converge to a specific equilibrium if they have already converged onto another equilibrium.
    
    After discovering the first eigenvalue~$\lambda_1 = -T_4/T_5$, we now
    consider the case where~$\lambda \neq -T_4/T_5$. In this case,
    the symmetric structure of the equation system implies that
    all eigenvector entries~$x_n$ associated with the same path~$r$ are equal.
    Hence, a reduction of the equation system to~$|R|$ equations
    is possible again:
    \begin{equation}
        \forall r \in R.\ \quad \left(\frac{-\lambda - 1}{I} + \frac{I-1}{I}\left(\frac{T_4}{T_5}-1\right)\right) X_r + \left(\frac{T_4+T_6}{T_5} - 1\right) \sum_{r'\in R\setminus \{r\}} X_{r'} = 0
        \label{eq:theo:sapp-homogeneous:stability-eq-system1}
    \end{equation}
    Again, this equation system admits different types of solutions.
    The first type is associated with the following eigenvalue:
    \begin{equation}
        \frac{-\lambda_2 - 1}{I} + \frac{I-1}{I}\left(\frac{T_4}{T_5}-1\right) = \frac{T_4+T_6}{T_5} - 1 \quad \implies \quad \lambda_2 = -\frac{T_4 + IT_6}{T_5}
    \end{equation} For such~$\lambda_2$, the equation system is solved  by all~$\mathbf{x}$ if~$T_4 + T_6 = -T_5$,
    and by all~$\mathbf{x}$ with~$\sum_{r \in R} X_r = 0$ otherwise.
    Moreover,~$\lambda_2$ is consistently negative.
    The second type of solution for the equation system in \cref{eq:theo:sapp-homogeneous:stability-eq-system1}
    is associated with the following eigenvalue:
    \begin{equation}
        \frac{-\lambda_3 - 1}{I} + \frac{I-1}{I}\left(\frac{T_4}{T_5}-1\right) = -(Q-1)\left(\frac{T_4+T_6}{T_5} - 1\right) \quad \implies \quad \lambda_3 = QI \left(\frac{T_4+T_6}{T_5} - 1\right) - \frac{T_4+IT_6}{T_5}
        \label{eq:theo:sapp-homogeneous:critical-eigenvalue}
    \end{equation} For such~$\lambda_3$, the equation system is solved by all~$\mathbf{x}$ with equal~$X_r$ across all paths~$r \in R$.
    By inspection of~$\lambda_3$, we confirm that the maximum~$\lambda_3$
    is negative:
    \begin{equation}
        \max_{\substack{\alpha_1,\alpha_0,\phi_1,\phi_0,\\\gamma_1,\rho,d,Q,I}} \lambda_3 
        = \max_{\phi_1,Q} 
        \lim_{\substack{\alpha_1,\alpha_0,\rho,d\\\rightarrow\infty}}\ 
        \lim_{\substack{\phi_0,\gamma_1\\\rightarrow0}}\ 
        \lim_{I\rightarrow 1}\ \lambda_3 < -\frac{1}{2}
    \end{equation}
    Hence, all eigenvalues of~$\mathbf{J}(\mathbf{A}^+)$
    for~$\hat{a}^+ > 0$ are negative, i.e., the equilibrium is asymptotically
    stable in this case.

\subsection{Case 2: Negative unrestricted equilibrium  ($\hat{a}^+ < 0$)}
  
    It remains to show that the equilibrium~$\mathbf{A}^+$ is also
    asymptotically stable for the case~$\hat{a}^+ < 0$ such that
    $\mathbf{A}^+ = \mathbf{0}$. This part of the proof is trivial:
    For~$\hat{a}^+ < 0$, the Jacobian~$\mathbf{J}(\mathbf{A}^+)$
    corresponds to the negative identity matrix, which 
    only has the negative eigenvalue~$\lambda=-1$.
    Hence, the proof is concluded.

\section{Proof of~\cref{thm:theo:sapp-homogeneous:intra-path:equilibrium-optimum}}
\label{sec:app:proofs:homogeneous--intra-path}
\subsection{NBS attribute}

    To characterize the NBS attribute~$a^{\circ}$,
    we first require an understanding of the aggregate profit
    of ISPs on the path:
    \begin{equation}
        \sum_{n \in r} \pi_n(\mathbf{A}) =
        d \frac{\alpha_1 \sum_{n\in r} a_n + \alpha_0}{1 + \alpha_1 \sum_{n\in r} a_n + \alpha_0}( I (\rho-\phi_0) - \phi_1 \sum_{n\in r} a_n )
        - \gamma_1 \sum_{n\in r} a_n - \gamma_0
    \end{equation}
    The aggregate profit can thus be considered a function
    of the sum~$a_r$ of ISP attributes on path, i.e., 
    $a_r := \sum_{n\in r} a_n$. By~\cref{thm:sapp:a_opt},
    the unrestricted optimal attribute sum~$a_r^{\circ}$ 
    is~$a_r^{\circ} = \max(0, \hat{a}_r^\circ)$, where:
    \begin{equation}
        \hat{a}_r^{\circ} = \frac{\sqrt{\frac{d}{d\phi1 + \gamma_1}\left(\phi_1(1+\alpha_0) + I\alpha_1(\rho-\phi_0)\right)} - (1+\alpha_0)}{\alpha_1}.
    \end{equation}
    
    Clearly, the Nash bargaining attributes~$\{a_n^{\circ}\}_{n\in r}$
    must sum to~$a_r^{\circ}$ in order to be optimal.
    Moreover, the Nash bargaining solution is fair
    for the cooperating entities, requiring equal profit
    for all ISPs in our context. As a result, the Nash bargaining
    solution stipulates a single attribute value~$a^{\circ}$,
    which is adopted by all ISPs. This NBS attribute~$a^{\circ}$
    is~$a^{\circ} = \max(0, \hat{a}^{\circ})$, where~$\hat{a}^{\circ} = \hat{a}_r^{\circ}/I$.

\subsection{Equilibrium attribute}

    The equilibrium attribute~$a^+$ is defined as in~\cref{thm:theo:sapp-homogeneous:equilibrium}, but
    can be considerably simplified for the case~$Q = 1$.
    In particular, the equilibrium attribute~$a^+$ for~$Q = 1$ 
    is~$a^+ = \max(0, \hat{a}^+)$, where:
    \begin{align}
        \hat{a}^+ = &\ \frac{\sqrt{T_{2}^{2}-4T_{1}T_{3}}-T_{2}}{2T_{1}},\\
        T_{1} = &\ I^2 \alpha_1^2, \\
        T_{2} = &\ 2I\alpha_1\left(1+\alpha_{0}\right) -\frac{d}{d\phi_1+\gamma_1} \alpha_1\phi_1\left(I-1\right), \text{ and}\nonumber\\
         T_{3} = &\ (1 + \alpha_0)^2 - \frac{d}{d\phi_1 + \gamma_1} \big(\phi_1(1+\alpha_0)+\alpha_1(\rho-\phi_0)\big).
    \end{align}

\subsection{Comparison of attributes}

    We show that~$a^+ \leq a^{\circ}$ by showing that~$\hat{a}^+ \leq \hat{a}^{\circ}$. This inequality can be rewritten as
    \begin{equation}
        \frac{\sqrt{T_{2}^{2}-4T_{1}T_{3}}-T_{2}}{2T_{1}} \leq a^{\circ}
        \iff T_{2}\hat{a}^{\circ} \geq -(T_{1}{a^{\circ}}^2 + T_{3})
        \label{eq:theo:sapp-homogeneous:single-path:1}
    \end{equation}
    
    The two sides of the second inequality in~\cref{eq:theo:sapp-homogeneous:single-path:1}
    expand to:
    \begin{align}
        T_{2}\hat{a}^{\circ} =&\  -2\left(1+\alpha_{0}\right)^{2}+2\left(1+\alpha_{0}\right)\sqrt{\frac{d}{d\phi_{1}+g_{1}}\left(\phi_{1}\left(1+\alpha_{0}\right)+I\alpha_{1}\left(r-\phi_{0}\right)\right)}\\
        &\ -\frac{d}{d\phi_{1}+g_{1}} \left(I-1\right) \alpha_{1} \phi_{1}\hat{a}^{\circ}\nonumber\\
        -(T_{1}{a^{\circ}}^2 + T_{3}) =&\  -2\left(1+\alpha_{0}\right)^{2}+2\left(1+\alpha_{0}\right)\sqrt{\frac{d}{d\phi_{1}+g_{1}}\left(\phi_{1}\left(1+\alpha_{0}\right)+I\alpha_{1}\left(r-\phi_{0}\right)\right)}\\
        &\ -\frac{d}{d\phi_{1}+g_{1}}\left(I-1\right)\alpha_{1}(r-\phi_{0}) \nonumber
    \end{align}
    Since these terms lend themselves to considerable simplification,
    \cref{eq:theo:sapp-homogeneous:single-path:1} reduces to:
    \begin{equation}
        \rho - \phi_1 \hat{a}^{\circ} - \phi_0 \geq 0
        \label{eq:theo:sapp-homogeneous:single-path:2}
    \end{equation}
    
    Interestingly,~$\hat{a}^{\circ}$ is guaranteed to satisfy this inequality.
    To see why, assume the opposite for the sake of contradiction: $\rho - \phi_1 \hat{a}^{\circ} - \phi_0 < 0$. 
    If~$\phi_1 = 0$, this inequality conflicts with the model assumption~$\rho - \phi_0 \geq 0$. If~$\phi_1 > 0$, 
    the same model assumption indicates that~$\hat{a}^{\circ} > (\rho-\phi_0)/\phi_1 \geq 0$.
    Hence, the profit function of any ISP~$n$ is negative:
    \begin{equation}
        \pi_n(\mathbf{\hat{A}^{\circ}}) = \underbrace{d \frac{\alpha_1 I \hat{a}^{\circ} + \alpha_0}{1 + \alpha_1 I \hat{a}^{\circ} + \alpha_0}}_{> 0}\underbrace{( \rho- \phi_1 \hat{a}^{\circ} - \phi_0 )}_{< 0}
        \underbrace{- \gamma_1 \hat{a}^{\circ} - \gamma_0}_{\leq 0}
    \end{equation} 
    However, this negative profit could be strictly improved
    by choosing the lower attribute value~$a' = (\rho-\phi_0)/\phi_1 < \hat{a}^{\circ}$. This profit improvement is a contradiction
    to the character of $\hat{a}^{\circ}$ as the profit-optimizing attribute
    value. Hence,~\cref{eq:theo:sapp-homogeneous:single-path:2} holds,
    and therefore also the proposition~$\hat{a}^+ \leq \hat{a}^{\circ}$ holds.
    This insight concludes the proof.

\section{Proof of~\cref{thm:beneficial-competition}}
\label{sec:app:proofs:homogeneous--competition-attributes}
\subsection{Equilibrium for competitive network~$\mathcal{N}_2$}

    We begin the proof by characterizing the equilibrium for the competitive network~$\mathcal{N}_2$,
    in which every ISP~$n$ optimizes the following profit function~$\pi_n$:
    \begin{equation}
        \pi_n(a_n) = d'\left(\sum_{q=1}^Q \frac{v_{r(m_{q1},m_{q2},n)}}{1+\sum_{r'\in R(m_{q1},m_{q2})}v_{r'}} \right) \left(\rho-\phi_1 a_n - \phi0 \right) - \gamma_1 a_n - \gamma_0
    \end{equation} where~$r(m_{q1},m_{q2},n)$ denotes the unique path connecting~$(m_{q1},m_{q2})$ via ISP~$n$.
    In the unrestricted equilibrium~$\mathbf{\hat{A}}^+$, 
    every ISP~$n$ has the optimal attribute value~$\hat{a}^{+}_n$ given competitor attributes~$\mathbf{\hat{A}}_{-n}$,
    which can be found by setting~$\partial\pi_n/\partial a_n = 0$:
    \begin{equation}
        d'\left(\sum_{q=1}^Q \frac{\alpha_1 \left(1+v_{-r(m_{q1},m_{q2},n)}\right)}{\left(1+\sum_{r'\in R(m_{q1},m_{q2})}v_{r'}\right)^2} \right) \left(\rho-\phi_1 a_n - \phi0 \right) - d'\phi_1 \left(\sum_{q=1}^Q \frac{v_{r(m_{q1},m_{q2},n)}}{1+\sum_{r'\in R(m_{q1},m_{q2})}v_{r'}} \right)  -\gamma_1 = 0
        \label{eq:theo:sapp-homogeneous:equilibrium-competitive}
    \end{equation}
    Since this equation is equivalent for every ISP~$n$, the equilibrium~$\hat{a}^+_n$ is identical
    for all ISPs~$n$, i.e., $\hat{a}^+_n = \hat{a}^+$. This simplification allows the following
    transformation of~\cref{eq:theo:sapp-homogeneous:equilibrium-competitive}:
    \begin{equation}
         d'Q\frac{\alpha_1 (1+(Q-1)(I\alpha_1 \hat{a}^+ + \alpha_0))}{\left(1+Q(I\alpha_1 \hat{a}^+ + \alpha_0)\right)^2}  \left(\rho-\phi_1 a_n - \phi0 \right) - d'Q\phi_1 \frac{I\alpha_1 \hat{a}^+ + \alpha_0}{1+Q(I\alpha_1 \hat{a}^+ + \alpha_0)} -\gamma_1 = 0
        \label{eq:theo:sapp-homogeneous:equilibrium-competitive}
    \end{equation}
    This equilibrium condition is identical to the equilibrium condition for a homogeneous
    parallel-path network with a single origin-destination pair and demand limit~$d = d'Q$.
    Hence, the unrestricted equilibrium value~$\hat{a}^+$ from~\cref{thm:theo:sapp-homogeneous:equilibrium}
    (with $d'Q$ substituted for~$d$) also applies to the competitive network~$\mathcal{N}_2$.

\subsection{Equilibrium for competition-free network~$\mathcal{N}_1$}
    
    Moreover, we note that a single sub-network (for one origin-destination pair) of the competition-free network~$\mathcal{N}_1$
    is equivalent to the network~$\mathcal{N}_2$ for~$Q = 1$.  Since the identical, isolated sub-networks
    of the competition-free network~$\mathcal{N}_1$ do not influence each other, the equilibrium attribute value~$\hat{a}^+$
    is thus equal in that whole network.
    
\subsection{Comparison of equilibria}
    
    Hence, if~$\hat{a}^+(Q)$ is considered the equilibrium attribute for the competitive network~$\mathcal{N}_2$,
    we can prove the proposition~$a^+(\mathcal{N}_2) \geq a^+(\mathcal{N}_1)$ for~$Q\geq 1$ by 
    showing~$\hat{a}^+(Q) \geq \hat{a}^+(1)$ for~$Q\geq 1$.
    To show this property, we solve the following inequality:
    \begin{align}
        &\hat{a}^+(Q) - \hat{a}^+(1) \geq 0 \label{eq:theo:sapp-homogeneous:competition-proof:start}\\
        \iff &\frac{\sqrt{T_2(Q)^2 - 4 T_1(Q) T_3(Q)} - T_2(Q)}{2 T_1(Q)} -  \hat{a}^+(1) \geq 0 \label{eq:theo:sapp-homogeneous:competition-proof:1}\\
        \iff &\sqrt{T_2(Q)^2 - 4 T_1(Q) T_3(Q)} \geq T_2(Q) + 2T_1(Q) \hat{a}^+(1)\label{eq:theo:sapp-homogeneous:competition-proof:2}\\
        \iff &T_2(Q)^2 - 4 T_1(Q) T_3(Q) \geq \left(T_2(Q) + 2T_1(Q) \hat{a}^+(1)\right)^2\label{eq:theo:sapp-homogeneous:competition-proof:3}
    \end{align}
    In~\cref{eq:theo:sapp-homogeneous:competition-proof:1}, the equilibrium constituent terms~$T_1$, $T_2$, and~$T_3$
    from~\cref{thm:theo:sapp-homogeneous:equilibrium} are considered functions of~$Q$.
    In~\cref{eq:theo:sapp-homogeneous:competition-proof:2}, the transformation is possible by
    the fact that~$T_1(Q) > 0$ for~$Q \geq 1$:
    \begin{equation}
        T_1(Q) = Q^{2}I^{2}\alpha_{1}^{2}-\frac{Qd'}{Qd'\phi_{1}+\gamma_{1}}\left(QI-1\right)\left(Q-1\right)I\alpha_{1}^{2}\phi_{1} > 0
        \iff Q > \frac{d'\phi_{1}}{\left(d'\phi_{1}(I+1) + I\gamma_{1}\right)}
    \end{equation} where the RHS in the last inequality is consistently below~$1$,
    and~$T_1(Q) > 0$ thus holds for all~$Q \geq 1$.
    Using lengthy rewriting, the inequality in \cref{eq:theo:sapp-homogeneous:competition-proof:3} can then be transformed
    into the following inequality containing a quadratic equation of~$Q$:
    \begin{equation}
        T_7 Q^2 + T_8 Q + T_9 \leq 0 \label{eq:theo:sapp-homogeneous:competition-proof:central}
    \end{equation} where
    \begin{align}
        T_7 =\ &\hat{a}^+(1)^{2}\alpha_{1}^{2}I\left(I\gamma_{1}+d\phi_{1}\left(I+1\right)\right) 
        + \hat{a}^+(1)\alpha_{1}\left(2I\alpha_{0}\gamma_{1}+\left(2I+1\right)d\phi_{1}\alpha_{0}-d\left(\rho-\phi_{0}\right)I\alpha_{1}\right)\\
        &+ \alpha_{0}^{2}\left(d\phi_{1}+\gamma_{1}\right)-d\alpha_{0}\alpha_{1}\left(\rho-\phi_{0}\right)\nonumber,\\
        T_8 =\ &-\hat{a}^+(1)^{2}dI\alpha_{1}^{2}\phi_{1}
        + \hat{a}^+(1)\alpha_{1}\left(2I\gamma_1+d\phi_{1}(I+1)-d\phi_{1}\alpha_{0}+d\left(\rho-\phi_{0}\right)I\alpha_{1}\right)\\
        &+ 2\alpha_{0}\gamma_1+\left(\alpha_{0}-1\right)d\alpha_{1}\left(\rho-\phi_{0}\right)+d\alpha_{0}\phi_{1}, \text{ and} \nonumber\\
        T_9 =\ &\gamma_1.
    \end{align}

    To solve~\cref{eq:theo:sapp-homogeneous:competition-proof:central}, 
    we make use of the following two properties.

    \begin{itemize}
        \item $Q = 1$ is a root of of~$\hat{a}^+(Q) - \hat{a}^+(1)$,
    which implies:
    \begin{equation}
        T_7 + T_8 + T_9 = 0 \iff T_7 + T_8 = -T_9.
        \label{eq:theo:sapp-homogeneous:competition-proof:root-property}
    \end{equation}
        \item The inspection of~$T_7$ yields the following insight,
    which we derived by means of the symbolic algebra system in \textsc{Matlab}:
    \begin{align}
        T_7 &\leq \lim_{\substack{d, \alpha_0\\\rightarrow 0}} T_7 = \gamma_1 = T_9
        \label{eq:theo:sapp-homogeneous:competition-proof:bound-property}
    \end{align}
    \end{itemize}

    Given the lower root~$\underline{Q}$ and the higher root~$\overline{Q}$ of the quadratic function
    in~\cref{eq:theo:sapp-homogeneous:competition-proof:central} (which are guaranteed to exist
    at least at~$Q = 1$ and are identical if~$T_7 =0$), 
    the inequality is solved by the following~$Q$:
    \begin{equation}
        Q \in \begin{cases}
            [\underline{Q},\ \overline{Q}] & \text{\circled{1} if } T_7 > 0\\
            (-\infty,\ \underline{Q}] \cup [\overline{Q},\ \infty) & \text{\circled{2} if } T_7 < 0\\
            (-\infty,\ \underline{Q}] & \text{\circled{3} if } T_7 = 0 \land T_8 > 0\\
             [\overline{Q},\ \infty) & \text{\circled{4} if } T_7 = 0 \land T_8 < 0\\
             (-\infty,\ \infty) & \text{\circled{5} if } T_7 = 0 \land T_8 = 0 \land T_9 \leq 0\\
             \emptyset & \text{\circled{6} if } T_7 = 0 \land T_8 = 0 \land T_9 > 0
        \end{cases}
        \label{eq:theo:sapp-homogeneous:competition-proof:solutions}
    \end{equation}
    
    This area of Q (leading to non-positive values of the quadratic
    function in~\cref{eq:theo:sapp-homogeneous:competition-proof:central})
    includes~$[1, \infty)$ in all cases:

    \begin{enumerate}
        \item $T_7 \neq 0$. (\cref{eq:theo:sapp-homogeneous:competition-proof:solutions}\circled{1} and
    \circled{2}):
    For~$T_7 \neq 0$,
    the property in~\cref{eq:theo:sapp-homogeneous:competition-proof:root-property} facilitates 
    finding the solutions~$(\underline{Q},\ \overline{Q})$:
    \begin{equation}
        (\underline{Q},\ \overline{Q}) 
        = \frac{-T_8 \pm \sqrt{T_8^2 - 4T_7T_9}}{2T_7}
        = \frac{-T_8 \pm \sqrt{(2T_7 + T_8)^2}}{2T_7}
        = \frac{-T_8 \pm (2T_7 + T_8)}{2T_7}
        \label{eq:theo:sapp-homogeneous:competition-proof:solutions:t7_neq_0}
    \end{equation}

    \begin{enumerate}
        \item $T_7 > 0$ (\cref{eq:theo:sapp-homogeneous:competition-proof:solutions}\circled{1}): For~$T_7 > 0$,
        we note that
        \begin{equation}
            2T_7 + T_8 \overset{\text{\cref{eq:theo:sapp-homogeneous:competition-proof:root-property}}}{=} T_7 - \gamma_1 \overset{\text{\cref{eq:theo:sapp-homogeneous:competition-proof:bound-property}}}\leq 0.
        \end{equation}

    Hence, the solutions from~\cref{eq:theo:sapp-homogeneous:competition-proof:solutions:t7_neq_0} are:
    \begin{equation}
         \underline{Q} =  \frac{-T_8 + (2T_7 + T_8)}{2T_7} = 1 \hspace{20pt} \overline{Q} =  \frac{-T_8 - (2T_7 + T_8)}{2T_7} = \frac{\gamma_1}{T_7} \geq 1
    \end{equation} where the higher solution~$\overline{Q}$ is spurious and has been introduced
    by the squaring operation in~\cref{eq:theo:sapp-homogeneous:competition-proof:3}. 

        \item $T_7 < 0$: (\cref{eq:theo:sapp-homogeneous:competition-proof:solutions}\circled{2}): For~$T_7 < 0$ (\cref{eq:theo:sapp-homogeneous:competition-proof:solutions}\circled{2}), the solutions are as follows:
    \begin{equation}
         \underline{Q} =  \frac{-T_8 - (2T_7 + T_8)}{2T_7} = \frac{\gamma_1}{T_7} < 0 \hspace{20pt} \overline{Q} =  \frac{-T_8 + (2T_7 + T_8)}{2T_7} =  1
    \end{equation} 
        
    \end{enumerate}
    
        \item $T_7 = 0$ (\cref{eq:theo:sapp-homogeneous:competition-proof:solutions}\circled{3}--\circled{6}): 
        For~$T_7 = 0$, it holds that~$T_8 = -\gamma_1 - T_7 = -\gamma_1 \leq 0$ and~$\overline{Q} = -T_9/T_8 = (-\gamma_1)/(-\gamma_1) = 1$.
        
    \begin{enumerate}
        \item $T_8 > 0$ (\cref{eq:theo:sapp-homogeneous:competition-proof:solutions}\circled{3}): The case~$T_8 > 0$ thus cannot arise.
        \item $T_8 < 0$ (\cref{eq:theo:sapp-homogeneous:competition-proof:solutions}\circled{4}): For~$T_8 < 0$, the proposition clearly holds.
        \item $T_8 = 0$ (\cref{eq:theo:sapp-homogeneous:competition-proof:solutions}\circled{5} and \circled{6}): For~$T_8 = 0$,
    the equality~$T_7 + T_8 = -T_9$ from~\cref{eq:theo:sapp-homogeneous:competition-proof:root-property} implies~$T_9 = 0$.
    Hence,
    the case 
    in~\cref{eq:theo:sapp-homogeneous:competition-proof:solutions}\circled{5}
    always arises if~$T_7 = T_8 = 0$, whereas the case in~\cref{eq:theo:sapp-homogeneous:competition-proof:solutions}\circled{6}
    never arises. 
    \end{enumerate}

    \end{enumerate}
    
    Since~\cref{eq:theo:sapp-homogeneous:competition-proof:start}
    thus always holds for~$Q\geq 1$, the proposition is proven.

\section{Proof of~\cref{thm:theo:sapp-homogeneous:competition:profits}}
\label{sec:app:proofs:homogeneous--competition-profits}

    To start the proof, we note that both 
    the equilibrium attribute sum~$a^{+}(\mathcal{N}_1)$
    and the NBS attribute sum~$a^{\circ}(\mathcal{N}_1)$
    for the competition-free network are found by analyzing
    a single path, since the isolated sub-paths in the 
    competition-free network do not influence each other.
    Hence, $a^{+}(\mathcal{N}_1)$ and~$a^{\circ}(\mathcal{N}_1)$
    are as in \cref{thm:theo:sapp-homogeneous:intra-path:equilibrium-optimum},
    which relates to the single-path context and
    thus states that~$a^{+}(\mathcal{N}_1) \leq a^{\circ}(\mathcal{N}_1)$.
    Therefore, the interval~$[a^+(\mathcal{N}_1), a^{\circ}(\mathcal{N}_1)]$
    is never empty.
    
    From the proof of~\cref{thm:beneficial-competition}, we know that the 
    proposition~$\pi^+(\mathcal{N}_2) \geq \pi^+(\mathcal{N}_1)$ is equivalent to
    the proposition~$\Delta\pi = \pi(Q,a^+(\mathcal{N}_2)) - \pi(1, a^+(\mathcal{N}_1)) \geq 0$, 
    where~$\pi(Q, a)$ is defined as follows:
    \begin{equation}
        \pi(Q, a) = Qd' \frac{I \alpha_1 a + \alpha_0}{1 + Q(I \alpha_1 a + \alpha_0)} (\rho - \phi_1 a - \phi_0) - \gamma_1 a - \gamma_0.
        \label{eq:theo:sapp-homogeneous:competition:increased-profits:profit}
    \end{equation}

    Clearly,~$\pi(Q,a^{\circ}(\mathcal{N}_1))$ is optimal for~$Q = 1$, 
    i.e., the NBS attribute sum is optimal in the competition-free network.
    Hence, it also holds that $\pi(1,a^{+}(\mathcal{N}_1)) \leq \pi(1,a^{\circ}(\mathcal{N}_1))$, 
    i.e., the equilibrium profit in the competition-free network is generally sub-optimal.
    Moreover, since~$\pi(Q,a)$ is consistently concave in~$a$ in the relevant
    regions, the 
    assumption~$a^+(\mathcal{N}_2) \in [a^+(\mathcal{N}_1), a^{\circ}(\mathcal{N}_1)]$
    implies
    \begin{equation}
        \pi(1, a^+(\mathcal{N}_1)) \leq \pi(1, a^+(\mathcal{N}_2)).
        \label{eq:theo:sapp-homogeneous:competition:increased-profits:condition}
    \end{equation}

    Given~\cref{eq:theo:sapp-homogeneous:competition:increased-profits:condition},
    we can lower bound the profit difference:
    \begin{equation}
       \Delta\pi = \pi(Q,a^+(\mathcal{N}_2)) - \pi(1, a^+(\mathcal{N}_1)) 
       \geq 
       \pi(Q,a^+(\mathcal{N}_2)) - \pi(1, a^+(\mathcal{N}_2))
       =: \underline{\Delta\pi}
    \end{equation}
    Hence, if~$\underline{\Delta\pi} \geq 0$ holds,
    the proof proposition~$\Delta\pi \geq 0$
    follows. At this point, we also note that~$a^+(\mathcal{N}_2) \in [a^+(\mathcal{N}_1), a^{\circ}(\mathcal{N}_1)]$
    is only a sufficient, but not a necessary condition for
    $\underline{\Delta\pi} \geq 0$; 
    hence, profit increases might also happen if~$a^+(\mathcal{N}_2) \notin [a^+(\mathcal{N}_1), a^{\circ}(\mathcal{N}_1)]$.
    
    We can reformulate the lower bound~$\underline{\Delta\pi}$ on the profit difference as follows:
    \begin{equation}
    \begin{split}
        \underline{\Delta\pi} = &\ \pi(Q,a^+(\mathcal{N}_2)) - \pi(1, a^+(\mathcal{N}_2))\\
        = &\ d' 
        \left(
         \frac{Q(I \alpha_1 a^+(\mathcal{N}_2) + \alpha_0)}{1 + Q(I \alpha_1 a^+(\mathcal{N}_2) + \alpha_0)}
         - 
         \frac{I \alpha_1 a^+(\mathcal{N}_2) + \alpha_0}{1 + I \alpha_1 a^+(\mathcal{N}_2) + \alpha_0}
        \right)
        \left(\rho - \phi_1 a^+(\mathcal{N}_2) - \phi_0\right)\\
        = &\ d' 
        \left(
         \frac{\left(Q-1\right)\left(I\alpha_{1}a^+(\mathcal{N}_2)+\alpha_{0}\right)}{(1 + Q(I \alpha_1 a^+(\mathcal{N}_2) + \alpha_0))(1 + I \alpha_1 a^+(\mathcal{N}_2) + \alpha_0)}
        \right)
        \left(\rho - \phi_1 a^+(\mathcal{N}_2) - \phi_0\right)
    \end{split}
        \label{eq:theo:sapp-homogeneous:competition:increased-profits:reformulation}
    \end{equation}

    Given~$d' > 0$ and~$Q \geq 1$, the first and second factor of~$\underline{\Delta\pi}$
    in~\cref{eq:theo:sapp-homogeneous:competition:increased-profits:reformulation}
    are non-negative. Hence,~$\underline{\Delta\pi} \geq 0$ is equivalent
    to~$\rho - \phi_1 a^+(\mathcal{N}_2) - \phi_0 \geq 0$.
    This latter condition also always holds, which is demonstrable by contradiction.
    Let~$\rho - \phi_1 a^+(\mathcal{N}_2) - \phi_0 < 0 \iff a^+(\mathcal{N}_2) > (\rho - \phi_0)/\phi_1$, 
    which makes the minuend in~$\pi(1, a^+(\mathcal{N}_2))$
    negative (cf.~\cref{eq:theo:sapp-homogeneous:competition:increased-profits:profit}).
    In that case, 
    all~$a' > a^+(\mathcal{N}_2)$ would lead to lower
    profit~$\pi(1,a')$. This observation contradicts the optimality of
    the NBS attribute sum~$a^{\circ}(\mathcal{N}_1)$
    regarding~$\pi(1,a')$, as~$a^{\circ}(\mathcal{N}_1) \geq a^+(\mathcal{N}_2)$.

    Hence, since~$\underline{\Delta\pi} \geq 0$, it holds that~$\Delta\pi \geq 0$ and
    the theorem proposition follows.


\section{Proof of~\cref{thm:heterogeneous:single-path:optimum}}
\label{sec:app:proofs:heterogeneous--intra-path-optimum}

In order to be a Nash bargaining solution, 
    the attribute values~$\mathbf{A}^{\circ}$ should both
    optimize the aggregate profit function~$\pi(\mathbf{A}) =\sum_{n \in r} \pi_n(\mathbf{A})$,
    and create a maximally equitable profit distribution across the ISPs~$n \in r$.
    This maximum fairness is achieved by optimizing the Nash bargaining product, i.e.:
    \begin{equation}
        \mathbf{A}^{\circ} = {\arg\max}_{\mathbf{A} \in \mathbb{R}_{\geq 0}} \Pi_{n \in r} \pi_{n}(\mathbf{A}) 
    \end{equation}

    This optimization of the Nash bargaining product must be performed subject
    to the constraints in~\cref{thm:heterogeneous:single-path:optimum} that 
    are associated with optimal aggregate profit.
    In the following, we characterize this aggregate-profit function,
    and show that the conditions stated
    in~\cref{thm:heterogeneous:single-path:optimum} are both
    sufficient and necessary in order for~$\mathbf{A}^{\circ}$
    to satisfy aggregate-profit optimality.

\subsection{Aggregate-profit function}
    The aggregate profit~$\pi(\mathbf{A})$ in our setting is:
    \begin{equation}
        \pi(\mathbf{A}) = \sum_{n \in r} \pi_n(\mathbf{A}) = d \frac{v_r(\mathbf{A})}{1+v_r(\mathbf{A})}\left(\sum_{n \in r} \rho_{n} - \phi_{n0}\right) - \sum_{n \in r} \left(\sum_{k \in  K} \gamma_{nk} a_{nk} + \gamma_{n0} \right)
        \label{eq:theo:sapp-constant:single-path:optimum:1}
    \end{equation}
    This aggregate-profit function has the following first and second derivative in any~$a_{nk}$:
    \begin{align}
        \frac{\partial}{\partial a_{nk}} \pi(\mathbf{A}) =&\ d\frac{\alpha_{nk}}{\left(1+v_r(\mathbf{A})\right)^2} \left(\sum_{n \in r} \rho_{n} - \phi_{n0}\right) - \gamma_{nk}\\
        \frac{\partial^2}{\partial a_{nk}^2} \pi(\mathbf{A}) =&\ -d \frac{2\alpha_{nk}^2}{\left(1+v_r(\mathbf{A})\right)^3} \left(\sum_{n \in r} \rho_{n} - \phi_{n0}\right)
    \end{align} As the second derivative is non-positive for all~$\mathbf{A} \in \mathbb{R}_{\geq 0}$,
    the aggregate-profit function is consistently concave in any~$a_{nk}$ on the valid domain~$\mathbb{R}_{\geq 0}$.
    Therefore, if the first derivative~$\partial/\partial a_{nk}\ \pi(\mathbf{A})$
    is negative for any~$a_{nk}$, all reductions of~$a_{nk}$ increase aggregate profit, 
    and all increases of~$a_{nk}$ 
    reduce the aggregate profit (The reverse holds
    for a positive first derivative).
    This condition on the first derivative is equivalent to
    the following condition, which is central for the proof:
    \begin{equation}
        \forall n \in r, k \in  K.\quad \frac{\partial}{\partial a_{nk}} \pi(\mathbf{A}) < 0
        \iff v_r(\mathbf{A}) >  \sqrt{\frac{\alpha_{nk}}{\gamma_{nk}}} \sqrt{d\sum_{n\in r} (\rho_n - \phi_{n0})} - 1
        \label{eq:theo:sapp-constant:single-path:optimum:1:profit-valuation}
    \end{equation}

\subsection{Sufficiency of conditions}

    After this characterization of the aggregate-profit function, 
    we now demonstrate that the conditions 
    in~\cref{thm:heterogeneous:single-path:optimum}
    are \emph{sufficient}, i.e., any~$\mathbf{A}^{\circ}$
    with the conditions optimizes the aggregate profit.
    Sufficiency can be demonstrated by performing
    the following case distinction:
    \begin{enumerate}
        \item $\forall (n, k) \in r \times K.\  \alpha_{r0} > \sqrt{\frac{\alpha_{nk}}{\gamma_{nk}}} \sqrt{d\sum_{n\in r} (\rho_n - \phi_{n0})} - 1$\\
        According to~\cref{thm:heterogeneous:single-path:optimum},
        all optimal attribute values~$\mathbf{A}^{\circ}$
        must be 0 in this case:
        \begin{equation}
            v_r(\mathbf{A}^{\circ}) = v_r^{\circ} \overset{\text{\cref{eq:heterogeneous:single-path:optimum:valuation}}}{=} \alpha_{r0} \iff \mathbf{A}^{\circ} = \mathbf{0}
        \end{equation}
        Moreover, the first derivative~$\partial/\partial a_{nk}\ \pi(\mathbf{A}^{\circ})$ for all~$n \in r$, $k \in  K$
        must be negative:
        \begin{equation}
            \forall n \in r, k \in  K.\ 
            v_r(\mathbf{A}^{\circ}) = \alpha_{r0} > 
            \sqrt{\frac{\alpha_{nk}}{\gamma_{nk}}} \sqrt{d\sum_{n\in r} (\rho_n - \phi_{n0})} - 1 \overset{\text{\cref{eq:theo:sapp-constant:single-path:optimum:1:profit-valuation}}}{\iff}  \frac{\partial}{\partial a_{nk}} \pi(\mathbf{A}) < 0
        \end{equation}
        Hence, the only way to further increase the aggregate
        profit~$\pi$ would be by reductions in any~$a_{nk}$.
        However, since every~$a_{nk} = 0$, such reductions
        are not possible given the restricted domain~$\mathbb{R}_{\geq 0}$.
        Hence,~$\mathbf{A}^{\circ} = \mathbf{0}$ is optimal.
        
        \item $\exists (n, k) \in r\times K.\  \alpha_{r0} \leq \sqrt{\frac{\alpha_{nk}}{\gamma_{nk}}} \sqrt{d\sum_{n\in r} (\rho_n - \phi_{n0})} - 1$\\
        In this case, the attributes~$(n^{\circ},k^{\circ}) \in r\times  K$
        with the maximal ratio~$\alpha_{n^{\circ}k^{\circ}}/\gamma_{n^{\circ}k^{\circ}}$ play
        a special role according to~\cref{thm:heterogeneous:single-path:optimum}.
        We denote the set of these attributes by~$K^{\circ}_r$:
        \begin{equation}
            K^{\circ}_r = \left\{(n^{\circ},k^{\circ})\ |\ (n^{\circ},k^{\circ}) = {\arg\max}_{(n,k) \in r\times K} \frac{\alpha_{nk}}{\gamma_{nk}}\right\}.
        \end{equation}
        This maximal ratio also determines the optimal path
        valuation~$v_r(\mathbf{A}^{\circ})$ according
        to~\cref{thm:heterogeneous:single-path:optimum}:
        \begin{equation}
            \forall (n^{\circ},k^{\circ}) \in K^{\circ}_r.\ v_r(\mathbf{A}^{\circ}) = \sqrt{\frac{\alpha_{n^{\circ}k^{\circ}}}{\gamma_{nk}}} \sqrt{d\sum_{n\in r} (\rho_{n} - \phi_{n0})} - 1
        \end{equation}
        In contrast, for all attributes~$(n^{\Oslash},k^{\Oslash}) \notin K^{\circ}_r$,
        the following condition holds:
        \begin{equation}
            \begin{split}
                \forall (n^{\Oslash},k^{\Oslash}) \notin K^{\circ}_r.\ &v_r(\mathbf{A}^{\circ}) =
                 \sqrt{\frac{\alpha_{n^{\circ}k^{\circ}}}{\gamma_{n^{\circ}k^{\circ}}}} \sqrt{d\sum_{n\in r} (\rho_{n} - \phi_{n0})} - 1 > \sqrt{\frac{\alpha_{n^{\Oslash}k^{\Oslash}}}{\gamma_{n^{\Oslash}k^{\Oslash}}}} \sqrt{d\sum_{n\in r} (\rho_{n} - \phi_{n0})} - 1\\ 
                \iff &\frac{\partial}{\partial a_{n^{\Oslash}k^{\Oslash}}} \pi(\mathbf{A}) < 0
            \end{split}
        \end{equation}
        Hence, the only way to increase the aggregate profit~$\pi$
        would be by reductions in any~$a_{n^{\Oslash}k^{\Oslash}}$ for~$(n^{\Oslash},k^{\Oslash}) \notin K^{\circ}_r$.
        However, since~$a_{n^{\Oslash}k^{\Oslash}} = 0\ \forall (n^{\Oslash},k^{\Oslash}) \notin K^{\circ}_r$
        by~\cref{thm:heterogeneous:single-path:optimum},
        such reductions are not possible, and hence~$\mathbf{A}^{\circ}$
        is optimal.
    \end{enumerate}

\subsection{Necessity of conditions}
 
    After demonstrating that the conditions 
    in~\cref{thm:heterogeneous:single-path:optimum}
    are sufficient for optimal aggregate profit,
    we now demonstrate that the conditions are also
    \emph{necessary}, i.e., no choice of attribute values~$\mathbf{A}^{\circ}$
    that violates these conditions can be optimal.
    For the sake of contradiction, we
    assume that some attribute values~$\mathbf{A}^{\circ}$
    are optimal while satisfying the following conditions:
    \begin{equation}
        \exists (n^{\Oslash},k^{\Oslash}) \notin K^{\circ}_r.\ a_{n^{\Oslash}k^{\Oslash}} > 0.
    \end{equation}
    A contradiction can be produced in all cases of the following case distinction:
    \begin{enumerate}
        \item $\forall (n, k) \in r \times K.\  \alpha_{r0} > \sqrt{\frac{\alpha_{nk}}{\gamma_{nk}}} \sqrt{d\sum_{n\in r} (\rho_n - \phi_{n0})} - 1$\\
        Since $a_{n^{\Oslash}k^{\Oslash}}^{\circ} > 0$ for the fixed attribute~$(n^{\Oslash},k^{\Oslash})$,
        the optimal path valuation~$v_r(\mathbf{A}^{\circ})$ 
        exceeds~$\alpha_{r0}$, and hence:
        \begin{equation}
            \forall (n,k) \in r \times  K.\quad 
            v_r(\mathbf{A}^{\circ}) > \alpha_{r0} > 
            \sqrt{\frac{\alpha_{nk}}{\gamma_{nk}}} \sqrt{d\sum_{n\in r} (\rho_n - \phi_{n0})} - 1
            \implies \frac{\partial}{\partial a_{n^{\Oslash}k^{\Oslash}}} \pi(\mathbf{A}) < 0
        \end{equation}
        The aggregate profit can thus be increased by reducing~$a_{n^{\Oslash}k^{\Oslash}}^{\circ}$,
        and such a reduction is also possible since~$a_{n^{\Oslash}k^{\Oslash}} > 0$.
        Hence, the attribute values~$\mathbf{A}^{\circ}$ are not optimal,
        which causes a contradiction.
        
        \item $\exists (n, k) \in r \times  K.\  \alpha_{r0} \leq \sqrt{\frac{\alpha_{nk}}{\gamma_{nk}}} \sqrt{d\sum_{n\in r} (\rho_n - \phi_{n0})} - 1$\\
        In that case, we perform a sub-case distinction
        on the value of~$v_r(\mathbf{A}^{\circ})$:
        \begin{enumerate}
            \item $v_r(\mathbf{A}^{\circ}) \leq 
            \sqrt{\frac{\alpha_{n^{\Oslash}k^{\Oslash}}}{\gamma_{n^{\Oslash}k^{\Oslash}}}} \sqrt{d\sum_{n\in r} (\rho_n - \phi_{n0})} - 1$\\
            Since~$a_{n^{\Oslash}k^{\Oslash}}/\gamma_{n^{\Oslash}k^{\Oslash}} <
            a_{n^{\circ}k^{\circ}}/\gamma_{n^{\circ}k^{\circ}}$, 
            it follows that:
            \begin{equation}
                v_r(\mathbf{A}^{\circ}) < \sqrt{\frac{\alpha_{n^{\circ}k^{\circ}}}{\gamma_{n^{\circ}k^{\circ}}}} \sqrt{d\sum_{n\in r} (\rho_n - \phi_{n0})} - 1 \iff \frac{\partial}{\partial a_{n^{\circ}k^{\circ}}} \pi(\mathbf{A}) > 0,
            \end{equation} which implies that the profit can be increased
            by increasing the value of~$a_{n^{\circ}k^{\circ}}\ \forall (n,k) \in K^{\circ}_r$, which contradicts the assumption that~$\mathbf{A}^{\circ}$
            is optimal.
            
            \item $v_r(\mathbf{A}^{\circ}) >
            \sqrt{\frac{\alpha_{n^{\Oslash}k^{\Oslash}}}{\gamma_{n^{\Oslash}k^{\Oslash}}}} \sqrt{d\sum_{n\in r} (\rho_n - \phi_{n0})} - 1$\\
            This condition implies:
            \begin{equation}
                \frac{\partial}{\partial a_{n^{\Oslash}k^{\Oslash}}} \pi(\mathbf{A}) < 0.
            \end{equation}
            Hence, the profit can be increased by reducing~$a_{n^{\Oslash}k^{\Oslash}}$,
            which is possible given~$a_{n^{\Oslash}k^{\Oslash}} >0$. Therefore, we again
            produce a contradiction to the optimality of~$\mathbf{A}^{\circ}$.
            
        \end{enumerate}
    \end{enumerate}
    
    We have thus identified the conditions on~$\mathbf{A}^{\circ}$ that are
        sufficient and necessary for optimal aggregate profit.
        Thereby, the theorem is proven.

\section{Proof of~\cref{thm:heterogeneous:single-path:equilibrium}}
\label{sec:app:proofs:heterogeneous--intra-path-equilibrium}

    Since the equilibrium conditions in~\cref{thm:heterogeneous:single-path:equilibrium} are highly similar
    to the optimality conditions in~\cref{thm:heterogeneous:single-path:optimum},
    the proof of~\cref{thm:heterogeneous:single-path:equilibrium} is analogous to
    the proof of~\cref{thm:heterogeneous:single-path:optimum}.
    The proof is analogous because the derivatives of the individual profit functions~$\pi_n$
    have equivalent properties to the derivatives of the aggregate-profit function~$\pi$
    from~\cref{eq:theo:sapp-constant:single-path:optimum:1}.
    In particular, every first derivative satisfies:
    \begin{align}
        \forall n \in r, k \in  K.\quad&\ \frac{\partial \pi_n(\mathbf{A})}{\partial a_{nk}} < 0
        \iff v_r(\mathbf{A}) > \sqrt{\frac{\alpha_{nk}}{\gamma_{nk}}} \sqrt{d (\rho_{n} - \phi_{n0})}- 1
    \end{align}
    Moreover, every individual profit function~$\pi_n$ is consistently concave in any relevant attribute~$a_{nk}$:
    \begin{equation}
        \forall n \in r, k \in  K.\  \frac{\partial^2}{\partial a_{nk}^2} \pi_n(\mathbf{A}) = -d \frac{2\alpha_{nk}^2}{\left(1+v_r(\mathbf{A})\right)^3} \left(\rho_{n} - \phi_{n0}\right)
    \end{equation}
    
    Building on these properties, the equilibrium conditions can be shown to be sufficient 
    and necessary analogously to~\cref{thm:heterogeneous:single-path:optimum}.

\section{Proof of~\cref{thm:heterogeneous:single-path:valuation}}
\label{sec:app:proofs:heterogeneous--intra-path-valuation}

    In order to show that~$v_r(\mathbf{A}^+) \leq v_r(\mathbf{A}^{\circ})$,
    it is enough to show that:
    \begin{equation}
        \sqrt{\frac{\alpha_{n^+k^+}}{\gamma_{n^+k^+}}} \sqrt{d (\rho_{n^+} - \phi_{n^+0})}- 1 \leq
        \sqrt{\frac{\alpha_{n^{\circ}k^{\circ}}}{\gamma_{n^{\circ}k^{\circ}}}} \sqrt{d\sum_{n\in r} (\rho_n - \phi_{n0})} - 1
    \end{equation}
    where~$(n^+,k^+) \in K^{+}_r$ and~$(n^{\circ}, k^{\circ}) \in K^{\circ}_r$. This inequality can be transformed
    into the following form:
    \begin{equation}
         \sqrt{\frac{\alpha_{n^+k^+}}{\gamma_{n^+k^+}}} \sqrt{\frac{d (\rho_{n^+} - \phi_{n^+0})}{d\sum_{n\in r} (\rho_n - \phi_{n0})}} \leq
        \sqrt{\frac{\alpha_{n^{\circ}k^{\circ}}}{\gamma_{n^{\circ}k^{\circ}}}}.
    \end{equation}
    
    Thanks to the following two insights, this inequality is always satisfied:
    \begin{equation}
        \sqrt{\frac{d (\rho_{n^+} - \phi_{n^+0})}{d\sum_{n\in r} (\rho_n - \phi_{n0})}} \leq 1 \hspace{30pt}
         \frac{\alpha_{n^+k^+}}{\gamma_{n^+k^+}} \leq \frac{\alpha_{n^{\circ}k^{\circ}}}{\gamma_{n^{\circ}k^{\circ}}} = \max_{(n,k) \in r\times K}  \frac{\alpha_{nk}}{\gamma_{nk}}
    \end{equation}
    
    Hence, the theorem holds.

\section{Proof of~\cref{thm:heterogeneous:equilibrium:two}}
\label{sec:app:proofs:heterogeneous--two-path-equilibrium}

\subsection{Analogy to \cref{thm:heterogeneous:single-path:equilibrium}}

    To start off, we once more characterize the derivatives of the individual-profit
    functions~$\pi_n$ for any attribute value~$a_{nk}$:
    \begin{align}
        \forall n \in N, k \in  K, r = r(n).\quad &\frac{\partial \pi_n(\mathbf{A})}{\partial a_{nk}} 
        = d \frac{\alpha_{nk}\left(1 + v_{\overline{r}}(\mathbf{A})\right)}{\left(1 + v_r(\mathbf{A}) + v_{\overline{r}}(\mathbf{A})\right)^2} (\rho_n - \phi_{n0}) - \gamma_{nk}\\
        \ &\frac{\partial^2 \pi_n(\mathbf{A})}{\partial a_{nk}^2} 
        = -d \frac{2\alpha_{nk}^2\left(1 + v_{\overline{r}}(\mathbf{A})\right)}{\left(1 + v_r(\mathbf{A}) + v_{\overline{r}}(\mathbf{A})\right)^3} (\rho_n - \phi_{n0})
    \end{align}
    Since the second derivative is never positive, every profit function~$\pi_n$ is consistently
    concave in the attribute values controlled by ISP~$n$. Hence, a negative
    first derivative~$\partial \pi(\mathbf{A}) / \partial a_{nk} < 0$ indicates 
    that~$a_{nk}$ must be reduced if the profit is to be increased.
    The case of a negative first derivative in~$a_{nk}$ can be expressed as follows (for $r = r(n)$):
    \begin{equation}
        \frac{\partial \pi_n(\mathbf{A})}{\partial a_{nk}} < 0 
        \iff v_{r}(\mathbf{A}) > \sqrt{\frac{\alpha_{nk}}{\gamma_{nk}}} \sqrt{ d\left(\rho_n - \phi_{n0}\right) } \sqrt{ 1 + v_{\overline{r}}(\mathbf{A}) } - (1 + v_{\overline{r}}(\mathbf{A}))
        \label{eq:theo:sapp-constant:competition:1}
    \end{equation}
    
    When thinking of~\cref{eq:theo:sapp-constant:competition:1} as
    an extension of~\cref{eq:theo:sapp-constant:single-path:optimum:1:profit-valuation}
    with~$v_{\overline{r}}$ as a fixed quantity, an analogous proof to the proof of~\cref{thm:heterogeneous:single-path:optimum}
    can be performed. 
    The extension by fixed~$v_{\overline{r}}$ does not change
    the finding that only attributes~$(n^{\circ}, k^{\circ}) \in K_r^{\circ}$
    might have non-zero values in equilibrium.
    However, the extension by~$v_{\overline{r}}$
    changes the equilibrium path valuation~$v_r^+$ 
    from~\cref{eq:heterogeneous:single-path:equilibrium:valuation} to:
    \begin{equation}
        v_r^+ = \max\left(\alpha_{r0},\ \sqrt{\frac{\alpha_{n^{\circ}k^{\circ}}}{\gamma_{n^{\circ}k^{\circ}}}} \sqrt{ d\left(\rho_{n^{\circ}} - \phi_{n^{\circ}0}\right) } \sqrt{ 1 + v_{\overline{r}} } - \left(1 + v_{\overline{r}}\right)\right).
    \end{equation}

    Crucially, this condition simultaneously holds for both paths~$r$ and~$\overline{r}$
    in a two-path scenario, creating an interdependence
    of the equilibrium path valuations:
    \begin{equation}
        \begin{split}
             \forall r \in R.\quad 
            v_r^+ &= \max\left(\alpha_{r0},\ \sqrt{\frac{\alpha_{n^{\circ}k^{\circ}}}{\gamma_{n^{\circ}k^{\circ}}}} \sqrt{ d\left(\rho_{n^{\circ}} - \phi_{n^{\circ}0}\right) } \sqrt{ 1 + v_{\overline{r}}^+ } - \left(1 + v_{\overline{r}}^+\right)\right)\\
            &= \max\left(\alpha_{r0},\ \psi_r \sqrt{ d } \sqrt{ 1 + v_{\overline{r}}^+ } - \left(1 + v_{\overline{r}}^+\right)\right) = \max\left(\alpha_{r0},\ \hat{v}_r^{*}\left(v_{\overline{r}}^+\right)\right),
        \end{split}
        \label{eq:theo:sapp-constant:competition:interdep}
    \end{equation}
    where~$\hat{v}_r^{*}(v_{\overline{r}})$ is the \emph{unrestricted best-response valuation} for path~$r$ given
    competing-path valuation~$v_{\overline{r}}$. 
    The \emph{characteristic ratio}~$\psi_r$ is reflected in~\cref{eq:theo:sapp-constant:competition:psi}.
    
    The remainder of the proof illustrates how to derive the
    equilibrium path valuations~$v_r^+$ and~$v_{\overline{r}}^+$.

\subsection{Unrestricted equilibrium~$\hat{v}_r^{+}$}

    Considering a relaxed setting in which the constraint~$\mathbf{A}^+ \in \mathbb{R}^{|N|\times |K|}_{\geq 0} \iff v_r^+ \geq \alpha_{r0}$
    is ignored, the unrestricted equilibrium path valuations~$\hat{v}_r^+$
    satisfy the following system of equations:
    \begin{equation}
        \forall r \in R.\quad \hat{v}_r^+ = \hat{v}_r^{\ast}(\hat{v}_{\overline{r}}^+).
        \label{eq:theo:sapp-constant:competition:unrestricted}
    \end{equation}
    In this relaxed setting, this system of two equations can be conventionally solved for
    the unrestricted equilibrium path valuations~$\hat{v}_{r}^+$, $r \in R$,
    resulting in the unique solution denoted in~\cref{eq:theo:sapp-constant:competition:eq-path-valuation}.

    Moreover, we make the following
    important observation:
                \begin{equation}
                    \hat{v}_{r}^{*}\left(\hat{v}_{\overline{r}}^{*}(v_{r})\right) \leq
                    v_{r}
                    \iff v_{r} \geq \hat{v}_{r}^+.
                    \label{eq:theo:sapp-constant:competition:map-lower}
                \end{equation}

\subsection{Restricted equilibrium~$v_r^+$}
  
    We now rely on the equilibrium gained by relaxation to characterize the equilibrium
    under the re-introduced constraint~$v_r^+ \geq \alpha_{r0}\ \forall r \in R$.
    In particular, we want to show that
    the calculation provided in~\cref{thm:heterogeneous:equilibrium:two} is correct:
    \begin{equation}
        \forall r \in R.\quad v_r^+ = \max\left(\alpha_{r0},\ \hat{v}_r^{*}\left(\max\left(\alpha_{\overline{r}0}, \hat{v}_{\overline{r}}^+\right)\right)\right)
        \label{eq:theo:sapp-constant:competition:computation}
    \end{equation}
    In other words, $v_r^+$ as calculated by~\cref{eq:theo:sapp-constant:competition:computation} should satisfy
    the equilibrium conditions on~$v_r^+$ in~\cref{eq:theo:sapp-constant:competition:interdep}.
    To satisfy these conditions,
    we consider all cases regarding~$\hat{v}_r^+$
    and~$\hat{v}_{\overline{r}}^+$:

    \begin{enumerate}
        \item $\hat{v}_r^+ \geq \alpha_{r0}$:
        \begin{enumerate}
            \item $\hat{v}_{\overline{r}}^+ \geq \alpha_{\overline{r}0}$:
            In that case,
            \cref{eq:theo:sapp-constant:competition:computation}
            suggests that
            \begin{equation}
                \begin{split}
                    &v_r^+ \overset{\text{1.(a)}}{=} \max\left(\alpha_{r0}, \hat{v}_r^{*}\left(\hat{v}_{\overline{r}}^+\right)\right)
                    \overset{\text{(\ref{eq:theo:sapp-constant:competition:unrestricted})}}{=} \max\left(\alpha_{r0}, \hat{v}_r^{+}\right)
                    \overset{\text{1.}}{=}
                    \hat{v}_r^+, \text{ and}\\
                     &v_{\overline{r}}^+ \overset{\text{1.}}{=} \max\left(\alpha_{\overline{r}0}, \hat{v}_{\overline{r}}^{*}\left(\hat{v}_{r}^+\right)\right)
                    \overset{\text{(\ref{eq:theo:sapp-constant:competition:unrestricted})}}{=} \max\left(\alpha_{\overline{r}0}, \hat{v}_{\overline{r}}^{+}\right)
                    \overset{\text{1.(a)}}{=}
                    \hat{v}_{\overline{r}}^+.
                \end{split}
                \label{eq:theo:sapp-constant:competition:1a}
            \end{equation}
            These values satisfy the
            equilibrium conditions 
            in~\cref{eq:theo:sapp-constant:competition:interdep}:
            \begin{equation}
                \begin{split}
                    &v_r^+ \overset{\text{(\ref{eq:theo:sapp-constant:competition:interdep})}}{=}
                    \max\left(\alpha_{r0},\ \hat{v}_r^{*}\left(v_{\overline{r}}^+\right)\right)
                    \overset{\text{(\ref{eq:theo:sapp-constant:competition:1a})}}{=}
                    \max\left(\alpha_{r0}, \hat{v}_r^{*}\left(\hat{v}_{\overline{r}}^+\right)\right)
                    \overset{\text{(\ref{eq:theo:sapp-constant:competition:unrestricted})}}{=} 
                    \max\left(\alpha_{r0}, \hat{v}_r^{+}\right)
                    \overset{\text{1.}} = \hat{v}_r^+,
                \end{split}
            \end{equation}
            and symmetrically for~$v_{\overline{r}}^+$.

            \item $\hat{v}_r^+ < \alpha_{\overline{r}0}$:
            In that case,
            \cref{eq:theo:sapp-constant:competition:computation}
            suggests that
            \begin{equation}
                \begin{split}
                    &v_r^+ \overset{\text{1.(b)}}{=} \max\left(\alpha_{r0}, \hat{v}_r^{*}\left(\alpha_{\overline{r}0}\right)\right)\\
                    &v_{\overline{r}}^+ \overset{\text{1.}}{=} \max\left(\alpha_{\overline{r}0}, \hat{v}_{\overline{r}}^{*}\left(\hat{v}_{r}^+\right)\right)
                    \overset{\text{(\ref{eq:theo:sapp-constant:competition:unrestricted})}}{=} \max\left(\alpha_{\overline{r}0}, \hat{v}_{\overline{r}}^{+}\right)
                    \overset{\text{1.(b)}}{=}
                    \alpha_{\overline{r}0}.
                \end{split}
                \label{eq:theo:sapp-constant:competition:1b}
            \end{equation}
            For that case, we perform
            another level of sub-case distinctions:
            \begin{enumerate}
                \item $\hat{v}_r^{*}(\alpha_{\overline{r}0}) \geq \alpha_{r0}$:
                In that case, \cref{eq:theo:sapp-constant:competition:1b} is simplified to
                \begin{equation}
                    \begin{split}
                        &v_r^+
                    \overset{\text{(\ref{eq:theo:sapp-constant:competition:1b})}}{=}
                    \max\left(\alpha_{r0}, \hat{v}_r^{*}\left(\alpha_{\overline{r}0}\right)\right)
                    \overset{\text{1.(b).i}}{=}
                    \hat{v}_r^{*}(\alpha_{\overline{r}0})\\
                        &v_{\overline{r}}^+ \overset{\text{(\ref{eq:theo:sapp-constant:competition:1b})}}{=}
                        \alpha_{\overline{r}0}
                    \end{split}
                    \label{eq:theo:sapp-constant:competition:1bi}
                \end{equation}

                Using~\cref{eq:theo:sapp-constant:competition:map-lower}
                and case condition 1.(b),
                we can also deduce:
                \begin{equation}
                    \hat{v}_{\overline{r}}^{*}\left(\hat{v}_{r}^{*}(\alpha_{\overline{r}0})\right) <
                    \alpha_{\overline{r}0}.
                    \label{eq:theo:sapp-constant:competition:alpha-bound}
                \end{equation}
                
                Then, we can again verify that these
            values satisfy the equilibrium
            conditions from~\cref{eq:theo:sapp-constant:competition:interdep}:
            \begin{equation}
                \begin{split}
                    &v_r^{+}
                    \overset{\text{(\ref{eq:theo:sapp-constant:competition:interdep})}}{=}
                    \max\left(\alpha_{r0},\ \hat{v}_r^{*}\left(v_{\overline{r}}^+\right)\right)\overset{\text{(\ref{eq:theo:sapp-constant:competition:1bi})}}{=} \max\left(\alpha_{r0}, \hat{v}_r^{*}\left(\alpha_{\overline{r}0}\right)\right)
                    \overset{\text{1.b.i}}{=}
                     \hat{v}_r^{*}(\alpha_{\overline{r}0})\\
                    &v_{\overline{r}}^+\overset{\text{(\ref{eq:theo:sapp-constant:competition:interdep})}}{=}
                    \max\left(\alpha_{\overline{r}0},\ \hat{v}_{\overline{r}}^{*}\left(v_{r}^+\right)\right)\overset{\text{(\ref{eq:theo:sapp-constant:competition:1bi})}}{=}
                    \max\left(\alpha_{\overline{r}0}, \hat{v}_{\overline{r}}^{*}\left( \hat{v}_r^{*}\left(\alpha_{\overline{r}0}\right)\right)\right)
                    \overset{\text{(\ref{eq:theo:sapp-constant:competition:alpha-bound})}}{=} \alpha_{\overline{r}0}.
                \end{split}
           \end{equation}

            \item $\hat{v}_r^{*}(\alpha_{\overline{r}0}) < \alpha_{r0}$:
            In that case, \cref{eq:theo:sapp-constant:competition:1b} is simplified to
                \begin{equation}
                    \begin{split}
                        &v_r^+
                    \overset{\text{(\ref{eq:theo:sapp-constant:competition:1b})}}{=}
                    \max\left(\alpha_{r0}, \hat{v}_r^{*}\left(\alpha_{\overline{r}0}\right)\right)
                    \overset{\text{1.(b).ii}}{=}
                    \alpha_{{r}0}\\
                        &v_{\overline{r}}^+ \overset{\text{(\ref{eq:theo:sapp-constant:competition:1b})}}{=}
                        \alpha_{\overline{r}0}
                    \end{split}
                    \label{eq:theo:sapp-constant:competition:1bii}
                \end{equation}

                Moreover, as proven in~\ref{sec:heterogeneous:single-path:equilibrium:1-b-ii},
                the current case implies
                \begin{equation}
                    \hat{v}_{\overline{r}}^{*}(\alpha_{r0}) \leq \alpha_{\overline{r}0}.
                    \label{eq:theo:sapp-constant:competition:alpha-bound-2}
                \end{equation}
                
                Based on these findings, the equilibrium conditions
                from~\cref{eq:theo:sapp-constant:competition:interdep}
                are satisfied:
                \begin{equation}
                \begin{split}
                    &v_r^{+}
                    \overset{\text{(\ref{eq:theo:sapp-constant:competition:interdep})}}{=}
                    \max\left(\alpha_{r0},\ \hat{v}_r^{*}\left(v_{\overline{r}}^+\right)\right)\overset{\text{(\ref{eq:theo:sapp-constant:competition:1bii})}}{=} \max\left(\alpha_{r0}, \alpha_{r0}\right)
                    = \alpha_{r0}\\
                    &v_{\overline{r}}^+\overset{\text{(\ref{eq:theo:sapp-constant:competition:interdep})}}{=}
                    \max\left(\alpha_{\overline{r}0},\ \hat{v}_{\overline{r}}^{*}\left(v_{r}^+\right)\right)\overset{\text{(\ref{eq:theo:sapp-constant:competition:1bii})}}{=}
                    \max\left(\alpha_{\overline{r}0}, \hat{v}_{\overline{r}}^{*}\left( \alpha_{r0}\right)\right)
                    \overset{\text{(\ref{eq:theo:sapp-constant:competition:alpha-bound-2})}}{=} \alpha_{\overline{r}0}.
                \end{split}
           \end{equation}
            \end{enumerate}
        \end{enumerate}

        \item $\hat{v}_r^+ < \alpha_{r0}$:
        \begin{enumerate}
            \item $\hat{v}_{\overline{r}}^+ \geq \alpha_{\overline{r}0}$:
            This case is symmetric to case 1.(b).

            \item $\hat{v}_{\overline{r}}^+ < \alpha_{\overline{r}0}$:
            In that case,
            \cref{eq:theo:sapp-constant:competition:computation}
            suggests that
            \begin{equation}
                \begin{split}
                    &v_r^+ \overset{\text{2.(b)}}{=} \max\left(\alpha_{r0}, \hat{v}_r^{*}\left(\alpha_{\overline{r}0}\right)\right)\\
                    &v_{\overline{r}}^+ \overset{\text{2.}}{=} \max\left(\alpha_{\overline{r}0}, \hat{v}_{\overline{r}}^{*}\left(\alpha_{r0}\right)\right)
                \end{split}
                \label{eq:theo:sapp-constant:competition:2b}
            \end{equation}
            
            Hence, we again need to perform
            another level of sub-case distinctions:

            \begin{enumerate}
                \item $\hat{v}_r^{*}(\alpha_{\overline{r}0}) \geq \alpha_{r0}$:
                For that case, we show in~\ref{sec:heterogeneous:single-path:equilibrium:2-b-i} that
                \begin{equation}
                    \hat{v}_{\overline{r}}^{*}(\alpha_{r0}) \leq \alpha_{\overline{r}0}.
                    \label{eq:theo:sapp-constant:competition:alpha-bound-3}
                \end{equation}
                Hence, \cref{eq:theo:sapp-constant:competition:2b}
                simplifies to:
                \begin{equation}
                \begin{split}
                    &v_r^+ \overset{\text{2.(b)}}{=} \max\left(\alpha_{r0}, \hat{v}_r^{*}\left(\alpha_{\overline{r}0}\right)\right) \overset{\text{2.(b).i}}{=} \hat{v}_r^{*}\left(\alpha_{\overline{r}0}\right)\\
                    &v_{\overline{r}}^+ \overset{\text{2.}}{=} \max\left(\alpha_{\overline{r}0}, \hat{v}_{\overline{r}}^{*}\left(\alpha_{r0}\right)\right)
                    \overset{\text{(\ref{eq:theo:sapp-constant:competition:alpha-bound-3})}}{=} \alpha_{\overline{r}0}
                \end{split}
                \label{eq:theo:sapp-constant:competition:2bi}
            \end{equation}

            Moreover, using~\cref{eq:theo:sapp-constant:competition:map-lower}
                and case condition 2.(b),
                we can again deduce:
                \begin{equation}
                    \hat{v}_{\overline{r}}^{*}\left(\hat{v}_{\overline{r}}^{*}(\alpha_{r0})\right) <
                    \alpha_{\overline{r}0}.
                    \label{eq:theo:sapp-constant:competition:alpha-bound-5}
                \end{equation}

            Based on these findings, the equilibrium conditions
                from~\cref{eq:theo:sapp-constant:competition:interdep}
                are satisfied:
                \begin{equation}
                \begin{split}
                    &v_r^{+}
                    \overset{\text{(\ref{eq:theo:sapp-constant:competition:interdep})}}{=}
                    \max\left(\alpha_{r0},\ \hat{v}_r^{*}\left(v_{\overline{r}}^+\right)\right)
                    \overset{\text{(\ref{eq:theo:sapp-constant:competition:2bi})}}{=}
                    \max\left(\alpha_{r0},\ \hat{v}_r^{*}\left(\alpha_{\overline{r}0}\right)\right)\overset{\text{2.(b).i}}{=}\hat{v}_r^{*}\left(\alpha_{\overline{r}0}\right)\\
                    &v_{\overline{r}}^+\overset{\text{(\ref{eq:theo:sapp-constant:competition:interdep})}}{=}
                    \max\left(\alpha_{\overline{r}0},\ \hat{v}_{\overline{r}}^{*}\left(v_{r}^+\right)\right)\overset{\text{(\ref{eq:theo:sapp-constant:competition:2bi})}}{=}
                    \max\left(\alpha_{\overline{r}0}, \hat{v}_{\overline{r}}^{*}\left( \hat{v}_r^{*}\left(\alpha_{\overline{r}0}\right)\right)\right)
                    \overset{\text{(\ref{eq:theo:sapp-constant:competition:alpha-bound-5})}}{=} \alpha_{\overline{r}0}
                \end{split}
           \end{equation}

                \item $\hat{v}_r^{*}(\alpha_{\overline{r}0}) < \alpha_{r0}$:
                In that case,
                \cref{eq:theo:sapp-constant:competition:2b}
                simplifies to:
                \begin{equation}
                \begin{split}
                    &v_r^+ \overset{\text{2.(b)}}{=} \max\left(\alpha_{r0}, \hat{v}_r^{*}\left(\alpha_{\overline{r}0}\right)\right) \overset{\text{2.(b).ii}}{=} \alpha_{r0}\\
                    &v_{\overline{r}}^+ \overset{\text{2.}}{=} \max\left(\alpha_{\overline{r}0}, \hat{v}_{\overline{r}}^{*}\left(\alpha_{r0}\right)\right)
                \end{split}
                \label{eq:theo:sapp-constant:competition:2bii}
            \end{equation}
            To further simplify~\cref{eq:theo:sapp-constant:competition:2bii},
            we perform another sub-case distinction:
            \begin{enumerate}
                \item $\hat{v}_{\overline{r}}^{*}(\alpha_{r0}) \geq \alpha_{\overline{r}0}$:
                This case is symmetric to case 2.(b).i.
                \item $\hat{v}_{\overline{r}}^{*}(\alpha_{r0}) < \alpha_{\overline{r}0}$:
                In that case, \cref{eq:theo:sapp-constant:competition:2bii}
                directly simplifies to:
                \begin{equation}
                \begin{split}
                    &v_r^+ \overset{\text{2.(b)}}{=} \max\left(\alpha_{r0}, \hat{v}_r^{*}\left(\alpha_{\overline{r}0}\right)\right) \overset{\text{2.(b).ii}}{=} \alpha_{r0}\\
                    &v_{\overline{r}}^+ \overset{\text{2.}}{=} 
                    \max\left(\alpha_{\overline{r}0}, \hat{v}_{\overline{r}}^{*}\left(\alpha_{r0}\right)\right)
                    \overset{\text{2.(b).ii.B}}{=}
                    \alpha_{\overline{r}0}
                \end{split}
                \label{eq:theo:sapp-constant:competition:2biiB}
            \end{equation}
            Clearly, the equilibrium conditions
                from~\cref{eq:theo:sapp-constant:competition:interdep}
                are satisfied by these values:
                \begin{equation}
                \begin{split}
                    &v_r^{+}
                    \overset{\text{(\ref{eq:theo:sapp-constant:competition:interdep})}}{=}
                    \max\left(\alpha_{r0},\ \hat{v}_r^{*}\left(v_{\overline{r}}^+\right)\right)
                    \overset{\text{(\ref{eq:theo:sapp-constant:competition:2biiB})}}{=}
                    \max\left(\alpha_{r0},\hat{v}_r^{*}\left(\alpha_{\overline{r}0}\right)\right)\overset{\text{2.(b).ii}}{=} \alpha_{r0}\\
                    &v_{\overline{r}}^+\overset{\text{(\ref{eq:theo:sapp-constant:competition:interdep})}}{=}
                    \max\left(\alpha_{\overline{r}0},\hat{v}_{\overline{r}}^{*}\left(v_{r}^+\right)\right)\overset{\text{(\ref{eq:theo:sapp-constant:competition:2biiB})}}{=}
                    \max\left(\alpha_{\overline{r}0}, \hat{v}_{\overline{r}}^{*}\left( \alpha_{r0}\right)\right)
                    \overset{\text{2.(b).ii.B}}{=} \alpha_{\overline{r}0}
                \end{split}
           \end{equation}
            \end{enumerate}
            \end{enumerate}
        \end{enumerate}
    \end{enumerate}

  Since the values calculated according to~\cref{eq:theo:sapp-constant:competition:computation}
  always satisfy the restricted-equilibrium
  conditions from~\cref{eq:theo:sapp-constant:competition:interdep},
  the theorem is proven.

\begin{figure*}
    \begin{minipage}{0.45\linewidth}
        \centering
        \def\alphaone{0.4}
\def\alphatwo{0.7}
\def\eqone{0.6}
\def\eqtwo{0.5}
\def\interone{0.0332}
\def\intertwo{0.7773}
\def\funtwopeak{0.25}
\def\funtwomax{(\alphatwo+0.25)}
\definecolor{amber}{rgb}{1.0, 0.49, 0.0}
\definecolor{azure}{rgb}{0.0, 0.5, 1.0}
\begin{tikzpicture}[declare function={
    fun1(\x) = (\eqone-1)/\eqtwo^2 * \x^2 + 1;
    fun1_up(\x) = pow(\eqtwo^2/(\eqone-1)*(\x-1), 0.5);
    nu_high = fun1_up(\alphaone);
    fun2(\x) = (\eqtwo-\funtwomax)/(\eqone-\funtwopeak)^2 * (\x-\funtwopeak)^2 + \funtwomax;
}]
    
    \begin{axis}%
    [
        xmin=-0.05,
        xmax=1,
        xtick={{fun1(\alphatwo)},\alphaone,\eqone,\interone},
        xticklabels={$\hat{v}_{r}^{*}(\alpha_{\overline{r}0})$,$\alpha_{r0}$, $\hat{v}_r^{+}$, $\tilde{v}_r^{+}$},
        ytick={\eqtwo, \alphatwo, nu_high, {fun2(\alphaone)},\intertwo,\funtwomax},
        yticklabels={$\hat{v}_{\overline{r}}^{+}$, $\alpha_{\overline{r}0}$, $\nu_{\overline{r}}$, $\hat{v}_{\overline{r}}^{*}(\alpha_{r0})$,$\tilde{v}_{\overline{r}}^{+}$,$\hat{v}_{\overline{r}}^{\uparrow}$},
        xtick style={draw=none},
        ytick style={draw=none},
        ymin=-0.05,
        ymax=1,
        x=5cm,
        axis y line=left,
        axis y line shift=-0.05,
        axis x line shift=-0.05,
        samples=200,
        domain=0:1,
        xlabel={$v_r$},
        ylabel={$v_{\overline{r}}$},
        x label style={at={(axis description cs:1,0)}},
        y label style={at={(axis description cs:0,0.1)}},
        axis x line=bottom,
    ]

        \draw[dotted,thick] (axis cs:\alphaone, 0) -- (axis cs:\alphaone,1);
        \draw[dotted,thick] (axis cs:0, \alphatwo) -- (axis cs:1,\alphatwo);

        \addplot[mark=none,color=azure] (x, {fun1_up(x)});
        \node[color=azure] at (axis cs:0.8,{fun1_up(0.8)+0.1}) {$\hat{v}_{r}^{*}(v_{\overline{r}})$};
        
        \addplot[mark=none,color=amber] (x, {fun2(x)});
        \node[color=amber] at (axis cs:0.55,{fun2(0.45)}) {$\hat{v}_{\overline{r}}^{*}(v_r)$};
        \draw[dotted] (0, {\funtwomax}) -- (\funtwopeak, {\funtwomax});

        \node[fill=black,circle] at (axis cs:\eqone, \eqtwo) {};
        \draw[dotted] (axis cs:\eqone, 0) -- (axis cs:\eqone, \eqtwo);
        \draw[dotted] (axis cs:0, \eqtwo) -- (axis cs:\eqone, \eqtwo);

        \draw[dotted] (axis cs:0, nu_high) -- (axis cs:\alphaone, nu_high);

        \draw[dotted] (axis cs:{fun1(\alphatwo)}, 0) -- (axis cs:{fun1(\alphatwo)}, \alphatwo);

        \draw[dotted] (axis cs:0, {fun2(\alphaone)}) -- (axis cs:\alphaone, {fun2(\alphaone)});

        \node[fill=black,circle] at (axis cs:\interone, \intertwo) {};
        \draw[dotted] (axis cs:\interone, 0) -- (axis cs:\interone, \intertwo);
        \draw[dotted] (axis cs:0, \intertwo) -- (axis cs:\interone, \intertwo);

    \end{axis}

\end{tikzpicture}
        \caption{Visualization of non-unique equilibrium~$(\hat{v}_r^+, \hat{v}_{\overline{r}}^+)$ in~\ref{sec:heterogeneous:single-path:equilibrium:1-b-ii}.}
        \label{fig:heterogeneous:single-path:equilibrium:1-b-ii}
    \end{minipage}
    \hfill 
    \begin{minipage}{0.45\linewidth}
        \centering
        \def\alphaone{0.4}
\def\alphatwo{0.3}
\def\eqone{0.25}
\def\eqtwo{0.13}
\def\interone{0.475}
\def\intertwo{0.24}
\def\interthree{0.335}
\def\interfour{0.635}
\def\funonepeak{0.4}
\def\funonemax{0.6}
\def\funtwopeak{0.37}
\def\funtwomax{(\alphatwo+0.25)}
\def\funthreepeak{0.48}
\def\funthreemax{(\alphatwo+0.65)}
\definecolor{amber}{rgb}{1.0, 0.49, 0.0}
\definecolor{azure}{rgb}{0.0, 0.5, 1.0}
\begin{tikzpicture}[declare function={
    fun1(\x) = (\eqone-\funonemax)/(\eqtwo-\funonepeak)^2 * (\x-\funonepeak)^2 + \funonemax;
    fun1_up(\x) = pow((\eqtwo-\funonepeak)^2/(\eqone-\funonemax)*(\x-\funonemax), 0.5) + \funonepeak;
    fun1_down(\x) = -pow((\eqtwo-\funonepeak)^2/(\eqone-\funonemax)*(\x-\funonemax), 0.5) + \funonepeak;
    nu_high = fun1_up(\alphaone);
    nu_low = fun1_down(\alphaone);
    fun2(\x) = (\eqtwo-\funtwomax)/(\eqone-\funtwopeak)^2 * (\x-\funtwopeak)^2 + \funtwomax;
    fun3(\x) = (\eqtwo-\funthreemax)/(\eqone-\funthreepeak)^2 * (\x-\funthreepeak)^2 + \funthreemax;
}]
    
    \begin{axis}%
    [
        xmin=-0.05,
        xmax=1,
        xtick={{fun1(\alphatwo)},\alphaone,\eqone},
        xticklabels={$\ \ \hat{v}_{r}^{*}(\alpha_{\overline{r}0})$,$\alpha_{r0}$, $\hat{v}_r^{+}$},
        ytick={\eqtwo, \alphatwo, nu_high, {fun2(\alphaone)}, {fun3(\alphaone)}, nu_low},
        yticklabels={$\hat{v}_{\overline{r}}^{+}$, $\alpha_{\overline{r}0}$, $\nu_{\overline{r}}^{\rightarrow}$, $\hat{v}_{\overline{r}}^{*}(\alpha_{r0})$,$\hat{v}_{\overline{r}}^{*}(\alpha_{r0})$, $\nu_{\overline{r}}^{\leftarrow}$},
        xtick style={draw=none},
        ytick style={draw=none},
        ymin=-0.05,
        ymax=1,
        x=5cm,
        axis y line=left,
        axis y line shift=-0.05,
        axis x line shift=-0.05,
        samples=200,
        domain=0:1,
        xlabel={$v_r$},
        ylabel={$v_{\overline{r}}$},
        x label style={at={(axis description cs:1,0)}},
        y label style={at={(axis description cs:0,1)}},
        axis x line=bottom,
    ]

        \draw[dotted,thick] (axis cs:\alphaone, 0) -- (axis cs:\alphaone,1);
        \draw[dotted,thick] (axis cs:0, \alphatwo) -- (axis cs:1,\alphatwo);

        \addplot[mark=none,color=azure] (x, {fun1_up(x)});
        \addplot[mark=none,color=azure] (x, {fun1_down(x)});
        \node[color=azure] at (axis cs:0.2,{fun1_up(0.2)+0.075}) {$\hat{v}_{r}^{*}(v_{\overline{r}})$};
        
        \addplot[mark=none,color=amber] (x, {fun2(x)});
        \node[color=amber] at (axis cs:0.17,{fun2(0.3)-0.05}) {$\hat{v}_{\overline{r}}^{*}(v_r)$};
        \addplot[mark=none,color=amber,dashed] (x, {fun3(x)});

        \node[fill=black,circle] at (axis cs:\eqone, \eqtwo) {};
        \draw[dotted] (axis cs:\eqone, 0) -- (axis cs:\eqone, \eqtwo);
        \draw[dotted] (axis cs:0, \eqtwo) -- (axis cs:\eqone, \eqtwo);

        \draw[dotted] (axis cs:0, nu_high) -- (axis cs:\alphaone, nu_high);
        \draw[dotted] (axis cs:0, nu_low) -- (axis cs:\alphaone, nu_low);

        \draw[dotted] (axis cs:{fun1(\alphatwo)}, 0) -- (axis cs:{fun1(\alphatwo)}, \alphatwo);

        \draw[dotted] (axis cs:0, {fun2(\alphaone)}) -- (axis cs:\alphaone, {fun2(\alphaone)});
        \draw[dotted] (axis cs:0, {fun3(\alphaone)}) -- (axis cs:\alphaone, {fun3(\alphaone)});

        \node[fill=black,circle] at (axis cs:\interone, \intertwo) {};
        \node[fill=black,circle] at (axis cs:\interthree, \interfour) {};

    \end{axis}

\end{tikzpicture}
        \caption{Visualization of non-unique equilibrium~$(\hat{v}_r^+, \hat{v}_{\overline{r}}^+)$ in~\ref{sec:heterogeneous:single-path:equilibrium:2-b-i}.}
        \label{fig:heterogeneous:single-path:equilibrium:2-b-i}
    \end{minipage}
\end{figure*}

\subsubsection{Upper bound on~$\hat{v}_{\overline{r}}^{*}(\alpha_{r0})$ for case 1.(b).ii}
\label{sec:heterogeneous:single-path:equilibrium:1-b-ii}

We can show that
                case 1.(b).ii implies
                \begin{equation}
                    \hat{v}_{\overline{r}}^{\ast}(\alpha_{r0}) \leq \alpha_{\overline{r}0}.
                    \label{eq:heterogeneous:single-path:equilibrium:1-b-ii:true}
                \end{equation}
                In particular, let us assume the opposite for the sake of contradiction, i.e., we assume
                \begin{equation}
                    \hat{v}_{\overline{r}}^{\ast}(\alpha_{r0}) > \alpha_{\overline{r}0}.
                    \label{eq:heterogeneous:single-path:equilibrium:1-b-ii:contradiction}
                \end{equation}
For this proof, we first investigate the functions~$\hat{v}_r^{*}$ and~$\hat{v}_{\overline{r}}^{*}$ more thoroughly,
and then produce a contradiction by
showing the existence of a second unrestricted 
equilibrium~$(\tilde{v}_r^+, \tilde{v}_{\overline{r}}^+) \neq (\hat{v}_r^+, \hat{v}_{\overline{r}}^+)$.
The proof idea is visualized in~\cref{fig:heterogeneous:single-path:equilibrium:1-b-ii}.

\paragraph{$\hat{v}_r^{*}$}
In case 1.(b).ii, we can more precisely
                characterize the function~$\hat{v}_r^{*}$
                based on the case conditions.
                In particular, we know that~$\hat{v}_r^{*}$
                evolves from value~$\hat{v}_r^{*}(\hat{v}_{\overline{r}}^+) =
                \hat{v}_{r}^{+} \geq \alpha_{r0}$
                (1.) at argument~$\hat{v}_{\overline{r}}^+$
                down to value~$\hat{v}_r^{*}(\alpha_{\overline{r}0}) < \alpha_{r0}$ (1.(b).ii) at
                argument~$\alpha_{\overline{r}0} > \hat{v}_{\overline{r}}^+$ (1.(b)).
                Hence, the intermediate-value theorem and the concavity of~$\hat{v}_{r}^{*}$ stipulate
                that
                \begin{equation}
                    \exists \nu_{\overline{r}} \in [\hat{v}_{\overline{r}}^+, \alpha_{\overline{r}0}). \quad
                    \hat{v}_r^{*}(\nu_{\overline{r}}) = \alpha_{r0}
                    \quad \text{and}
                    \quad \forall v_{\overline{r}} > \nu_{\overline{r}}.\ \hat{v}_r^{*}(v_{\overline{r}}) < \alpha_{r0}.
                    \label{eq:heterogeneous:single-path:equilibrium:1-b-ii:nu-overline-r}
                \end{equation}

\paragraph{$\hat{v}_{\overline{r}}^{*}$}
The assumption in \cref{eq:heterogeneous:single-path:equilibrium:1-b-ii:contradiction} 
suggests that~$\hat{v}_{\overline{r}}^{*}$
reaches a value above~$\alpha_{\overline{r}0}$
at argument~$\alpha_{r0}$.
Hence, the maximum of
$\hat{v}_{\overline{r}}^{*}$
is also at least~$\alpha_{\overline{r}0}$:
\begin{equation}
    \hat{v}_{\overline{r}}^{\uparrow}
    = \max_{v_r}
    \hat{v}_{\overline{r}}^{*}(v_r) \geq \alpha_{\overline{r}0} \overset{\text{(\ref{eq:heterogeneous:single-path:equilibrium:1-b-ii:nu-overline-r})}}{>} \nu_{\overline{r}}.
    \label{eq:heterogeneous:single-path:equilibrium:1-b-ii:maximum}
\end{equation}

\paragraph{Contradiction}
Based on these properties of $\hat{v}_{r}^{*}$
and~$\hat{v}_{\overline{r}}^{*}$,
we now show that there exist
$\tilde{v}_r^{+} < \alpha_{r0}$ and $\tilde{v}_{\overline{r}}^+ > \nu_{\overline{r}}$,
with the unrestricted-equilibrium 
properties~$\tilde{v}_r^+ = \hat{v}_r^{*}(\tilde{v}_{\overline{r}}^+)$ and
$\tilde{v}_{\overline{r}}^+ = \hat{v}_{\overline{r}}^{*}(\tilde{v}_r^+)$.
To verify the existence of these valuations,
note that the value~$\tilde{v}_{\overline{r}}^+$
satisfies the condition:
\begin{equation}
    \tilde{v}_{\overline{r}}^+ = 
    \hat{v}_{\overline{r}}^{*}(\tilde{v}_{r}^+) =
    \hat{v}_{\overline{r}}^{*}(\hat{v}_r^{*}(\tilde{v}_{\overline{r}}^+)) \iff 
    \hat{v}_{\overline{r}}^{**}(\tilde{v}_{\overline{r}}^+) := \tilde{v}_{\overline{r}}^+ - \hat{v}_{\overline{r}}^{*}(\hat{v}_r^{*}(\tilde{v}_{\overline{r}}^+)) = 0.
    \label{eq:heterogeneous:single-path:equilibrium:1-b-ii:reflector}
\end{equation}

We now evaluate the `reflector' function~$\hat{v}_{\overline{r}}^{**}$
at two arguments~$v_{\overline{r}}$, namely~$\nu_{\overline{r}}$ and~$\hat{v}_{\overline{r}}^{\uparrow}$.
For~$v_{\overline{r}} = \nu_{\overline{r}}$,
it holds that
\begin{equation}
    \hat{v}_{\overline{r}}^{**}(\nu_{\overline{r}})
     \overset{\text{(\ref{eq:heterogeneous:single-path:equilibrium:1-b-ii:reflector})}}{=}  \nu_{\overline{r}} - \hat{v}_{\overline{r}}^{*}(\hat{v}_r^{*}(\nu_{\overline{r}}))
    \overset{\text{(\ref{eq:heterogeneous:single-path:equilibrium:1-b-ii:nu-overline-r})}}{=}
    \nu_{\overline{r}} - \hat{v}_{\overline{r}}^{*}(\alpha_{r0})
     \overset{\text{(\ref{eq:heterogeneous:single-path:equilibrium:1-b-ii:contradiction})}}{<}
     \nu_{\overline{r}} - \alpha_{\overline{r}0}
      \overset{\text{(\ref{eq:heterogeneous:single-path:equilibrium:1-b-ii:nu-overline-r})}}{<}
      0
\end{equation}

For~$v_{\overline{r}} = \hat{v}_{\overline{r}}^{\uparrow} > \nu_{\overline{r}}$,
it holds that
\begin{equation}
    \hat{v}_{\overline{r}}^{**}(\hat{v}_{\overline{r}}^{\uparrow})
     \overset{\text{(\ref{eq:heterogeneous:single-path:equilibrium:1-b-ii:reflector})}}{=}  \hat{v}_{\overline{r}}^{\uparrow} - \hat{v}_{\overline{r}}^{*}(\hat{v}_{r}^{*}(\hat{v}_{\overline{r}}^{\uparrow}))
    \overset{\text{(\ref{eq:heterogeneous:single-path:equilibrium:1-b-ii:maximum})}}{=} 
     0.
\end{equation}

Since~$\hat{v}_{\overline{r}}^{**}$ is continuous,
the intermediate-value theorem stipulates that a~$\tilde{v}_{\overline{r}}^+ \in (\nu_{\overline{r}},  v_{\overline{r}}^{\uparrow}]$
exists that satisfies~$\hat{v}_{\overline{r}}^{**}(\tilde{v}_{\overline{r}}^+) = 0$.
Then, since~$\tilde{v}_{\overline{r}}^+ > \nu_{\overline{r}}$, it follows that~$\tilde{v}_r^+ = \hat{v}_r^{*}(\tilde{v}_{\overline{r}}^+) < \alpha_{r0}$ by~\cref{eq:heterogeneous:single-path:equilibrium:1-b-ii:nu-overline-r}.

Since the unrestricted equilibrium 
valuations~$(\hat{v}_r^{+}, \hat{v}_{\overline{r}}^{+})$ are unique,
it must hold that~$\hat{v}_{r}^+ = \tilde{v}_{r}^+$.
However, the case condition~$\hat{v}_{r}^+ \geq \alpha_{r0}$ conflicts with the
derived condition~$\tilde{v}_{r}^+ < \alpha_{r0}$.
We thus arrive at a contradiction,
which invalidates~\cref{eq:heterogeneous:single-path:equilibrium:1-b-ii:contradiction}
and confirms~\cref{eq:heterogeneous:single-path:equilibrium:1-b-ii:true}.

\subsubsection{Upper bound on~$\hat{v}_{\overline{r}}^{*}(\alpha_{r0})$ for case 2.(b).i}
\label{sec:heterogeneous:single-path:equilibrium:2-b-i}

The following proof is similar in structure and goal as the proof
in~\ref{sec:heterogeneous:single-path:equilibrium:1-b-ii},
namely to prove~$\hat{v}_{\overline{r}}^{*}(\alpha_{r0}) \leq \alpha_{\overline{r}0}$
by assuming
\begin{equation}
    \hat{v}_{\overline{r}}^{\ast}(\alpha_{r0}) > \alpha_{\overline{r}0}.
    \label{eq:heterogeneous:single-path:equilibrium:2-b-i:contradiction}
\end{equation}
The arguments of the proof are visualized in~\cref{fig:heterogeneous:single-path:equilibrium:2-b-i}.

\paragraph{$\hat{v}_r^{*}$}
In case 2.(b).i, the strict concavity of~$\hat{v}_r^{*}$,
together with the knowledge of~$\hat{v}_{r}^{*}(\alpha_{\overline{r}0}) \geq \alpha_{r0}$ 
(2.(b).i), imply:
\begin{equation}
    \begin{split}
        \exists\ \nu_{\overline{r}}^{\leftarrow}, \nu_{\overline{r}}^{\rightarrow}.\quad 
    &\alpha_{\overline{r}0} \in [\nu_{\overline{r}}^{\leftarrow}, \nu_{\overline{r}}^{\rightarrow}] \text{ and }
    \hat{v}_{r}^{*}(\nu_{\overline{r}}^{\leftarrow}) = \alpha_{r0}
    \text{  and  }
    \hat{v}_{r}^{*}(\nu_{\overline{r}}^{\rightarrow}) = \alpha_{r0}
    \text{  and  }\\
    &\forall v_{\overline{r}} \in [\nu_{\overline{r}}^{\leftarrow}, \nu_{\overline{r}}^{\rightarrow}].\ 
    \hat{v}_r^{*}(v_{\overline{r}}) \geq \alpha_{r0}
    \text{  and  }
    \forall v_{\overline{r}} \notin [\nu_{\overline{r}}^{\leftarrow}, \nu_{\overline{r}}^{\rightarrow}].\ 
    \hat{v}_r^{*}(v_{\overline{r}}) < \alpha_{r0}.
    \end{split}
    \label{eq:heterogeneous:single-path:equilibrium:2-b-i:nu-overline-r}
\end{equation}

\paragraph{$\hat{v}_{\overline{r}}^{*}$}
Given~\cref{eq:heterogeneous:single-path:equilibrium:2-b-i:contradiction},
we find the maximum of~$\hat{v}_{\overline{r}}^{*}$:
\begin{equation}
    \hat{v}_{\overline{r}}^{\uparrow}
    = \max_{v_r}
    \hat{v}_{\overline{r}}^{*}(v_r) 
    \geq \hat{v}_{\overline{r}}^{\ast}(\alpha_{r0})
    \overset{\text{(\ref{eq:heterogeneous:single-path:equilibrium:2-b-i:contradiction})}}{>} \alpha_{\overline{r}0}.
    \label{eq:heterogeneous:single-path:equilibrium:2-b-i:maximum}
\end{equation}

\paragraph{Contradiction}
Similar as in~\ref{sec:heterogeneous:single-path:equilibrium:1-b-ii},
we show the existence of~$(\tilde{v}_r^+, \tilde{v}_{\overline{r}}^+) \neq (\hat{v}_r^+, \hat{v}_{\overline{r}}^+)$,
which satisfy the unrestricted-equilibrium 
properties~$\tilde{v}_r^+ = \hat{v}_r^{*}(\tilde{v}_{\overline{r}}^+)$ and
$\tilde{v}_{\overline{r}}^+ = \hat{v}_{\overline{r}}^{*}(\tilde{v}_r^+)$,
and thus contradict the uniqueness
of the unrestricted equilibrium valuations~$(\hat{v}_r^+, \hat{v}_{\overline{r}}^+)$.
To that end, we again introduce a reflector
function~$\hat{v}_{\overline{r}}^{**}$
with
\begin{equation}
    \hat{v}_{\overline{r}}^{**}(\tilde{v}_{\overline{r}}^+) =
    \tilde{v}_{\overline{r}}^+ - 
    \hat{v}_{\overline{r}}^{*}(\hat{v}_{r}^{*}(\tilde{v}_{\overline{r}}^{+})) = 0.
    \label{eq:heterogeneous:single-path:equilibrium:2-b-i:reflector}
\end{equation}
However, unlike in~\ref{sec:heterogeneous:single-path:equilibrium:1-b-ii},
we additionally have to consider
the relative position of~$\nu_{\overline{r}}^{\rightarrow}$ from~\cref{eq:heterogeneous:single-path:equilibrium:2-b-i:nu-overline-r} and~$\hat{v}_{\overline{r}}^{*}(\alpha_{r0})$.

\begin{itemize}
    \item $\hat{v}_{\overline{r}}^{*}(\alpha_{r0}) \geq \nu_{\overline{r}}^{\rightarrow}$:
    For that case, we evaluate the reflector function~$\hat{v}_{\overline{r}}^{**}$
at arguments~$\nu_{\overline{r}}^{\rightarrow}$ from~\cref{eq:heterogeneous:single-path:equilibrium:2-b-i:nu-overline-r} and~$\hat{v}_{\overline{r}}^{\uparrow}\geq \nu_{\overline{r}}^{\rightarrow}$:
\begin{equation}
    \begin{split}
    &\hat{v}_{\overline{r}}^{**}(\nu_{\overline{r}}^{\rightarrow})
    \overset{\text{(\ref{eq:heterogeneous:single-path:equilibrium:2-b-i:reflector})}}{=} \nu_{\overline{r}}^{\rightarrow} - \hat{v}_{\overline{r}}^{*}(\hat{v}_{r}^{*}(\nu_{\overline{r}}^{\rightarrow}))
    \overset{\text{(\ref{eq:heterogeneous:single-path:equilibrium:2-b-i:nu-overline-r})}}{=}
    \nu_{\overline{r}}^{\rightarrow} - \hat{v}_{\overline{r}}^{*}(\alpha_{r0}) 
     \leq 0\\
     &\hat{v}_{\overline{r}}^{**}(\hat{v}_{\overline{r}}^{\uparrow})
        \overset{\text{(\ref{eq:heterogeneous:single-path:equilibrium:2-b-i:reflector})}}{=} \hat{v}_{\overline{r}}^{\uparrow} - 
    \hat{v}_{\overline{r}}^{*}(\hat{v}_{r}^{*}(\hat{v}_{\overline{r}}^{\uparrow}))
    \overset{\text{(\ref{eq:heterogeneous:single-path:equilibrium:2-b-i:maximum})}}{\geq} 0
    \end{split}
\end{equation}
     By the intermediate value theorem,
    there must thus exist 
    a~$\tilde{v}_{\overline{r}}^+ \in [\nu_{\overline{r}}^{\rightarrow}, \hat{v}_{\overline{r}}^{\uparrow}]$
    with~$\hat{v}_{\overline{r}}^{**}(\tilde{v}_{\overline{r}}^+) = 0$.
    Since~$\tilde{v}_{\overline{r}}^+ \in [\nu_{\overline{r}}^{\rightarrow}, \hat{v}_{\overline{r}}^{\uparrow}]$,
    \cref{eq:heterogeneous:single-path:equilibrium:2-b-i:nu-overline-r} implies that~$\tilde{v}_{\overline{r}}^+ \geq \nu_{\overline{r}}^{\rightarrow} \geq \alpha_{\overline{r}0}$, which conflicts
    with~$\hat{v}_{\overline{r}}^+ < \alpha_{\overline{r}0}$ from
    case condition~2.(b) and the uniqueness of
    the unrestricted equilibrium.
    
    \item $\hat{v}_{\overline{r}}^{*}(\alpha_{r0}) < \nu_{\overline{r}}^{\rightarrow}$:
    For that case, we evaluate the reflector function~$\hat{v}_{\overline{r}}^{**}$
at arguments~$\nu_{\overline{r}}^{\leftarrow}$ from~\cref{eq:heterogeneous:single-path:equilibrium:2-b-i:nu-overline-r} and~$\hat{v}_{\overline{r}}^{\uparrow}$
from~\cref{eq:heterogeneous:single-path:equilibrium:2-b-i:maximum}:
\begin{equation}
    \begin{split}
    &\hat{v}_{\overline{r}}^{**}(\nu_{\overline{r}}^{\leftarrow})
    \overset{\text{(\ref{eq:heterogeneous:single-path:equilibrium:2-b-i:reflector})}}{=} \nu_{\overline{r}}^{\leftarrow} - \hat{v}_{\overline{r}}^{*}(\hat{v}_{r}^{*}(\nu_{\overline{r}}^{\leftarrow}))
    \overset{\text{(\ref{eq:heterogeneous:single-path:equilibrium:2-b-i:nu-overline-r})}}{=}
    \nu_{\overline{r}}^{\leftarrow} - \hat{v}_{\overline{r}}^{*}(\alpha_{r0}) 
    \overset{\text{(\ref{eq:heterogeneous:single-path:equilibrium:2-b-i:contradiction})}}{<} 
     \nu_{\overline{r}}^{\leftarrow} - \alpha_{\overline{r}0}
     \overset{\text{(\ref{eq:heterogeneous:single-path:equilibrium:2-b-i:nu-overline-r})}}{\leq} 0\\
     &\hat{v}_{\overline{r}}^{**}(\hat{v}_{\overline{r}}^{\uparrow})
        \overset{\text{(\ref{eq:heterogeneous:single-path:equilibrium:2-b-i:reflector})}}{=} \hat{v}_{\overline{r}}^{\uparrow} - 
    \hat{v}_{\overline{r}}^{*}(\hat{v}_{r}^{*}(\hat{v}_{\overline{r}}^{\uparrow}))
    \overset{\text{(\ref{eq:heterogeneous:single-path:equilibrium:2-b-i:maximum})}}{\geq} 0
    \end{split}
\end{equation}
    By the intermediate value theorem,
    there must thus exist 
    a~$\tilde{v}_{\overline{r}}^+ \in [\nu_{\overline{r}}^{\leftarrow}, \hat{v}_{\overline{r}}^{\uparrow}]$
    with~$\hat{v}_{\overline{r}}^{**}(\tilde{v}_{\overline{r}}^+) = 0$.
    Since~$\tilde{v}_{\overline{r}}^+ \in [\nu_{\overline{r}}^{\leftarrow}, \hat{v}_{\overline{r}}^{\uparrow}] \subset [\nu_{\overline{r}}^{\leftarrow}, \nu_{\overline{r}}^{\rightarrow}]$,
    \cref{eq:heterogeneous:single-path:equilibrium:2-b-i:nu-overline-r} implies that~$\tilde{v}_r^+ = \hat{v}_r^+(\tilde{v}_{\overline{r}}^+) \geq \alpha_{r0}$, which conflicts
    with~$\hat{v}_r^+ < \alpha_{r0}$ from
    case condition~2 and the uniqueness of
    the unrestricted equilibrium.
    
\end{itemize}

\section{Proof of~\cref{thm:heterogeneous:equilibrium:stability}}
\label{sec:app:proofs:heterogeneous--two-path-stability}

\subsection{Proof idea}

    To confirm the asymptotic stability of an
    equilibrium~$\mathbf{A}^+$, we demonstrate
    that the Jacobian matrix~$\mathbf{J}(\mathbf{A})$
    of the process in~\cref{eq:theo:sapp-constant:process}
    is negative definite when evaluated at 
    the equilibrium~$\mathbf{A}^+$.
    On a high level, the Jacobian 
    matrix~$\mathbf{J}(\mathbf{A})$ is
    defined as follows:
    \begin{equation}
        \forall (n,k),\ (n',k') \in N\times K. \quad 
        J_{I(n,k),I(n',k')}(\mathbf{A}) = \frac{\partial}{\partial a_{n'k'}}
        \left(a^{*}_{nk}(\mathbf{A}_{-nk}) - a_{nk}\right)
        \label{eq:theo:sapp-constant:stability:jacobian}
    \end{equation} where~$I(n,k)$ is an index
    corresponding to attribute~$(n,k)$.
    The derivatives of the restricted best-response~$a^{*}_{nk}$ 
    in any attribute prevalence~$a_{n'k'}$ are as follows:
    \begin{equation}
        \frac{\partial}{\partial a_{n'k'}} a^{*}_{nk}(\mathbf{A})
        = \frac{\partial}{\partial a_{n'k'}} \max(0, \hat{a}^{*}_{nk}(\mathbf{A}))
        = \begin{cases}
            \frac{\partial}{\partial a_{n'k'}} \hat{a}^{*}_{nk}(\mathbf{A})
            & \text{if }  \hat{a}^{*}_{nk}(\mathbf{A}) \geq 0,\\
            0 & \text{otherwise.}
        \end{cases}
        \label{eq:theo:sapp-constant:stability:restricted-best-response-derivative}
    \end{equation}

     To show this negative definiteness of~$\mathbf{J}^+ = \mathbf{J}(\mathbf{A}^+)$, we demonstrate
     that every eigenvalue~$\lambda$ of~$\mathbf{J}^+$ has 
     a negative real part~$\text{Re}(\lambda)$.
     To find the eigenvalues~$\lambda$ of~$\mathbf{J}^+$,
     we solve the equation system~$\mathbf{J}^+\mathbf{x} = \lambda \mathbf{x}$
     for~$\lambda \in \mathbb{C}$, $\lambda \neq 0$, and~$\mathbf{x} \in \mathbb{C}^{|N||K|}$, $\mathbf{x} \neq \boldsymbol{0}$.
     This equation system can be represented in the following form:
     \begin{equation}
         \forall (n,k) \in N \times  K. \quad 
         \left(J^+_{I(n,k),I(n,k)} - \lambda\right) x_{I(n,k)}
         + \sum_{\substack{(n',k') \in N\times K.\\(n',k') \neq (n,k)}}
         J^+_{I(n,k),I(n',k')} x_{I(n',k')} = 0
         \label{eq:theo:sapp-constant:stability:eig-eq:0}
     \end{equation}

    \subsection{Simplification of \cref{eq:theo:sapp-constant:stability:eig-eq:0}}

    To concretize~\cref{eq:theo:sapp-constant:stability:eig-eq:0}, 
    we instantiate the Jacobian matrix~$\mathbf{J}^+$.
    As~$a^{*}_{nk}$ is independent of~$a_{nk}$, the diagonal entries of~$\mathbf{J}^+$
    are:
    \begin{equation}
        \forall (n,k) \in N \times  K. \quad J^+_{I(n,k),I(n,k)} \overset{\text{(\ref{eq:theo:sapp-constant:stability:jacobian})}}{=} \frac{\partial}{\partial a_{n'k'}}
        \left(a^{*}_{nk}(\mathbf{A}_{-nk}) - a_{nk}\right) = -1,
    \end{equation}

    Now, we consider the entries not on the diagonal of~$\mathbf{J}^+$,
    i.e., $J^+_{I(n,k),I(n',k')}$ for all~$(n,k) \neq (n',k')$.
    First, we specifically consider the rows of~$\mathbf{J}^+$ 
    associated with attributes~$(n,k) \in L^+$, where~$L^+$
    is the set of attributes
    which must have zero prevalence~$a_{nk}^+ = 0$ in equilibrium:
    \begin{equation}
        L^+ = N \times  K \setminus (K_r^+ \cup K_{\overline{r}}^+).
    \end{equation}
    The following inequality holds on
    the equilibrium path valuation~$v_r^+$ for each path~$r$
    (cf.~\cref{thm:heterogeneous:single-path:equilibrium,thm:heterogeneous:equilibrium:two}):
    \begin{equation}
        \forall (n,k) \in L^+. \quad v_{r(n)}^+ \geq \psi_r \sqrt{d} \sqrt{1 + v_{\overline{r}(n)}^+} 
        - (1 + v_{\overline{r}(n)}^+) > \sqrt{\frac{\alpha_{nk}}{\gamma_{nk}}
        d(\rho_{n} - \phi_{n0})}\sqrt{1 + v_{\overline{r}(n)}^+} 
        - (1 + v_{\overline{r}(n)}^+)
    \end{equation}
    Then, remember the following equivalence
    from~\cref{thm:heterogeneous:single-path:equilibrium}
    for any attribute~$(n,k)$:
    \begin{equation}
        v_{r(n)}^+ > \sqrt{\frac{\alpha_{nk}}{\gamma_{nk}}
        d(\rho_{n} - \phi_{n0})}\sqrt{1 + v_{\overline{r}(n)}^+} 
        - (1 + v_{\overline{r}(n)}^+) \iff
        \frac{\partial \pi_{n}(\mathbf{A}^+)}{\partial a_{nk}} < 0.
    \end{equation}
    Together with the concavity of~$\pi_n$, we thus
    note that the attribute value~$a_{nk}$ needs to 
    be decreased to optimize the profit~$\pi_n$ in~$a_{nk}$.
    Given~$a^+_{nk} = 0$, we note that the
    unrestricted
    best-response attribute~$\hat{a}^{*}_{nk}$
    for~$(n,k) \in L^+$ is
    thus negative in the equilibrium:
    \begin{equation}
        \forall (n,k) \in L^+. \quad 
        \hat{a}^{*}(\mathbf{A}^+_{-nk}) < 0
    \end{equation}
    Given the definition of the Jacobian entries 
    in~\cref{eq:theo:sapp-constant:stability:jacobian}
    and the derivative of the restricted best response~$a^{*}_{nk}$ in
    \cref{eq:theo:sapp-constant:stability:restricted-best-response-derivative},
    we note that:
    \begin{equation}
        \forall (n,k) \in L^+.\ \forall (n',k') \neq (n,k). \quad
        J^+_{I(n,k),I(n',k')} = 0.
    \end{equation}
    The eigenvalue equation system in~\cref{eq:theo:sapp-constant:stability:eig-eq:0} can thus be written as:
    \begin{align}
            \forall (n,k) \in K_r^+ \cup K_{\overline{r}}^+. \quad 
            &-(\lambda + 1) x_{I(n,k)} 
            + \sum_{\substack{(n',k') \in N\times  K.\\ (n',k') \neq (n,k)}} J^+_{I(n,k),I(n',k')} x_{I(n',k')} 
            = 0 \label{eq:theo:sapp-constant:stability:eig-eq:1}\\
            \forall (n,k) \in L^+.\quad &-(\lambda + 1) x_{I(n,k)}  
            = 0 \label{eq:theo:sapp-constant:stability:eig-eq:2}
    \end{align}

    Interestingly, the equation system in~~\cref{eq:theo:sapp-constant:stability:eig-eq:1,eq:theo:sapp-constant:stability:eig-eq:2} can be
    considerably simplified in our proof, which
    can be shown by a case distinction on~$L^+ =
    (N \times  K) \setminus (K_r^+ \cup K_{\overline{r}}^+)$,
    i.e., the set of attributes that certainly have
    zero prevalence in the equilibrium.

    \begin{itemize}
        \item $\boldsymbol{L^+ = \emptyset:}$
        In this case, the equation system
    can be simplified in two respects. First,
    we note that~$L^+ = \emptyset$
    implies that no equations in the form of~\cref{eq:theo:sapp-constant:stability:eig-eq:2} exist in the equation system.
    Second, we note that we only investigate 
    networks with a unique equilibrium,
    i.e., non-zero equilibrium prevalence is possible for only one 
    attribute on each path ($|K_r^+| = |K_{\overline{r}}^+| = 1$).
    Hence, if~$L^+ = \emptyset$,
    we know that ~$K_r^+ \cup K_{\overline{r}}^+$ covers both
    of the two attributes of the network, one on each path:
    \begin{align}
        &K_r^+ \cup K_{\overline{r}}^+ = N \times  K = \{(n(r), k(r)), (n(\overline{r}), k(\overline{r}))\}
    \end{align} where~$(n(r), k(r))$ is the single attribute
    with possibly non-zero equilibrium prevalence on path~$r$.
    These insights allow to reduce the
    equation system in~\cref{eq:theo:sapp-constant:stability:eig-eq:1} to 
    only two equations (No equations like \cref{eq:theo:sapp-constant:stability:eig-eq:2} exist):
    \begin{equation}
        -(\lambda + 1) x_r
        +  J^+_{\overline{r}} x_{\overline{r}}
            = 0 \hspace{50pt}
            -(\lambda + 1) x_{\overline{r}}
            +  J^+_{r} x_{r}
            = 0 
            \label{eq:theo:sapp-constant:stability:eig-eq:3}
    \end{equation}
    where we have abbreviated:
    \begin{equation}
        J^+_r = J^+_{I(n(r),k(r)),I(n(\overline{r}),k(\overline{r}))} \hspace{50pt}
        x_r =  x_{I(n(r),k(r))}
    \end{equation}
    Since the eigenvector~$\mathbf{x}$ in the current case
    only has the two entries~$x_r$ and~$x_{\overline{r}}$,
    we require~$\mathbf{x} = (x_r, x_{\overline{r}})^\top \neq \boldsymbol{0}$.
    We find that~$\lambda = -1$ is an eigenvalue of
    the system in~\cref{eq:theo:sapp-constant:stability:eig-eq:3}
    if and only if~$J_r^+ = 0$ or~$J_{\overline{r}}^+ = 0$, i.e., at least one
    of the two relevant Jacobian entries is zero.
    Since~$\lambda = -1$
    would preserve negative definiteness of~$\mathbf{J}^+$,
    we do not need to consider this case further.

    \item $\boldsymbol{L^+ \neq \emptyset:}$
    If~$L^+ \neq \emptyset$, the equation system
    contains equations
    in the form of~\cref{eq:theo:sapp-constant:stability:eig-eq:2}. 
    Then,~$\lambda = -1$
    may be an eigenvalue of~$\mathbf{J}^+$, which would
    preserve negative definiteness of~$\mathbf{J}^+$;
    hence, this case is not further considered.
    Conversely, if~$\lambda = -1$ is not a solution
    of the system, the equations in the form of~\cref{eq:theo:sapp-constant:stability:eig-eq:2} imply that~$x_{I(n,k)} = 0$ 
    for all~$(n,k) \in L^+$.
    This insight then again allows the simplification
    to the equation system in~\cref{eq:theo:sapp-constant:stability:eig-eq:3}.
    Crucially, since~$x_{I(n,k)} = 0$ for all $(n,k) \in L^+$,
    it must hold that~$(x_{r}, x_{\overline{r}})^\top \neq \boldsymbol{0}$
    such that~$\mathbf{x} \neq \boldsymbol{0}$, i.e., such that $\mathbf{x}$ 
    is a valid eigenvector.
    
    \end{itemize} 

\subsection{Solution of \cref{eq:theo:sapp-constant:stability:eig-eq:3}}

    In summary, we only need to consider the equation system in~\cref{eq:theo:sapp-constant:stability:eig-eq:3} and the case~$\lambda \neq -1$.
    Furthermore, we require~$(x_r, x_{\overline{r}})^\top \neq \boldsymbol{0}$.
    Without loss of generality, let~$r$ be the path with~$x_r \neq 0$.
    Then, we can perform the following transformation:
    \begin{align}
        -(\lambda + 1) x_r
        +  J^+_{\overline{r}} x_{\overline{r}}
            = 0 \implies& \lambda + 1 = J_{\overline{r}} \frac{x_{\overline{r}}}{x_r}\label{eq:theo:sapp-constant:stability:eig-eq:4}\\
        -(\lambda + 1) x_{\overline{r}}
        +  J^+_{r} x_{r}
            = 0 \implies& x_{\overline{r}} =\frac{J_{r}}{\lambda + 1} x_r
            \label{eq:theo:sapp-constant:stability:eig-eq:5}
    \end{align}
    Inserting~\cref{eq:theo:sapp-constant:stability:eig-eq:5} into
    \cref{eq:theo:sapp-constant:stability:eig-eq:4} yields
    a quadratic equation in~$\lambda$:
    \begin{equation}
        (\lambda + 1)^2 - J^+_r J^+_{\overline{r}} = 0 \implies
        \lambda_{1,2} = -1 \pm \sqrt{J^+_r J_{\overline{r}}^+}
        \label{eq:theo:sapp-constant:stability:eig-vals}
    \end{equation}

    If~$J_r^+ J_{\overline{r}}^+ = 0$, then~$\lambda_{1,2} = -1$
    produces a contradiction to the assumption~$\lambda \neq -1$,
    which implies that no eigenvalue~$\lambda \neq -1$ exists.

    If~$J_r^+ J_{\overline{r}}^+ \neq 0$, then
    we know that
    \begin{equation}
        \text{Re}(\lambda_{1,2}) < 0 \iff \text{Re}\left(\sqrt{J_r^+ J_{\overline{r}}^+}\right) < 1 \iff J_r^+ J_{\overline{r}}^+ < 1.
        \label{eq:theo:sapp-constant:stability:stability-condition}
    \end{equation}

\subsection{Bounding of~$\lambda_{1,2}$}
    
    To verify that the condition in~\cref{eq:theo:sapp-constant:stability:stability-condition} always holds,
    we first find~$J_{r}^+$ for any path~$r$:
    \begin{equation}
        J_r^+ = \begin{cases}
            \frac{\alpha_{n(\overline{r})k(\overline{r})}}{\alpha_{n(r)k(r)}}\left(\frac{\psi_r \sqrt{d}}{2\sqrt{1 + v_{\overline{r}}^+}} - 1 \right) &
            \text{if } \hat{a}_{n(r)k(r)}^{*}\left(\mathbf{A}^+_{-n(r)k(r)}\right) \geq 0,\\
            0 & \text{otherwise.}
        \end{cases}
        \label{eq:theo:sapp-constant:stability:jacobian-entry}
    \end{equation}
    Given~$J_r^+ J_{\overline{r}}^+ \neq 0$, stability requires:
    \begin{align}
        &J_r^+ J_{\overline{r}}^+ \overset{\text{(\ref{eq:theo:sapp-constant:stability:jacobian-entry})}}{=} \left(\frac{\psi_r \sqrt{d}}{2\sqrt{1 + v_{\overline{r}}^+}} - 1 \right)\left(\frac{\psi_{\overline{r}} \sqrt{d}}{2\sqrt{1 + v_{r}^+}} - 1 \right) \overset{\text{(\ref{eq:theo:sapp-constant:stability:stability-condition})}}{<} 1\\
        \iff &\psi_r \psi_{\overline{r}} d - 2 \sqrt{d} \left(\psi_r \sqrt{1+v_{\overline{r}}^+}
        + \psi_{\overline{r}} \sqrt{1+v_{r}^+}\right) < 0
        \label{eq:theo:sapp-constant:stability:inequality}
    \end{align}
    Moreover, we know that the restricted equilibrium valuation~$v_r^+$ for
    each path~$r$ corresponds to the unrestricted equilibrium
    valuation~$\hat{v}_r^+$ from \cref{thm:heterogeneous:equilibrium:two}:
    \begin{equation}
        \begin{split}
             &\forall r \in R.\ J_r^+ J_{\overline{r}}^+ \neq 0 \implies \forall r \in R.\ \hat{a}_{n(r)k(r)}^{*}(\mathbf{A}^+_{-n(r)k(r)}) \geq 0\\
        \implies &\forall r \in R.\ \hat{v}_r^+ \geq \alpha_{r0} \implies
        \forall r \in R.\ 
        v_r^+ = \hat{v}_r^+.
        \label{eq:theo:sapp-constant:stability:restricted-unrestricted}
        \end{split}
    \end{equation}
    Hence, we can expand (symmetrically for~$\psi_{\overline{r}} \sqrt{1+v_{\overline{r}}^+}$):
    \begin{equation}
        \begin{split}
            \psi_{r} \sqrt{1 + v_{r}^+} &\overset{\text{(\ref{eq:theo:sapp-constant:stability:restricted-unrestricted})}}{=}\ \psi_{r} \sqrt{1 + \hat{v}_{r}^+}\\
            &\overset{\text{Th\ref{thm:heterogeneous:equilibrium:two}}}{=}\ \psi_{r} \sqrt{\frac{\psi_r^3\psi_{\overline{r}}}{(\psi_r^2 + \psi_{\overline{r}}^2)^2} \left(\sqrt{d\left(\psi_r^{2}+\psi_{\overline{r}}^{2}\right)+\frac{1}{4}\psi_r^{2}\psi_{\overline{r}}^{2}d^{2}}+\frac{d}{2}\psi_r\psi_{\overline{r}}\right)+\frac{\psi_{r}^{2}}{\psi_r^{2}+\psi_{\overline{r}}^{2}}}\\
            &=\ 
            \psi_{r}^2 \sqrt{\frac{\psi_r\psi_{\overline{r}}}{(\psi_r^2 + \psi_{\overline{r}}^2)^2} \left(\sqrt{d\left(\psi_r^{2}+\psi_{\overline{r}}^{2}\right)+\frac{1}{4}\psi_r^{2}\psi_{\overline{r}}^{2}d^{2}}+\frac{d}{2}\psi_r\psi_{\overline{r}}\right)+\frac{1}{\psi_r^{2}+\psi_{\overline{r}}^{2}}}
        \end{split}
        \label{eq:theo:sapp-constant:stability:expansion}
    \end{equation}

    We use this equality to rewrite the inequality in \cref{eq:theo:sapp-constant:stability:inequality}:
    \begin{equation}
        \begin{split}
            &\psi_r \psi_{\overline{r}} d - 2 \sqrt{d} \left(\psi_r \sqrt{1+v_{\overline{r}}^+}
        + \psi_{\overline{r}} \sqrt{1+v_{r}^+}\right)\\
        \overset{\text{(\ref{eq:theo:sapp-constant:stability:expansion})}}{=}\ &\psi_r \psi_{\overline{r}} d - 2 \sqrt{d} \left(\psi_r^2 + \psi_{\overline{r}}^2\right) \sqrt{\frac{\psi_r\psi_{\overline{r}}}{(\psi_r^2 + \psi_{\overline{r}}^2)^2} \left(\sqrt{d\left(\psi_r^{2}+\psi_{\overline{r}}^{2}\right)+\frac{1}{4}\psi_r^{2}\psi_{\overline{r}}^{2}d^{2}}+\frac{d}{2}\psi_r\psi_{\overline{r}}\right)+\frac{1}{\psi_r^{2}+\psi_{\overline{r}}^{2}}}\\
        =\ &\psi_r \psi_{\overline{r}} d 
        - \sqrt{2d^2\psi_r^2\psi_{\overline{r}}^2 +   4d \left(\psi_r^2 + \psi_{\overline{r}}^2\right)^2 \left(\frac{\psi_r\psi_{\overline{r}}}{(\psi_r^2 + \psi_{\overline{r}}^2)^2}\sqrt{d\left(\psi_r^{2}+\psi_{\overline{r}}^{2}\right)+\frac{1}{4}\psi_r^{2}\psi_{\overline{r}}^{2}d^{2}}+\frac{1}{\psi_r^{2}+\psi_{\overline{r}}^{2}}\right) }\\
        <\ &\left(1 - \sqrt{2}\right) d \psi_r \psi_{\overline{r}} < 0
        \end{split}
    \end{equation}
    The last upper bound holds because we can exclude~$\psi_r = 0$
    for any path~$r$ in the current case~$v_r^+ = \hat{v}_r^+$,
    as~$\psi_r = 0$ produces the contradiction~$v_r^+ < v_r^+$:
    \begin{equation}
        \begin{split}
            &v_r^+ \overset{\text{(\ref{eq:theo:sapp-constant:stability:restricted-unrestricted})}}{=} \hat{v}_r^+ \overset{\text{(\ref{eq:theo:sapp-constant:competition:unrestricted})}}{=} \psi_r \sqrt{d} \sqrt{1+\hat{v}^+_{\overline{r}}} - (1 + \hat{v}_{\overline{r}}^+) \overset{\psi_r = 0}{=}  - (1 + \hat{v}_{\overline{r}}^+) \overset{\text{(\ref{eq:theo:sapp-constant:stability:restricted-unrestricted})}}{=} - (1 + v_{\overline{r}}^+)\\ \overset{\text{Th\ref{thm:heterogeneous:equilibrium:two}}}{\leq} &- (1+\alpha_{\overline{r}0}) \overset{\alpha_{\overline{r}0} \geq 0}{<} 0
        \overset{\alpha_{r0} \geq 0}{\leq} \alpha_{r0} \overset{\text{Th\ref{thm:heterogeneous:equilibrium:two}}}{\leq} v_r^+.
        \end{split}
    \end{equation}

    In summary, we now have shown that~$J_r^+ J_{\overline{r}}^+ < 1$,
    which ensures a negative real part~$\text{Re}(\lambda_{1,2}) < 0$ (\cref{eq:theo:sapp-constant:stability:stability-condition})
    of the eigenvalues~$\lambda_{1,2}$ from~\cref{eq:theo:sapp-constant:stability:eig-vals}, 
    and thus confirms that~$\mathbf{J}^+$ is negative definite.
    Since~$\mathbf{J}^+$ is negative definite,
    the equilibrium~$\mathbf{A}^+$ from~\cref{thm:heterogeneous:equilibrium:two}
    is asymptotically stable with respect to the process
    in~\cref{eq:theo:sapp-constant:process},
    which concludes the proof.

\section{Proof of~\cref{thm:theo:sapp-constant:competition:good}}
\label{sec:app:proofs:heterogeneous--competition-beneficial}

    For the competitive network~$\mathcal{N}_4$, the unrestricted equilibrium network valuation~$\hat{V}^+$ is
    \begin{equation}
        \hat{V}^+(\mathcal{N}_4) = \hat{v}_r^+ + \hat{v}_{\overline{r}}^+ =
        \frac{\psi_r\psi_{\overline{r}}}{\psi_r^{2}+\psi_{\overline{r}}^{2}}\left(\sqrt{d\left(\psi_r^{2}+\psi_{\overline{r}}^{2}\right)+\frac{1}{4}\psi_r^{2}\psi_{\overline{r}}^{2}d^{2}}+\frac{d}{2}\psi_r\psi_{\overline{r}}\right)-1,
    \end{equation} where~$\hat{v}_r^+$ and~$\hat{v}_{\overline{r}}^+$ are as in~\cref{thm:heterogeneous:equilibrium:two}.
    The restricted equilibrium network valuation~$V^+(\mathcal{N}_4)$ is equal to~$\hat{V}^+(\mathcal{N}_4)$
    if~$\hat{v}_r^+ \geq \alpha_{r0}$ and~$\hat{v}_{\overline{r}}^+ \geq \alpha_{\overline{r}0}$.
    These unrestricted equilibrium path valuations are monotonically increasing in the demand limit~$d$
    (cf.~\cref{thm:heterogeneous:equilibrium:two}). Hence, if~$d$ is high enough, it holds 
    that~$V^+(\mathcal{N}_4) =  \hat{V}^+(\mathcal{N}_4)$.
    
    For the competition-free network~$\mathcal{N}_3$, the unrestricted equilibrium network valuation~$\hat{V}^+$ is
    \begin{equation}
        \hat{V}^+(\mathcal{N}_3) = \psi_r \sqrt{d_r} - 1 + \psi_{\overline{r}} \sqrt{d_{\overline{r}}} - 1.
    \end{equation}
    Among all demand distributions~$(d_r, d_{\overline{r}})$ with $d_r + d_{\overline{r}} = d$,
    the demand distribution maximizing~$\hat{V}^+(\mathcal{N}_3)$
    can be found as follows:
    \begin{equation}
        \frac{\partial}{\partial d_r} \left(\psi_r \sqrt{d_r} + \psi_{\overline{r}} \sqrt{d-d_r} - 2\right) = 0
        \iff d_r = \frac{\psi_r^2}{\psi_r^2+\psi_{\overline{r}}^2} d,
    \end{equation} where the maximum character of this value is ensured by a consistently
    non-positive second derivative of~$\hat{V}^+(\mathcal{N}_3)$ in~$d_r$.
    In the following, we thus consider only the maximum unrestricted equilibrium network valuation~$\hat{V}^+(\mathcal{N}_3)$:
    \begin{equation}
        \hat{V}^+(\mathcal{N}_3) = \psi_r \sqrt{\frac{\psi_r^2}{\psi_r^2+\psi_{\overline{r}}^2} d} + \psi_{\overline{r}} \sqrt{\frac{\psi_{\overline{r}}^2}{\psi_r^2+\psi_{\overline{r}}^2} d} - 2
        = \sqrt{ d (\psi_r^2+\psi_{\overline{r}}^2) } - 2.
    \end{equation}
    Again, for~$V^+(\mathcal{N}_3) =  \hat{V}^+(\mathcal{N}_3)$, ~$d$ must be high enough
    such that
    \begin{equation}
        \psi_r \sqrt{\frac{\psi_r^2}{\psi_r^2+\psi_{\overline{r}}^2} d} -1 \geq \alpha_{r0} \hspace{15pt} \text{and} \hspace{15pt}  
        \psi_{\overline{r}} \sqrt{\frac{\psi_{\overline{r}}^2}{\psi_r^2+\psi_{\overline{r}}^2} d} -1 \geq \alpha_{\overline{r}0}.
    \end{equation}

    If~$d$ is high enough, the difference of the equilibrium network valuations is thus:
    \begin{equation}
         \Delta V^+ = V^+(\mathcal{N}_4) - V^+(\mathcal{N}_3) = 
        \frac{\psi_r\psi_{\overline{r}}}{\psi_r^{2}+\psi_{\overline{r}}^{2}}\left(\sqrt{d\left(\psi_r^{2}+\psi_{\overline{r}}^{2}\right)+\frac{1}{4}\psi_r^{2}\psi_{\overline{r}}^{2}d^{2}}+\frac{d}{2}\psi_r\psi_{\overline{r}}\right) - \sqrt{ d (\psi_r^2+\psi_{\overline{r}}^2) } + 1
    \end{equation}
    Clearly, this difference is eventually positive when increasing the demand limit~$d$, meaning that
     $V^+(\mathcal{N}_4)$ exceeds $V^+(\mathcal{N}_3)$ for high enough~$d$:
     \begin{equation}
         \lim_{d\rightarrow\infty} \Delta V^+ = \infty.
     \end{equation}
     This last insight proves the theorem.

\section{Proof of~\cref{thm:theo:sapp-constant:competition:bad}}
\label{sec:app:proofs:heterogeneous--competition-bad}

    The following proof is constructive, i.e., we demonstrate how to choose~$(\psi_r, \psi_{\overline{r}})$
    and $(\alpha_{r0}, \alpha_{\overline{r}0})$ such that~$V^{+}(\mathcal{N}_4) < V^{+}(\mathcal{N}_3)$ holds given
    demand distribution~$(d_r, d_{\overline{r}})$.
    In this construction, the goal is to create a scenario where the competitive network~$\mathcal{N}_4$
    will be at minimum valuation~$\alpha_{r0} + \alpha_{\overline{r}0}$, but the competition-free network
    has a path~$r$ with an equilibrium valuation~$v_r^+$ exceeding the minimum path valuation~$\alpha_{r0}$.
    
    Regarding the path-characteristic ratios~$(\psi_r, \psi_{\overline{r}})$,
    our first step consists of choosing the ratios such that the competitive network
    is at minimum valuation, i.e., such that~$v_r^+ = \alpha_{r0}$
    and~$v_{\overline{r}}^+ = \alpha_{\overline{r}0}$.
    To do so, we first determine~$\psi_{\overline{r}}$ such 
    that~$\hat{v}_{\overline{r}}^+ \leq \alpha_{\overline{r}0}$ for all~$\psi_r$,
    which is achieved by~$\psi_{\overline{r}} = 0$:
    \begin{equation}
       \lim_{\psi_{\overline{r}} \rightarrow 0}  \hat{v}_{\overline{r}}^+ 
       \overset{\text{Th\ref{thm:heterogeneous:equilibrium:two}}}{=} -1 \overset{\alpha_{\overline{r}0} \geq 0}{<} \alpha_{\overline{r}0}.
    \end{equation}
    
    Having selected~$\psi_{\overline{r}}$ such that~$\hat{v}_{\overline{r}}^+ \leq \alpha_{\overline{r}0}$,
    it holds that~$v_{\overline{r}}^+ = \alpha_{\overline{r}0}$ by~\cref{thm:heterogeneous:equilibrium:two}. As a result, the
    equilibrium path valuation for path~$r$ in the competitive network~$\mathcal{N}_4$
    is:
    \begin{equation}
        v_r^+(\mathcal{N}_4) \overset{\text{Th\ref{thm:heterogeneous:equilibrium:two}}}{=} \max\left(\alpha_{r0},\ \psi_r\sqrt{d}\sqrt{1+\alpha_{\overline{r}0}} - (1+\alpha_{\overline{r}0})\right).
    \end{equation}
    To ensure that~$v_r^+$ is minimal (i.e., equals $\alpha_{r0}$), the following condition must hold:
    \begin{equation}
        \alpha_{r0} \geq \psi_r\sqrt{d}\sqrt{1+\alpha_{\overline{r}0}} - (1+\alpha_{\overline{r}0})
        \iff \psi_r \leq \frac{1+\alpha_{r0} + \alpha_{\overline{r}0}}{\sqrt{d}\sqrt{1+\alpha_{\overline{r}0}}}
        \label{eq:theo:sapp-constant:competition:bad:1}
    \end{equation}
    If~$\psi_r $ is chosen according to~\cref{eq:theo:sapp-constant:competition:bad:1}, 
    the equilibrium network valuation in the competition-free network is minimal, 
    i.e.,~$V^+(\mathcal{N}_4) = \alpha_{r0} + \alpha_{\overline{r}0}$. 
    
    To let~$V^+(\mathcal{N}_3)$ of the competition-free network exceed~$V^+(\mathcal{N}_4)$ of the competitive network,
    we further need to choose~$\psi_r$ such that~$\hat{v}_r^+(\mathcal{N}_3) > \alpha_{r0}$.
    This condition can be transformed in the following fashion:
    \begin{equation}
        \psi_{r}\sqrt{d_r} - 1 > \alpha_{r0} \iff \psi_r > \frac{1+\alpha_{r0}}{\sqrt{d_r}}
    \end{equation}
    
    To allow a selection of~$\psi_r$ that achieves~$v_r^+(\mathcal{N}_4) = \alpha_{r0}$
    but~$v_r^+(\mathcal{N}_3) > \alpha_{r0}$, it must thus hold that
    \begin{equation}
        \frac{1+\alpha_{r0}}{\sqrt{d_r}} < \frac{1+\alpha_{r0} + \alpha_{\overline{r}0}}{\sqrt{d}\sqrt{1+\alpha_{\overline{r}0}}}
        \iff \alpha_{r0} < \frac{\sqrt{d_r}\alpha_{\overline{r}0}}{\sqrt{d}\sqrt{1+\alpha_{\overline{r}0}}-\sqrt{d_{r}}}-1,
    \end{equation}
    
    This condition always holds when choosing~$\alpha_{r0} = 0$ and~$\alpha_{\overline{r}0} > d_{\overline{r}}/d_{r}$:
    \begin{equation}
        \begin{split}
            &\frac{
            \sqrt{d_r}
            \alpha_{\overline{r}0}
        }{
            \sqrt{d}
            \sqrt{1+\alpha_{\overline{r}0}}
            -\sqrt{d_{r}}
        } - 1
        \overset{\alpha_{\overline{r}0} > d_{\overline{r}}/d_r}{>}
        \frac{
            \sqrt{d_r}
            \frac{
                d_{\overline{r}}
            }{
                d_r}
        }{
            \sqrt{d}
            \sqrt{1 +
                \frac{
                    d_{\overline{r}}
                }{
                    d_r
                }
            }
            -
            \sqrt{d_{r}}
        } - 1
        \overset{d_r + d_{\overline{r}} = d}{=}
        \frac{
            \sqrt{d_r}
            \frac{
                d_{\overline{r}}
            }{
                d_r}
        }{
            \sqrt{d}
            \sqrt{
                \frac{
                    d
                }{
                    d_r
                }
            }
            -
            \sqrt{d_{r}}
        } - 1
        \\
        \overset{\frac{\sqrt{d_r}}{d_r} = \frac{1}{\sqrt{d_r}}}{=}
         &=
         \frac{
            \frac{
                d_{\overline{r}}
            }{
                \sqrt{d_r}}
        }{
            \frac{
                    d
            }{
                    \sqrt{d_r}
            }
            -
            \frac{
                    d_r
            }{
                    \sqrt{d_r}
            }
        } - 1
         \overset{d_r + d_{\overline{r}} = d}{=} 
         \frac{
            \frac{
                d_{\overline{r}}
            }{
                \sqrt{d_r}}
        }{
            \frac{
                    d_{\overline{r}}
            }{
                    \sqrt{d_r}
            }
        } - 1 = 0 \overset{\alpha_{r0} = 0}{=} \alpha_{r0}
        \end{split}
    \end{equation}
    
    Hence, $(\psi_r, \psi_{\overline{r}})$ and $(\alpha_{r0}, \alpha_{\overline{r}0})$ can  be chosen
    such that~$V^+(\mathcal{N}_3) > V^+(\mathcal{N}_4)$, which concludes the proof.


\end{document}